# Skyrmions in Magnetic Multilayers

Wanjun Jiang[1,2,*]
[1]State Key Laboratory of Low-Dimensional Quantum Physics,
and Department of Physics, Tsinghua University, Beijing, China 100084
[2]Collaborative Innovation Center of Quantum Matter, Beijing, China 100084
*Correspondence should be addressed to: jiang_lab@tsinghua.edu.cn

Gong Chen* and Kai Liu
Department of Physics, University of California, Davis, CA, USA, 95616
*Correspondence should be addressed to: gchenncem@gmail.com

Jiadong Zang*
Department of Physics, University of New Hampshire, Durham, NH, USA, 03824
*Correspondence should be addressed to: Jiadong.zang@unh.edu

Suzanne G. E. te Velthuis and Axel Hoffmann*
Materials Science Division, Argonne National Laboratory, Lemont, IL, USA, 60439
*Correspondence should be addressed to: hoffmann@anl.gov

Abstract

Symmetry breaking together with strong spin-orbit interaction give rise to many exciting phenomena within condensed matter physics. A recent example is the existence of chiral spin textures, which are observed in magnetic systems lacking inversion symmetry. These chiral spin textures, including domain walls and magnetic skyrmions, are both fundamentally interesting and technologically promising. For example, they can be driven very efficiently by electrical currents, and exhibit many new physical properties determined by their real-space topological characteristics. Depending on the details of the competing interactions, these spin textures exist in different parameter spaces. However, the governing mechanism underlying their physical behaviors remain essentially the same. In this review article, the fundamental topological physics underlying these chiral spin textures, the key factors for materials optimization, and current developments and future challenges will be discussed. In the end, a few promising directions that will advance the development of skyrmion based spintronics will be highlighted.

This review article is organized as follows:
1. Topological physics of magnetic skyrmions
    1.1 Origin of spin topology
    1.2 Real space topological physics
    1.3 Topological distinction of bubble-like spin textures

**Introduction**

As one of the oldest topics in physics, magnetism has a long and illustrative history that is full of remarkable achievements [1-3]. Many fundamental discoveries of magnetic phenomena have dramatically changed human life, as evidenced by the invention of the compass, permanent magnets, electromagnetic power machines, magnetic recording, magnetic random access memory (MRAM) and numerous other innovative applications. Traditionally, a main focus of studying magnetic materials was to characterize their magnetic responses to external magnetic field and temperature. Common figures among others of merit for describing magnetic materials include Curie temperature, coercivity, saturation magnetization, anisotropy, and universality class of phase transition [4].

Recent advancements in nanotechnology resulted in concomitant progress in magnetism, with two developments being particularly influential in nanomagnetic systems: controlling magnets via electric (field/current) excitations [5-8] and the discovery of topological spin textures [9]. Electric control of magnetism is made possible by utilizing the coupling between electron spin and its orbital motion. This can be understood based on the electromagnetic duality in special relativity, namely the interconversion between magnetic field ($B$) and electric field ($E$): $B = (v_e \times E)/c^2$, where $v_e$ is the electron velocity. On the other hand, the relativistic coupling between spin and its orbit motion, the spin orbit coupling (SOC), can be written as the following:

$$H_{SO} = \frac{e^2}{2m_e^2 c^2 r^3} \sigma \cdot \mathcal{L} \qquad (1)$$

where $\sigma$ is the electron spin angular momentum and $\mathcal{L}$ is the orbital angular momentum, $c$ is the speed of light, $e$ is the unit electronic charge, $r$ is the radius of the orbit, and $m_e$ is the mass of electron. It thus demonstrates an attractive route by which the motion of the electron, controlled by electric fields, can produce an effective magnetic field acting on the spins. Many emergent phenomena, such as multiferroics properties [10], spin Hall effect (SHE) [11-13], perpendicular magnetic anisotropy (PMA) [14], spin transfer torque [15-17], spin-orbit torque (SOT) [18], Rashba-Edelstein effect [19], etc., are all closely associated with the SOC, and lead to various approaches of electric control of magnetism that have significantly advanced spintronics from both fundamental research and applied perspectives.

In magnetic materials, due to the competition between different energy contributions such as magnetic anisotropies, dipole interactions, or exchange interactions, there naturally exist nanoscale noncollinear spin textures such as magnetic domain walls and magnetic vortices [20, 21]. In particular, two different types of magnetic domain wall can be observed, one is called a Néel wall, where spins inside the wall rotate as cycloidal spirals within the domain wall region, and the other is called a Bloch wall, where spin rotates as helical spirals [21]. Manipulation of these nanoscale noncollinear spin textures by electric currents is one of the major developments in spintronics due to their potential applications as information carrier in low-power nanoelectronics and data storage. This is enabled by the exchange coupling between spins of conduction electrons and localized spin textures that results in a transfer of spin angular momentum, the so-called spin transfer torque (STT) mechanism [15, 16, 22-24]. Under the influence of STT, dynamics of spin textures, such as magnetic domain walls, can be driven by spin-polarized electric currents, providing for an efficient approach for electric control and manipulation of local magnetizations. Unfortunately, the threshold current required for domain wall motion is typically as high as $10^6 A/cm^2$, and is hard to be reduced because of intrinsic pinning effects that are introduced by inevitable material impurities and defects [25, 26].

It has been theoretically suggested that localized/stable spin textures could solve the intrinsic pinning issue by avoiding local structural imperfections due to the presence of Magnus forces from local pining sites, which results in a significant reduction of the threshold for the depinning current. This context leads to the second aspect of contemporary magnetism – topological spin textures [27]. However, looking for stable and mobile spin textures is highly nontrivial. For example, magnetic vortices are well-known spin textures, but they are only stabilized in geometrically confined magnetic nanostructures such as nanodisks and therefore immobile simply because of their confinement [2, 28].

Furthermore, the necessary condition for the existence of stable localized spin texture in conventional ferromagnets is ruled out by Derrick's scaling argument [29]. Suppose $\mathbf{m}(\mathbf{r})$ is a local solution that minimizes the magnetic energy $E = \int [(\nabla \mathbf{m})^2 + f(\mathbf{m})]d^3r \equiv I_1 + I_2$, where the first term is the Heisenberg exchange interaction, and $f(\mathbf{m})$ in the second term contains the magnetic anisotropies. Under a scaling transformation $\mathbf{m}(\mathbf{r}) \rightarrow \mathbf{m}(\lambda \mathbf{r})$, where $\lambda$ is an arbitrary constant and which is minimized at $\lambda = 1$, the energy changes as $E(\lambda) = I_1/\lambda + I_2/\lambda^3$, where $\lambda$ is an arbitrary constant. Stability of the solution (when $\lambda = 1$) thus requires $dE/d\lambda = 0$ and $d^2E/d^2\lambda > 0$. The existence of a topological excitation $I_1 = \int (\nabla \mathbf{m})^2 d^3r < 0$ is clearly invalid.

The way out of the Derrick's criteria is to introduce a spin exchange interaction term that involves spatial derivatives other than second order. This can be easily done by including the first order derivative, which explicitly requires the associated material system exhibiting a broken (spatial) inversion symmetry [because $dE/dx \neq dE/d(-x)$]. This circumstance is, however, absent in most ferromagnets, but not in non-centrosymmetric magnets and/or interfacially asymmetric multilayers [30-33]. Microscopically, the broken (space) inversion symmetry in these material systems leads to a noncollinear Dzyaloshinskii-Moriya interaction (DMI) that can be written as [33-37]:

$$E_{DMI} = -\mathbf{D}_{ij} \cdot (\mathbf{S}_i \times \mathbf{S}_j) \quad (2)$$

between neighboring atomic spins $\mathbf{S}_i$ and $\mathbf{S}_j$. The DMI vector $\mathbf{D}_{ij}$ is dictated by the crystal symmetry. Under an inversion operation where the midpoint of two sites, $\mathbf{S}_i$ and $\mathbf{S}_j$ are exchanged, the energy consequently flips sign, which demonstrates the broken inversion symmetry. In the continuous limit, $\mathbf{S}_j = \mathbf{S}_i + (\mathbf{r}_{ji} \cdot \nabla)\mathbf{S}_i + O(r_{ji}^2)$. Substitution into *Eq.* (2) leads $E_{DM} = -\mathbf{D}_{ij} \cdot \mathbf{S}_i \times (\mathbf{r}_{ji} \cdot \nabla)\mathbf{S}_i$ which only has the first order derivative as desired. The energy of the DMI is minimized if two spins are lying in the plane normal to $\mathbf{D}_{ij}$ and perpendicular to each other, in contrast to the Heisenberg exchange, which favors parallel/antiparallel spin configurations. As a result, neighboring spins thus extend a finite angle with respect to each other. This is one of the origins of the noncollinear stable/localized spin textures including helical, conical and Bloch-type skyrmion phases in non-centrosymmetric bulk magnets [38], and also spin spiral, chiral magnetic domain walls, and Néel-type skyrmions in magnetic multilayers [39].

On the other hand, the strength of DMI can be approximately linked to the strength of SOC which in first approximation is expected to scale with $Z^4$ ($Z$ is the atomic number). Note however that the strength of SOC in the transition metals is far smaller than the $Z^4$ relation and depends on the detailed band structures [40]. Nevertheless, a large DMI is typically found in materials with heavy element, where the strength of SOC is enhanced. In magnetic thin films, beyond DMI, large SOC also gives rise to many interesting phenomena such as SHE, SOT, PMA, etc. In particular, multilayered material systems with a PMA and interfacial DMI will be discussed in this review article, emphasizing on the stabilization, manipulation and implementation of room-temperature chiral magnetic skyrmions [41-44].

## 1. Topological physics of magnetic skyrmions

The emergence of magnetic skyrmion, a new type of localized stable spin texture, is a direct consequence of evading Derrick's argument by including an antisymmetric DMI term. The skyrmion spin texture is named after the British physicist Tony Skyrme who constructed a topological configuration of a 4-dimensional vector field in 3+1 space-time [45]. Although the existence of the original skyrmion-like objects is still under debate in high energy physics, its derivative in spin space, has attracted considerable attention in the condensed matter physics community. For example, skyrmion spin textures were theoretically proposed to explain the quantum Hall effect [46-48] and very recently, experimentally realized in the inversion asymmetric bulk magnets [31, 49-52] and magnetic multilayers [41-44, 53].

Unlike in well-studied ferromagnets such as Ni, Fe, Co and their alloys, where the symmetric Heisenberg exchange interaction governs the underlying magnetism/physics, in the inversion asymmetric magnets, the presence of an antisymmetric DMI component favors the formation of swirling spin textures, as schematically shown in Fig. 1A. These swirling spin textures were first discovered by identifying a hexagonal symmetry from the reciprocal space in the so-called "A" phase of bulk MnSi single crystal, through a small angle neutron scattering (SANS) experiment [54]. Similar to the Bloch type magnetic domain wall in ferromagnetic materials, inside the wall (the region where magnetization vector transits from being upward to downward), the spin structure rotates like a helical spiral, thus this type of spin structure in chiral bulk magnets was named Bloch-type skyrmion [38, 54-57]. Its existence was later directly observed by utilizing a spin sensitive real-space imaging technique, Lorentz Transmission Electron Microscopy (Lorentz TEM), as shown in Fig. 1B [38, 55, 56, 58].

### 1.1 Origin of the spin topology

A topology class is mathematically characterized by homotopy theory, which describes the equivalence between two maps under smooth deformation. In the context of noncollinear spin textures, these maps are the spatial distribution of magnetic moments. Each topological class is characterized by an integer $Q$ called skyrmion number, which is the number of times spins wind around the unit sphere [59]. This can also be done by wrapping the spin unit vector ($\boldsymbol{m}$) around a unit sphere, as shown in Fig. 1C, where the nontrivial topological number $Q$ was defined as follows [57]:

$$Q = {1}/{4\pi} \int \boldsymbol{m} \cdot (\partial_x \boldsymbol{m} \times \partial_y \boldsymbol{m})\, dxdy \qquad (3)$$

The skyrmion number (or topological charge) $Q$ of each configuration also characterizes the number of magnetic monopoles therein. The corresponding spin configuration on the unit sphere is similar to the magnetic vortex and therefore, this type of skyrmion in chiral B20 bulk magnets is also named a vortex-like skyrmion. As a matter of fact, magnetic vortices in magnetic nanostructures have half of the topological skyrmion number $Q = \pm\frac{1}{2}$. This can be determined again by projecting each individual spin of magnetic vortices onto the unit sphere which occupies (upper/lower) half of the unit sphere that correspondingly yields half of the skyrmion number. As a result, the skyrmion number/topological charge can be used to identify the topological distinction of different types of spin texture.

Robust magnetic skyrmions were first experimentally observed in bulk MnSi single crystals that lack inversion symmetry due to its cubic B20 crystal structure (space group $P2_13$) [38, 54]. In contrast to conventional ferromagnets (Fe, Co, Ni and their alloys), in which collinear Heisenberg exchange interaction dictates ferromagnetism (first term on the right side of *Eq*. 4), noncentro-symmetric B20 compounds contain an additional non-collinear DMI component (second term on the right side of *Eq*. 4). This DMI term competes with the Heisenberg exchange interaction and Zeeman energy from the applied external magnetic field **B** (third term on the right side of *Eq*. 4). As a result, chiral magnets accommodate various complex spin states, including helical, conical, skyrmion, polarized ferromagnetic, and disordered paramagnetic states. The Hamiltonian of B20 magnets can be expressed as follows [38, 54, 57, 59]:

$$H = \sum_{<i,j>} -J\mathbf{S}_i \cdot \mathbf{S}_j + \mathbf{D}_{ij} \cdot (\mathbf{S}_i \times \mathbf{S}_j) - \sum_i g\mu_B \mathbf{B}\mathbf{S}_i \qquad (4)$$

where $J$ is the strength of collinear Heisenberg exchange interaction between the neighboring atomic spins, $\mathbf{D}_{ij}$ is the anisotropic DMI vector, $g$ is the Landé factor, and $\mu_B$ is the Bohr magneton.

The competition between these energy terms results in a complex phase diagram as a function of temperature and magnetic field to be established, shown in Fig. 2A. It is clear to see that the regime where skyrmions can exist is very narrow (labeled as the "A" phase). Note that "A" stands for anomalous. It survives in the temperature range of 27 – 30 K and at a high magnetic field around 0.2 T, beyond which magnetic skyrmions transform into other complex magnetic states. While the skyrmion phase in most chiral magnets appears at low temperature, it is noted that progresses have recently been made in extending bulk skyrmion materials to ≈ 280 K (in FeGe [58, 60, 61]) and even above/at room temperature (in $\beta$-Mn-type Co-Zn-Mn alloys) [62, 63].

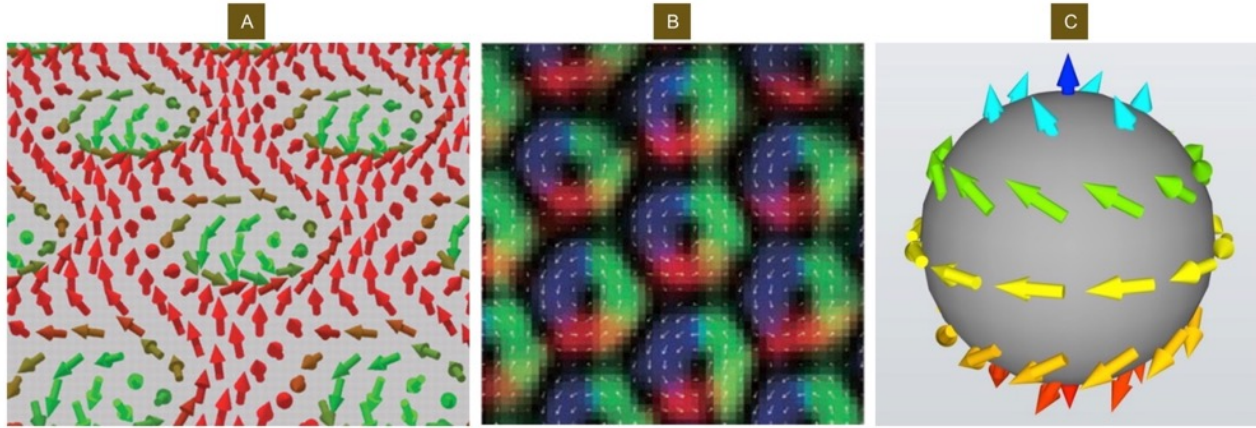

Figure 1. Illustration of skyrmion spin textures. (A) is a schematic illustration of skyrmion spin textures in a B20 bulk magnet. Reproduced with permission [54]. Copyright 2009, AAAS. Note that the original schematic plot was mirrored to make it consistent with the experimental observation shown in (B), which was acquired by using a Lorentz TEM with the superimposed spin orientation. Reproduced with permission [38]. Copyright 2010, Nature Publishing Group. (C) is the topological presentation of a magnetic skyrmion by wrapping each individual spin onto a unit sphere, the solid angle contained enables the topological charge $Q = -1$ to be determined.

## 1.2 Real space topological physics

As a result of their spin topological properties, magnetic skyrmions provide a unique platform to study many intriguing real-space topological transport phenomena that could bridge topology and magnetism together [38, 52, 54, 57, 64-66], as shown in Figs. 2B and 2C. Specifically, electric current provides an efficient avenue for driving the motion of topological skyrmions in metallic skyrmion materials. This electrically driven motion is facilitated by transfer of the spin angular momentum [17, 52, 58, 67]. The motion occurs at a very low current density (on the order of $10^2$ A/cm$^2$), indicative of negligible pinning effects from material imperfection in single crystalline samples [68]. Note that this current density is about 4 orders smaller than current densities need for electrically driven domain wall motion in conventional Permalloy (NiFe) nanostructures [26]. Because of its small size and efficient electrical manipulation, magnetic skyrmions thus also attract considerable attention for possible applications in which the presence/absence of skyrmions as the binary information digits (1/0) can be used in future memory devices such as skyrmion racetrack memory [32].

In a metallic skyrmion material, an applied electric current generates STT between conduction electrons and local magnetic moments due to the exchange of spin angular momenta between them [22, 23]. In contrast, for magnetic multilayers with interfacial DMI that will be focused on later, the electrical current passing through the heavy metal layer injects a perpendicular spin current into the adjacent magnetic thin layer via the spin Hall effect originating from the strong spin-orbit interaction [12, 13]. This is also known as spin-orbit torque and has been demonstrated as an efficient avenue for controlling magnetization dynamics [69, 70]. Both of these two effects give rise to a collective motion of magnetic skyrmions. The detailed magnetization dynamics ($m$) can be studied by using a generalized Landau-Lifshitz-Gilbert (LLG) equation:

$$\dot{\bm{m}} = -\gamma \bm{m} \times \bm{H}_{eff} + \alpha \bm{m} \times \dot{\bm{m}} + \frac{g\mu_B}{2eM_s}[(\bm{j}_1 \cdot \nabla)\bm{m} + \frac{\theta_{SH}}{t_f}\bm{m} \times (\bm{m} \times (\hat{z} \times \bm{j}_e))] \tag{5}$$

where $\gamma$ is the gyromagnetic ratio, and $\alpha$ is the magnetic damping parameter. $\bm{H}_{eff} = -\partial F/\partial(M_s \bm{m})$, with $F$ the magnetic free energy and $M_s$ the saturation magnetization, is the effective field containing the external magnetic field and spin interactions. The first term in the square bracket is the conventional STT where $\bm{j}_1$ is electrical current density. The second term comes from the spin Hall spin torque, where $\theta_{SH}$ is the spin Hall angle, $\hat{z}$ is the plane normal, $t_f$ is the film thickness, and $\bm{j}_e$ is the density of electron current floating in the heavy metal layer. Note that *Eq.* (5) does not include a field-like torque due to its small influence on magnetization dynamics in typical material systems. By accessing the correct material specific parameters, this equation also provides a formulism for micromagnetic simulation studies that in turn have helped to establish an in-depth understanding of the dynamics of magnetic skyrmions.

It is worth mentioning that the current $\bm{j}_1$ in the STT term of the LLG equation [$(\bm{j}_1 \cdot \nabla)\bm{m}$] is not necessarily the electric current, but can also be a heat or magnonic current [71, 72]. As a spin wave (magnon) is the superposition of flipped spins, both thermally excited and microwave excited magnons carry collinear spins with respect to the local magnetization texture. As a result, similar to the STT generated by electric currents, the propagation of magnons (hence magnonic spin current) can in principle provide a source of STT on the local magnetic moments, and drives the motion of skyrmions even in chiral magnetic insulator [73]. Note that a SOT-like magnonic spin torque originating from DMI could also play an important role [74, 75]. The experimental feasibility has been demonstrated in a magnetic insulator yttrium iron garnet (YIG), where domain walls can be driven by thermally excited magnons [76]. The associated experimental realization in insulating skyrmion materials remains to be explored in the future.

On the other hand, the nontrivial topology of the skyrmion spin texture leads to an elegant structure of emergent electromagnetism of conduction electrons [65]. Due to the large Hund's rule exchange coupling, between conduction electrons and local magnetization vectors, which is generally on the scale of eV, the spin of the former always tries to follow adiabatically the spatial distribution of the latter. Therefore, although an electron traversing the skyrmion has the same incoming and outgoing spin states, its spin orientation actually rotates spatially, as if it precesses about an effective (spatially varying) magnetic field. In other words, the conduction electron "feels" an emergent virtual magnetic field ($b_z$) arising from skyrmion spin texture [57, 60, 64, 65, 77, 78]. This physical picture can be formulated by considering the Hamiltonian of a single conduction electron, given by $\hat{H} = (-i\hbar \nabla)^2 / 2m - J_H \sigma \cdot m$, where $J_H$ is the strength of the Hund's rule coupling, and $m$ is local magnetic moment. Due to large $J_H$, the electron spin must be parallel with $m$, described by the spinor state $|m>$ satisfying $\sigma \cdot |m> = |m>$. Consequently, the projected Hamiltonian is $\hat{H}_{eff} = \hat{P}\hat{H}\hat{P} = (-i\hbar \nabla - ea/c)^2/2m - J_H$, where the projection operator $\hat{P} = |m><m|$. It demonstrates that an emergent vector potential $a = (i\hbar/2e)<m|\nabla|m>$ is acting on conduction electrons. The effective magnetic field thus reads as:

$$b_z = \nabla \times a = \frac{\hbar c}{2e} \hat{z} m \cdot (\partial_x m \times \partial_y m) \tag{6}$$

which is closely related to the definition of topological charge $Q$ that is given by *Eq.* 2. Therefore, the quantization of topological charge $Q$ indicates the flux quantization of effective magnetic field of unit sphere $S_u$, that is, $\Phi_0 = \int b_z \cdot dS_u = (hc/e) \cdot Q$.

This emergent magnetic field provides an additional contribution to the experimentally measured Hall resistivity $\rho_{xy}$ which appears as a bump or dip during the hysteretic measurements, which can be written as follows:

$$\rho_{xy} = \rho_{xy}^{OHE} + \rho_{xy}^{AHE} + \rho_{xy}^{THE} \tag{7}$$

where the first term on the right side of *Eq.* 7 is the ordinary Hall resistivity that measures the carrier density and type, the second term on the right side is the anomalous Hall resistivity ($\rho_{xy}^{AHE} = 4\pi R_s M$, $R_s$ is the anomalous Hall coefficient and $M$ is the magnetization orientation, respectively) [79]; the third term is the so-called topological Hall resistivity $\rho_{xy}^{THE}$ that is proportional to the strength of emerging magnetic field $\rho_{xy}^{THE} \propto \langle b_z \rangle$, where $\langle b_z \rangle = \langle \Phi_0/\pi R^2 \rangle$ is associated with the flux quanta $\Phi_0$ contained in the skyrmion of diameter $R$. The strength of the emergent magnetic field can be as large as $\langle b_z \rangle \approx 100$ Tesla given for a skyrmion of 10 nm in diameter, which thus provides a unique platform to study the high magnetic field response of electrons [80]. Note that a contribution to the emergent magnetic field and consequently topological Hall effect that is caused by the inversion symmetry breaking combined with spin-orbit interaction has been theoretically proposed [81]. At the same time, based on Faraday's law ($E_t = d\Phi/dt$ – a time dependent flux change creates an electromotive force/potential), the translational motion of magnetic skyrmion produces a time-dependent electric field $\langle E_t \rangle$ that leads to an emergent electromagnetic induction and extra contribution to Hall resistivity. These aspects have been observed in bulk samples [64, 77], epitaxial thin films [78], and even nanowires [82].

It is emphasized here that the appearance of characteristic bumps or dips in the Hall measurements of chiral magnets is a direct consequence of the noncollinear topological spin textures [64, 78, 83]. One should, however, note that complications can arise from the electronic band structures [84], as well as a sign change of the dominant scattering responsible for the anomalous Hall effect [85] that influences the electronic Hall transport equivalently, resulting in similar transport features. Namely, the appearance of a bump/dip features of Hall effect alone may not hold the strict, original definition of the topological Hall effect in the skyrmion phase, and thus cannot serve as an unambiguous evidence for identifying noncollinear topological spin textures [86, 87].

Another interesting phenomenon is the Hall effect of topological charge – the skyrmion Hall effect. As mentioned previously, the well-defined spin topology of magnetic skyrmion results in a unit topological charge $Q = \pm 1$. It is assumed that a magnetic skyrmion is a rigid quasi-particle. Then its motion upon driven by a lateral homogeneous (electric/thermal) currents can be expressed by using a modified Thiele's equation [88-90]. This motion exhibits a well-defined transverse component as a result of the topological Magnus force [57], similar to that of a Lorentz force acting on moving charge carriers in conductor when subjected to a perpendicular magnetic field [64, 65]. This phenomenon, which has not yet been experimentally demonstrated in chiral bulk magnets, will be discussed in the section 3.6 of this review article.

Stimulated by the nontrivial spin topology of magnetic skyrmions, many other new and exciting topological transport phenomena have been theoretically proposed [91-94] and experimentally confirmed [64, 72, 95, 96]. For example, the relation between skyrmion and Hall transport in DMI magnets can be very complex [71, 97] for which the Onsager reciprocal relation might be worthy of investigation in the future [98]. The interaction between skyrmion spin texture and heat (magnonic) current has been predicted and observed. This is known as the topological Nernst effect in metallic chiral magnets where the carriers are electrons [99], and as the topological magnon Hall effect in insulating chiral magnets where the carriers are magnons [100]. In the latter case, magnons interact with the emergent magnetic field of skyrmions that produces a finite deflection of magnons with respect to their original propagation direction. As a result, magnons accumulates at the edge of devices and hence generate a measurable temperature difference along the transverse direction. Indeed, control of magnetic skyrmion by spin wave has been numerically confirmed [101]. Creation of artificial monopoles were also proposed and envisioned to explain the annihilation of skyrmions [57, 66, 102-104]. Many other exotic phenomena such as a Majorana bound state [105] in skyrmion materials/superconductor hybrids, and topological spin drag effect driven by skyrmion diffusion [106] have also been theoretically proposed.

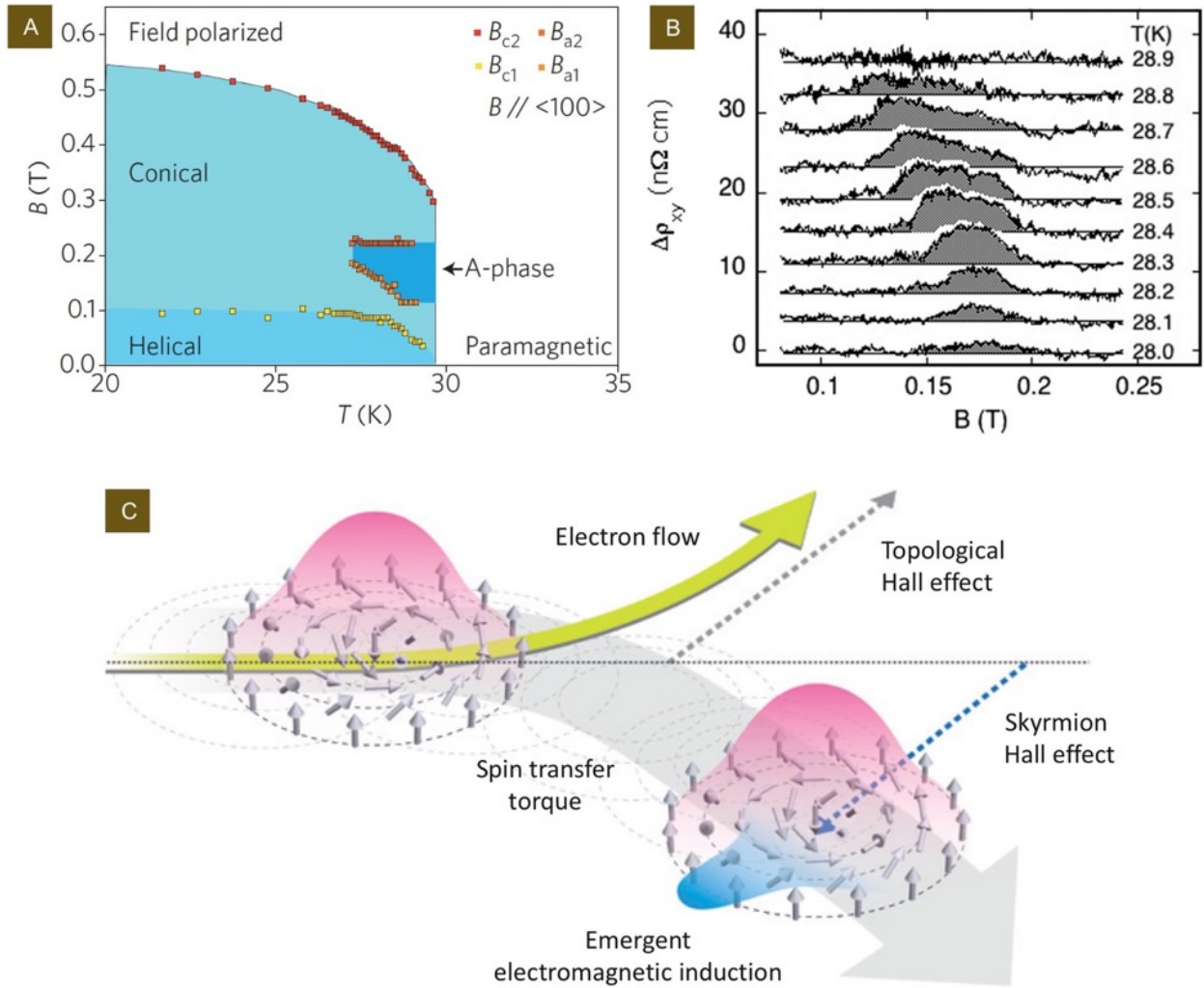

Fig 2. Phase diagram and emerging topological physics of skyrmion materials. (A) is the magnetic phase diagram as a function of temperature and magnetic field in MnSi which contains five different spin states, conical, helical, skyrmion, fully polarized ferromagnetic, and disordered paramagnetic state. The narrow regime in the temperature-magnetic field plane, marked as A-phase is the skyrmion phase. Reproduced with permission [32]. Copyright 2013, Nature Publishing Group. (B) is the extracted topological Hall contribution in the skyrmion phase that was done by subtracting the ordinary Hall, anomalous Hall components from the measured total Hall resistivity. It only exists in the A phase with a nontrivial spin topology of spin textures, namely in the skyrmion phase. Reproduced with permission [64]. Copyright 2009, American Physical Society. (C) summarizes the intriguing topological transport phenomena associated with the skyrmion spin geometrical phase, including the spin transfer torque that produces the motion of a skyrmion, topological Hall effect of electron, skyrmion Hall effect due to the Magnus force experienced by the topological charge, and the emergent electromagnetic induction arising from the time dependent translational motion of a magnetic skyrmion. Reproduced with permission [27]. Copyright 2013, Nature Publishing Group.

## 1.3 Topological distinction of bubble-like spin textures

Conventional bubble-like spin textures have been extensively studied for over four decades in many material systems such as insulating yttrium iron garnets [21, 107] and single-crystalline Co (0001) films [108]. These materials are centro-symmetric and therefore a chiral DMI component does not exist. Magnetic bubbles in these materials were solely stabilized by strong dipole interactions that give rise to a rich collection of spin textures with non-uniform spin chiralities/topologies, and consequently arbitrary topological skyrmion numbers. The size of these magnetic bubbles depends on the thickness of film, can be as large as 10 µm (thus exhibit larger inner cores of uniform perpendicular magnetization and narrow domain wall boundary). As schematically shown in Fig. 3, depending on how spins rotate inside the domain wall, the accompanied topological skyrmion number $Q$ and spin chirality ($\chi$) of these (neighboring) magnetic bubbles can be very different, varying from +1, 0, -1 [20, 109]. This has been experimentally confirmed in hexa-ferrites in which the spin chiralities of neighboring magnetic bubbles are random, as determined by using a spin sensitive Lorentz TEM technique [110], shown in Fig. 3B. Some of these magnetic bubbles exhibit an extra twist as compared with the simple skyrmion spin texture [107, 111]. In fact, detailed spin configurations of magnetic bubbles can be even more complex by having alternating Bloch and Néel segments along the wall structure. The dynamics of these magnetic bubbles driven by the spin transfer torque can also be very complex, depending on the specific topological skyrmion numbers [112]. This highlights the governing role of chiral DMI that gives rise to a uniform spin topology of the involved spin textures over the entire system, and also distinguishes magnetic skyrmion from general concepts of bubble-like spin textures.

The key anti-symmetric DMI that stabilizes magnetic skyrmions in chiral magnets implies that attempts in distinguishing magnetic skyrmion, skyrmion bubble and magnetic bubble solely based on the size of the inner core of spin textures is not justified [113, 114]. One should carefully examine the spin topology and spin chirality of the involved spin textures, rather than the size as it does not have direct bearing on the value of topological charge. For example, while the size of spin textures varies from 10 nm to 10 µm, the associated topological charge can remain the same [109, 113]. As a result, they belong to the same topological class [59].

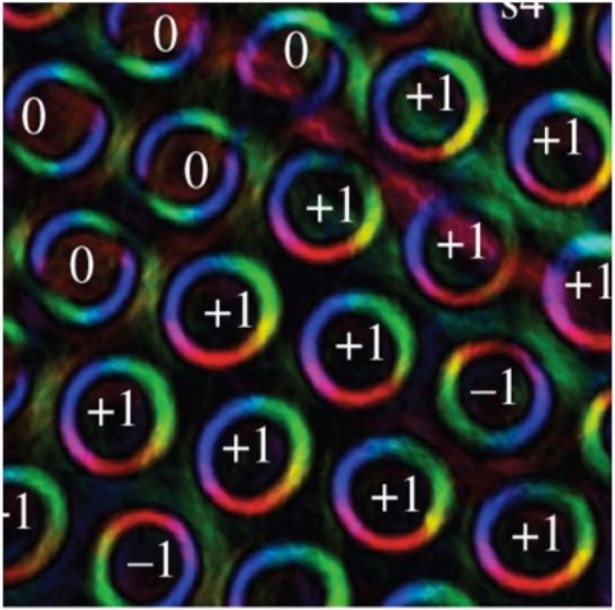

Fig 3. Dipole interaction stabilized magnetic bubbles with various spin chiralities ($\chi$). This is experimentally acquired by using a Lorentz transmission electron microscopy (TEM) in Sc-doped ferrites of composition $BaFe_{0.79}Sc_{0.16}Mg_{0.05}O_{19}$. Reproduced with permission [110]. Copyright 2012, The Royal Society.

Topological spin textures have also been observed in other material systems without DMI. In parallel to the extensively studied bulk chiral magnets, most recently, by utilizing the interlayer coupling between a PMA film (such as Co/Pt or Co/Pd multilayer) and a patterned in-plane nanomagnet in the magnetic vortex state, artificial skyrmions were theoretically proposed, and experimentally stabilized in the absence of DMI [115-118]. They, however, intriguingly exhibit the same spin topology as quantified by a nontrivial topological skyrmion number $Q = \pm 1$. Thus, in principle, topological transport phenomena can be observed in this material system. It should be emphasized here that these artificial Bloch-type skyrmions can be stabilized in the absence of external magnetic field. These artificial skyrmions are so far localized objects that were physically defined by nano-patterning. This will be discussed in section 3.8.

The recent observation of room-temperature biskyrmion phases in Heusler alloy NiMnGa compound [119, 120] and layered manganite $La_{2-2x}Sr_{1+2x}Mn_2O_7$ [121] are also intriguing. Note that a topological charge $Q = 2$ of these bound skyrmions is observed in these materials (thus underlying the name of biskyrmion). Crystal structures of these materials are centrosymmetric and hence a substantial DMI component is generally absent, it is thus interesting to theoretically formulate how the dipolar interaction together with the low crystal symmetry result in this noncollinear biskyrmion phase. Another interesting thin-film skyrmion system is the interfacially symmetric Fe/Gd magnetic multilayer [122, 123], in which Bloch-type magnetic skyrmions and their densely packed triangular lattices were stabilized by the competition between long-range dipole interactions and short-range exchange interaction (similar to that of conventional magnetic bubble materials). Through a systematic study of layer thickness, repetition, and alloying composition of the magnetic multilayer structure and their temperature/magnetic field dependence, Montoya *et al.*, have reported a successful control of a skyrmion lattice in Fe/Gd multilayers. Due to the absence of interfacial or bulk DMI, this material system exhibits two different types of magnetic skyrmion with opposite spin chiralities. Beyond stabilization of magnetic skyrmions by antisymmetric DMI, other exotic mechanisms such as four-spin interaction [124], and frustrated exchange interaction [125-127] and nonmagnetic impurities [128, 129] have also been proposed.

As previously mentioned, topologically protected magnetic skyrmions in metallic chiral magnets can efficiently be manipulated by a spin transfer torque using the spin-polarized electric current [24, 130]. Due to their nanoscale features, magnetic skyrmions were thus envisioned as information carriers for future non-volatile, low-power consumption, and high density spintronic memory and logic devices. Similar to the concept of domain wall based race-track memory [26], skyrmion race-track memory is one of the application examples that will be discussed in section 3.3.

## 2. Interfacial chiral magnetism

The material choice of bulk chiral magnets with a broken inversion symmetry is still limited. This limits practical spintronics applications and also impedes the exploration of topological transport at room temperature. In contrast, recent advances in nanomagnetism offer a new opportunity in low-dimensional magnetic heterostructures with a broken interfacial inversion symmetry [12, 32, 40, 131]. These inversion asymmetric heterostructures can be readily synthesized by using standard techniques such as magnetron sputtering, electron beam evaporation, pulsed laser deposition, or molecular beam epitaxy techniques. These multilayered films can also be easily grown onto commercially available Si substrates at a large scale that can be potentially integrated with mature semiconductor technology. This novel material system expands the material paradigm of chiral spin textures, enables the observation of room-temperature chiral domain walls and magnetic skyrmions (with sizes demonstrated down to a few tens of nanometer), provides novel and more energy-efficient driving mechanism through the usage of spin-orbit torques. In this part, first the formation of chiral spin spirals, chiral magnetic domain walls, and chiral magnetic skyrmions in magnetic heterostructures will be discussed. Second, the diverse physical origins for spin chirality will be presented. Third, different experimental methods that have been utilized to quantify the interfacial chiral magnetism are introduced. The unique advantage of chiral spin textures in magnetic multilayers will be further elaborated in section 2.4.

**2.1 From spin spirals to chiral domain walls**
In this section, the influence of the interfacial DMI on spin spirals and chiral domain walls in thin film systems will be discussed, with the aim of introducing the fundamental mechanisms for the formation of the Néel-type (cycloidal-type) spin spirals, domain walls and magnetic skyrmions.

The formation of nanoscale spin spirals can be attributed to the interplay among three key ingredients. They are interfacial DMI, Heisenberg exchange interaction, and magnetic anisotropy [35]. The DMI vector at the interface, $\boldsymbol{D}_{12}$, is perpendicular to the vector $\boldsymbol{r}_{12}$ connecting two neighboring spins within the interface plane. Therefore, depending on the sign of the vector $\boldsymbol{D}_{12}$, the interfacial DMI prefers a non-collinear cycloidal spin configuration with either right-handed chirality (↑→↓←↑) or left-handed chirality (↑←↓→↑). The direction of arrows indicates the related spin orientation. In contrast, dominant Heisenberg exchange interaction [$J(\boldsymbol{S}_1 \cdot \boldsymbol{S}_2)$] leads to either ferromagnetic (↑↑↑↑) or antiferromagnetic (↑↓↑↓) spin configurations, depending on the sign of the exchange constant $J$. Another competing energy is magnetic anisotropy ($K sin^2\varphi$), where $K$ is the magnetic anisotropy constant and $\varphi$ is the angle between the easy axis of the anisotropy and spins. This term determines the propagation direction of the spin spirals. For instance, spin cycloids (Néel-type) propagate normal to the hard axis of the magnetic anisotropy (thus in the film plane for the PMA system discussed here).

The experimental observation of chiral interfacial magnetism was first reported in one monolayer Mn film grown onto W(110) single crystal, where a left-handed cycloidal spin spiral (↓↑↘↖⇆↙↗↓↑↘↖⇆↙↗↓↑) with a period ~12 nm was identified by using a spin-polarized scanning tunneling microscopy (SP-STM) [39]. In this material system, the interfacial DMI strength is sufficiently strong to overcome the antiferromagnetic state, *i.e.*, $D_{12}^2 > 4JK$. Later, the same left-handed cycloidal spin spirals with a 2.2 nm period were observed in one monolayer Mn film grown onto a W(001) single crystal. Its relatively short period, compared to 12 nm in Mn/W(110) system is attributed to the softened exchange interaction [132]. This highlights the important role of competing interactions in controlling chiral spin textures.

For a moderate DMI strength range, *i.e.*, $D_{12}^2 < 4JK$, but close to this energy threshold, the material system can accommodate another interesting spin texture called inhomogeneous spin spirals, where the angle between neighboring spins alternates periodically along the spin spiral. This has been confirmed in two monolayer Fe films grown onto a W(110) crystal, in which right-handed chirality inhomogeneous cycloidal spin spirals were observed [133].

For thicker thin films with an in-plane magnetic anisotropy, the majority of spins are aligned along the in-plane easy axis. The orientation of the spin rotation plane within the wall depends on the stray field energy cost associated with the magnetic domain wall geometry: the width of the domain wall and height of the wall (equivalent to the thickness of the film $t_f$). For bulk materials, the Bloch type of domain wall is the ground state, while this may change when the thickness of the thin film becomes thinner than the width of the domain wall. Such a thickness-dependent Néel to Bloch wall transition with an increasing film thickness has been observed before [21]. In the presence of an interfacial DMI, neither the Bloch walls (helical-like) nor the Néel walls (cycloidal-like) in in-plane magnetized systems is expected to be chiral. This is because the vector of the interfacial DMI, $\boldsymbol{D}_{12}$ is usually aligned within the film plane and normal to the distance vector $\boldsymbol{r}_{12}$. As a result, $\boldsymbol{S}_1 \times \boldsymbol{S}_2$ is expected to be normal to $\boldsymbol{D}_{12}$ for both cases, which causes the DMI energy to vanish. For instance, $\boldsymbol{S}_1 \times \boldsymbol{S}_2$ is parallel to $\boldsymbol{r}_{12}$ for a Bloch wall configuration (note that $\boldsymbol{D}_{12}$ is normal to $\boldsymbol{r}_{12}$), and $\boldsymbol{S}_1 \times \boldsymbol{S}_2$ is perpendicular to the film plane for an in-plane Néel wall configuration (note that $\boldsymbol{D}_{12}$ is usually aligned within the film plane) [30].

For material system with a PMA, two possible rotation axes within the wall regions are both aligned within the film plane. In terms of the stray field energy cost, the competition is now related to the width as well as the length of the wall; the latter is usually treated infinitely long in a thin film. Therefore, a Bloch wall is generally considered as the ground state without the presence of interfacial DMI. Note that the length of the wall can possibly be narrower than the width of the wall in pattern structures, for instance, the Bloch to Néel transition has been observed in magnetic nanowires by decreasing the nanowire width (equivalent to the length of the wall) [134].

Next the spin chirality in PMA systems is discussed. First, in these systems the energy of the interfacial DMI for the Bloch wall configuration vanishes ($\boldsymbol{S}_1 \times \boldsymbol{S}_2$ is parallel to $\boldsymbol{r}_{12}$). Second, the Néel wall configuration results in a finite DMI energy, *i.e.*, $\boldsymbol{S}_1 \times \boldsymbol{S}_2$ is parallel or anti-parallel to $\boldsymbol{D}_{12}$, and the system can exhibit a homo-chiral spin structures in the presence of a sufficiently strong DMI, where spins favors a rotation sense in a way such that $\boldsymbol{S}_1 \times \boldsymbol{S}_2$ is parallel to $\boldsymbol{D}_{12}$. Because the stray field energy generally favors a Bloch wall configuration and the DMI favors Néel wall configuration, the domain wall configuration in thin films may be either of achiral Bloch type or chiral Néel type based on the interplay between the two energy terms.

As discussed above, Néel-type walls containing both out-of-pane and in-plane components are expected to be the only chiral spin rotating configuration in the presence of the interfacial DMI. Figs. 4A - B show sketches of chiral Néel walls with both right-handed chirality and left-handed in PMA systems. Experimentally, chiral Néel walls in thin films were directly observed using spin-polarized low-energy electron microscopy (SPLEEM) that is capable of quantifying 3-dimensional spin textures [135-137]. Figs. 4C-E shows SPLEEM images of the surface of an Fe/Ni bilayer taken along out-of-plane and two in-plane orthogonal directions, respectively. Combining these three images as a compound image (Fig. 4F) highlights the boundary and magnetization direction of the domain walls, from which the formation of right-handed chiral Néel walls is confirmed, based on the fact that spins within domain walls are normal to the domain boundary direction and always pointing from grey domains (+z) to black domain (-z). It is noted that the Néel-type feature can also be inferred by using Nitrogen-vacancy diamond microscopy [138], photoemission electron microscopy (PEEM) [44], or by Lorentz TEM [139].

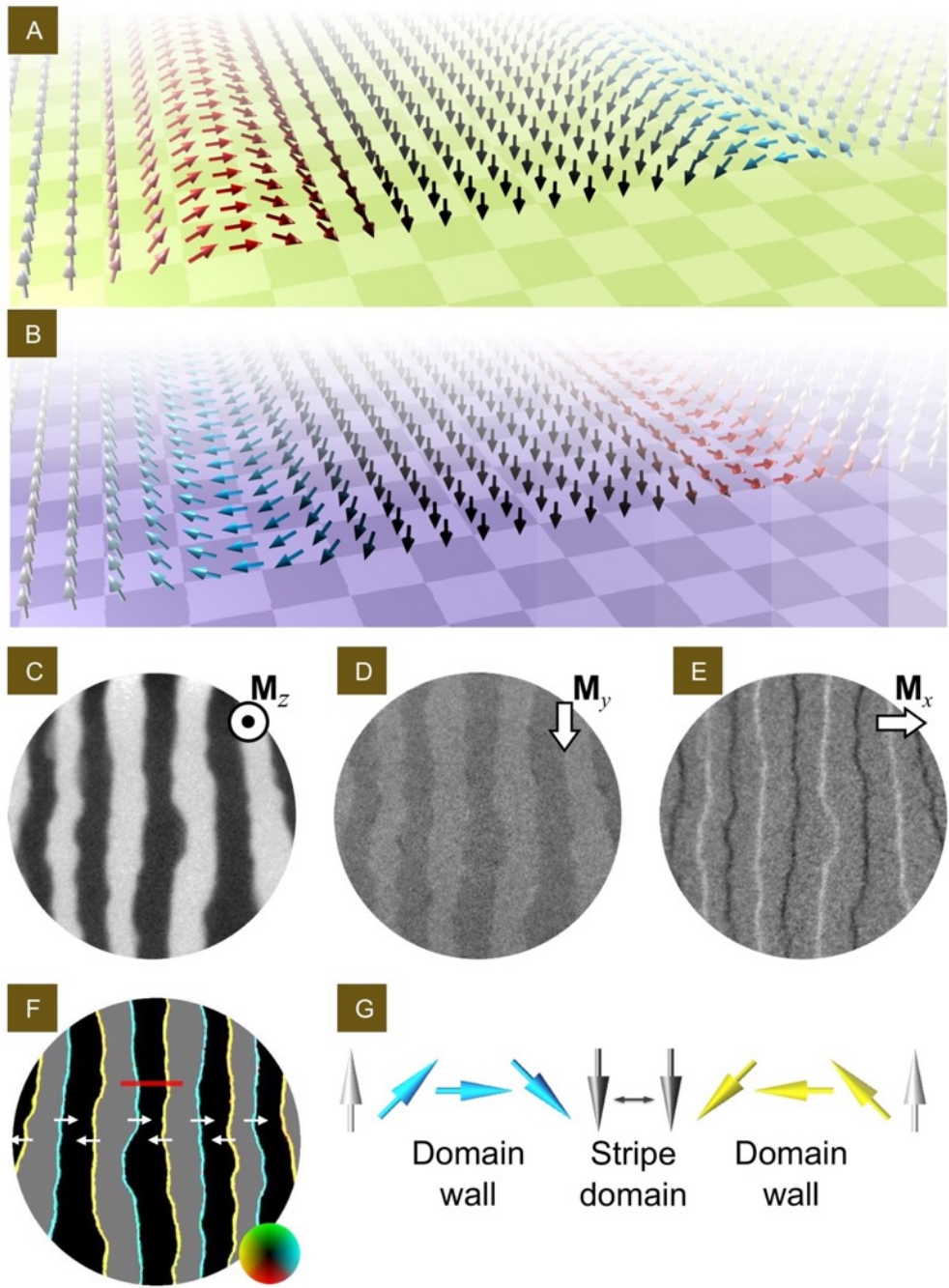

Figure 4. Chiral Néel domain walls. (A) is a sketch of a right-handed chiral Néel wall, with arrows representing the orientation of spin blocks. (B) is the sketch of left-handed chiral Néel wall. Reproduced with permission [135]. Copyright 2013, Nature Publishing Group. (C)-(E) are SPLEEM images of a Fe/Ni bilayer grown on Cu(001) single crystal substrate, mapping three orthogonal magnetization components $M_z$, $M_y$, $M_x$ respectively. The top right symbols define the spin polarization directions of the incident electron, white/black contrast in the images corresponds to magnetization components parallel/antiparallel to the spin polarization directions. The field of view is 10μm. (F) is a SPLEEM compound image constructed from SPLEEM images in (C)-(E). (G) is the cross-sectional sketch of the right-handed chiral Néel wall structure along the red line in (F). Reproduced with permission [136]. Copyright 2013, American Physical Society.

## 2.2 Physical origin of the chiral interfacial DMI

In this section, the fundamental reasons that underlie the chiral interfacial DMI in asymmetric magnetic heterostructures will be addressed. It is commonly accepted that SOC is crucial to establish the interfacial DMI. However, an experimentally measurable quantity or parameter that closely connects the strength and sign of SOC to that of interfacial DMI is not established, if there is any. This is a controversial topic currently under extensive debate that reflects the highly complex nature of the transition metal interfaces. Indeed, by using circular dichroism and x-ray resonant magnetic scattering, early work done by Dürr *et al.* have revealed the existence of chiral spin structures in 40-nm thick FePd films (while capped by 2-nm thick Pd) grown on a MgO (001) substrate where the interface contribution is minimal compared to the bulk [140]. In the following, a few diverse views of the interfacial DMI physics will be discussed, based on both experimental measurements and theoretical calculations.

Experimentally, by investigating the heavy metal (HM) dependent current-driven dynamics of the chiral domain wall in HM/FM/I heterostructures, Torrojen *et al.* have proposed that electron negativity changing the charge localization at the interface is responsible for the emergence of spin chirality [141]. On the other hand, Ryu *et al.* have proposed that magnetic proximity induces moments in the HM layer (this typically occurs in Pt, Pd and Ir that are close to the Stoner limit), which dictates the emergence of an interfacial DMI in these inversion asymmetric multilayers [142].

Through first-principles calculations, Yang *et al.* have shown in a bilayer Pt/Co that the interfacial DMI strength predominantly depends on the interfacial Co layer, and it *does not* extend from the interface [143]. The strength of interfacial DMI between the interface Co spins is directly related to the SOC energy of the interfacial Pt atoms, and excludes the dominant contribution from magnetic proximity effect to the interfacial DMI in asymmetric heterostructures.

More recently, by using first-principles calculations and noting the critical role of $3d$ orbital occupations and their corresponding spin-flip mixing processes [144], Belabbes *et al.* have revealed that the degree of orbital hybridization at the $3d$-$5d$ interfaces, and the $3d$ band lineup dictated by the atomic Hund's first rule, directly control the sign and the strength of interfacial DMI. This important finding thus adds key insights to the complex interfacial DMI physics, in addition to the commonly accepted broken inversion symmetry and strong spin orbit interaction.

Their results revealed that the ratio of $E_{DMI}/m_s^2$ (DMI energy divided by the square of the spin magnetic moment $m_s$) varies as a function of $3d$ transition metals. Namely, the Mn overlayer has the highest ratio amongst $3d$ metals, and it gradually decreases on both sides of Mn along the row of $3d$ transition metal in the periodic table. This is further confirmed by a similar dependence of spin magnetic moment per atom versus the adlayer (red line in Fig. 5A). At the same time, based on Hund's first rule, they have also recognized the important contribution of spin hopping between occupied and unoccupied states to establish the interfacial DMI [145, 146]. *I.e.*, the band lineup of $3d$ spin channels is critical to the DMI. For instance, manganese has five filled $3d$ orbitals with all spin up states, where the spin-up (spin-down) channels are totally occupied (unoccupied), so that all possible transitions between these states contribute to the DMI, as shown in Fig. 5B. In contrast, Vanadium and Nickel have both spin-up and spin-down channels that are largely unoccupied and occupied, respectively, and consequently the transition of $3d$ electrons do not contribute to the interfacial DMI. Another interesting example is Au, despite its large spin orbit interaction, the interfacial DMI at the Au/$5d$ interface is almost vanished, as a result of the complete filling of $d$ shell of Au substrate. It thus suggests that interfacial DMI also depends on the $d$ orbital hybridization of the involved interface as it influences the interlayer electron hopping and hence the magnetic coupling.

The results from the first-principles calculations are consistent with the experimentally measured DMI values and sign. For instance, opposite DMI between Pt/Co and Fe/Ir interfaces [43, 53], or the derived values that Co/Ir is three times larger than Ni/Ir but with opposite sign [147]. This approach may capture the underlying complex SOC physics that is responsible for the interfacial DMI. As it has already been pointed out that the interfacial DMI emerges as a result of the complex interplay between the following three ingredients: (1) the ratio of spin polarization of the $3d$/$5d$ interfacial atom and their band filling, (2) the strength of SOC of the involved $5d$ transition metal substrate, and (3) the interfacial inversion symmetric breaking. Efforts along these directions are highly desirable in the future, as they could provide a guiding principle for the miniaturization of room-temperature skyrmion by maximizing the interfacial DMI [143, 144, 148, 149].

It should be emphasized here that SOC plays different roles in the SHE [150], as compared with the interfacial DMI. There is no obvious connection between the strength and sign of SHE and that of the interfacial chiral DMI. This is also suggested by observing the opposite spin Hall angles in Pt and Ta thin films [151]. The spin chirality of magnetic domain walls in their related magnetic hetero-structures, however, remains the same being left-handed [151, 152]. Ir provides another affirmative evidence, which has a negligible SHE [150] but a significant (negative) interfacial DMI [43], as opposed to Pt, and Ta [42, 135, 147, 151]. This can be simply understood as SHE is a bulk effect, while interfacial DMI is only generated at the interface.

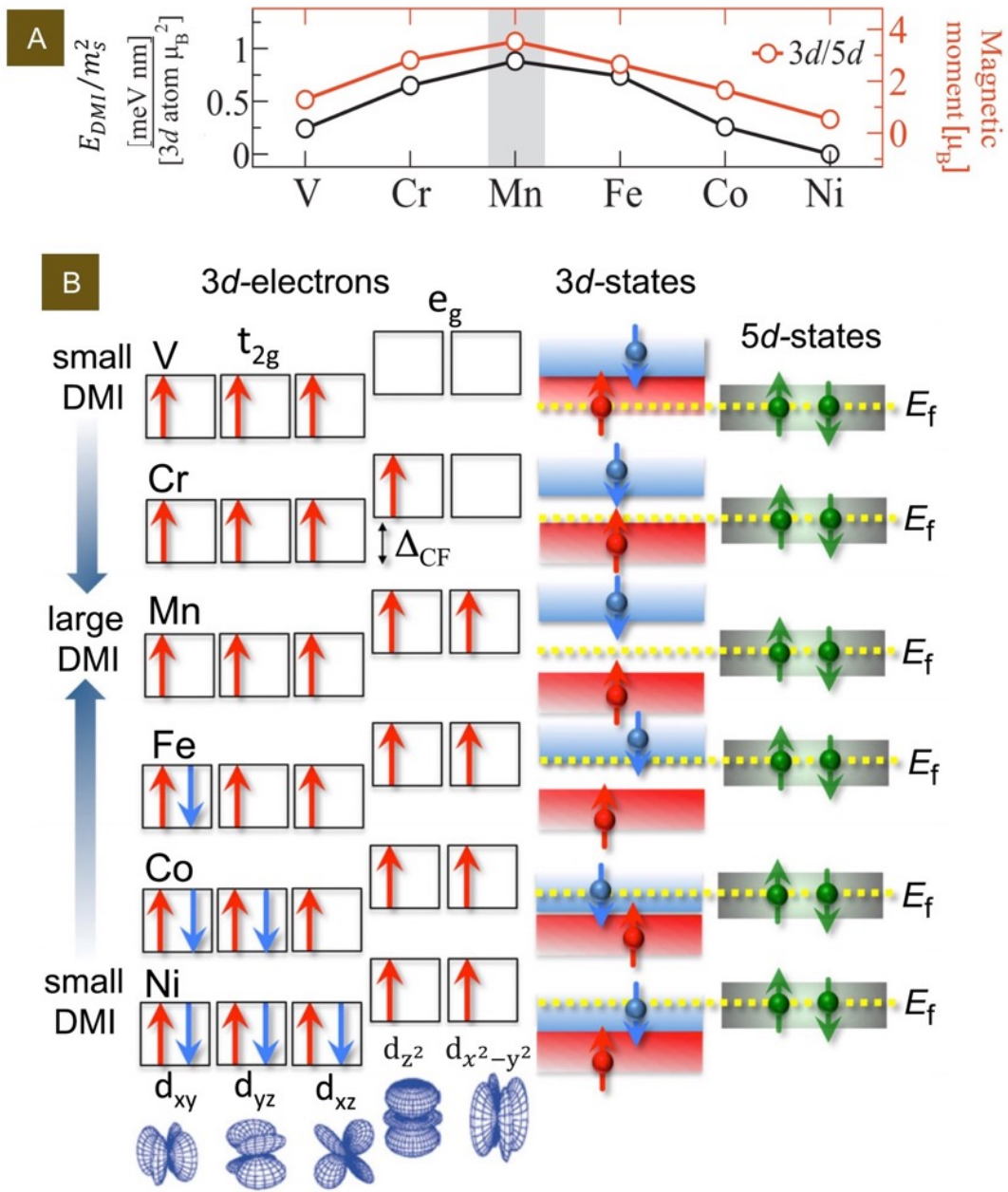

Figure 5 (A) Dependence of $E_{DMI}/m_s^2$ (black line) as well as local magnetic moment per atom (red line) averaged over $3d$/$5d$ interfaces vs. $3d$ adlayer, where $E_{DMI}$ is the DMI energy and $m_s$ is spin magnetic moment. All values are averaged based on the data of $5d$ metals from tungsten to platinum. (B) Shown on the left are the filling of $3d$ orbitals of the typical transition metal elements guided by Hund's first rule. On the right is the spin-split band positions of $3d$ states with respect to the $5d$ state of the W substrate. Reproduced with permission [144]. Copyright 2016, American Physical Society.

## 2.3 Measurement of the interfacial DMI

Another interesting and challenging topic is how to experimentally probe the strength and sign of interfacial DMI vector $\boldsymbol{D}_{ij}$. A direct measurement of the chiral exchange interaction in these multilayers is challenging. It can be indirectly measured and inferred by detecting its influence on spin wave propagation [153], and magnetic domain wall dynamics [42], etc. In the following, experimental efforts using various techniques to study those effects will be introduced, including spin-polarized electron energy loss spectroscopy (SPEELS), Brillouin light scattering (BLS) [154-158], SPLEEM [135, 136], magneto-optical Kerr effect (MOKE) microscopy [159, 160] or magnetometry studies of asymmetric hysteresis loops [161].

Early experimental efforts to quantify the interfacial DMI was done using SPEELS [153], where a spin-flip scattering process [162] was applied to excite spin waves in a bilayer Fe film grown on W(110). The authors found that the presence of interfacial DMI can break the degeneracy of spin waves, which results in an asymmetric spin wave dispersion relation. The DMI strength $D_{DMI} = |\boldsymbol{D}_{ij}|$ can be obtained by fitting the dependence of the energy asymmetry as a function of wave-vector transfer.

To detect the DMI in heterostructures, long-wavelength spin waves excited by microwaves can also be used due to their longer lifetime in the nanosecond range and micron-scale coherence length, compared to SPEELS investigations of exchange dominated spine-waves, where the lifetime is on the order of 10 fs, and the attenuation lengths is on the order of 1 nm. This experiment can be done by using a wavevector-resolved BLS in a backscattering geometry [163]. If chiral interfacial DMI is present in a heterostructures, then the degeneracy of the spin wave spectrum is lifted, which results in an asymmetric spin wave dispersion relation, Fig. 6A [164, 165]. The interfacial DMI produced a frequency difference - $\Delta f(k)$ of the counter-propagating (forward and backward moving) spin waves at equal but opposite wavevectors and can thus be written as [154-158]:

$$\Delta f(k) = \frac{2\gamma}{\pi M_s} D_{DMI} k \qquad (8)$$

where $\gamma$ is gyromagnetic ratio, $M_s$ is the saturation magnetization, $k$ is the spin wave propagation vector, and $D_{DMI}$ is the interfacial DMI constant. The slope of frequency difference $\Delta f(k)$ as a function of $k$ thus enables the value of $D_{DMI}$ to be directly determined, as shown in Fig. 6B. This method has an advantage compared to other methods, since it depends solely on the saturation magnetization $M_s$ that can be accurately measured. This experiment can be done either by using a wavevector-resolved BLS in a backscattering geometry, or by using an electrical detection method where spin waves were excited by a coplanar waveguide [166].

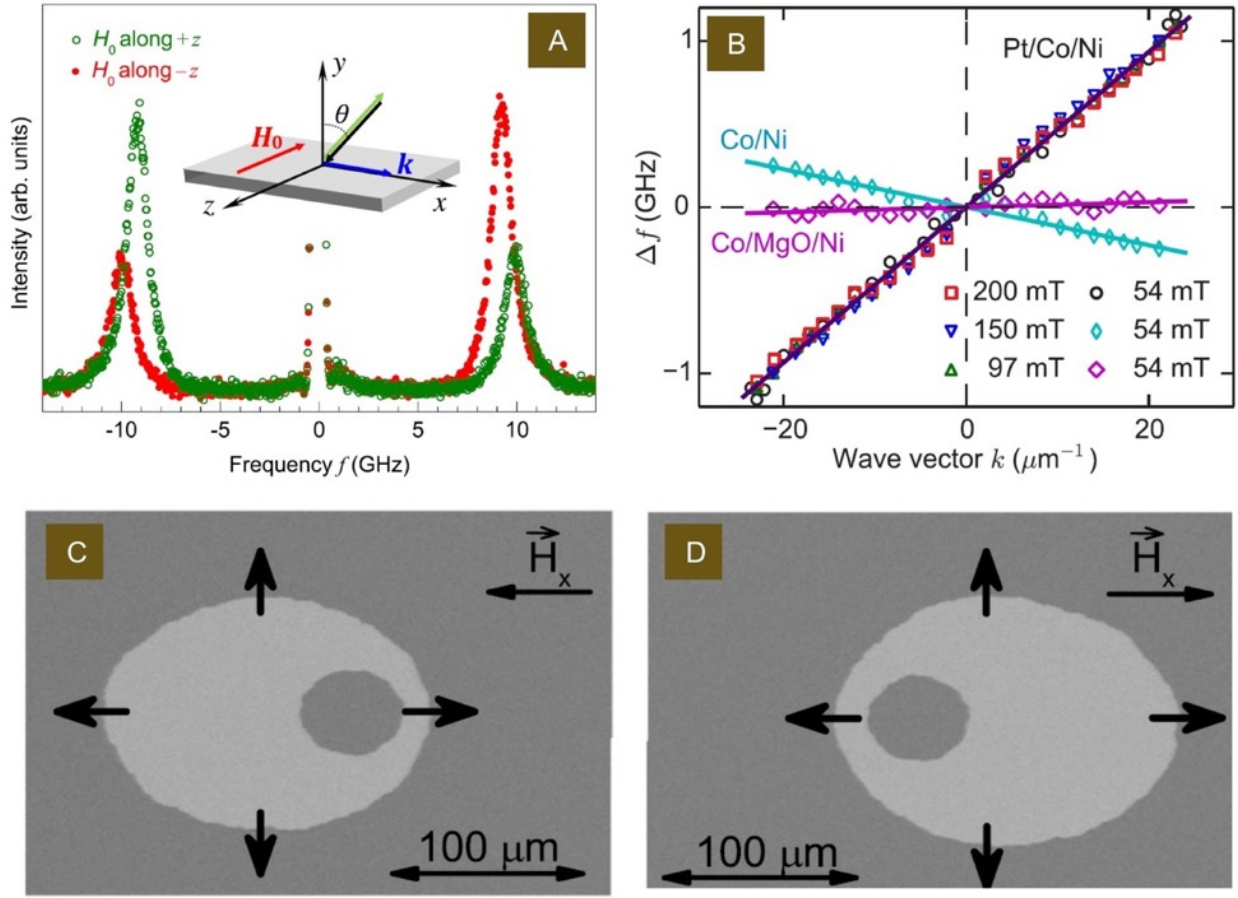

Fig 6. Origin and determination of the interfacial DMI in HM/FM bilayers. (A) shows the typical BLS spectra carried out for opposite in-plane magnetic fields ($H = \pm 97\,\text{mT}$), where one can see asymmetric (both frequency and intensity) magnon peaks in the Stokes and anti-Stokes portions of a typical spectrum. (B) is the frequency difference of counter-propagating spin-waves as a function of wave vector which clearly exhibit a linear dependence. By changing the layer stack structure, the slope changes drastically, reflecting the significant contribution of interfacial DMI from the involved HM Pt layer that is part of the bilayer. Reproduced with permission [163]. Copyright 2015, American Physical Society. (C)-(D) show the asymmetrical expansion of big chiral bubbles under the influence of in-plane magnetic fields ($H_x$). Reproduced with permission [167]. Copyright 2014, American Physical Society.

The DMI can also be quantified by studying the configurational transition of magnetic domain walls, which is associated with the energy competition between the interfacial DMI and the stray field energy of magnetic domain walls, details of which are described in section 2.1. Using SPLEEM, the film thickness dependent transition from chiral Néel wall (dominated by the DMI) and non-chiral Bloch wall (dominated by the stray field) can be directly imaged. Note that the DMI energy and stray field energy is comparable near the transition thickness, and experimentally this transition thickness can be found by gradually changing the total thickness of the magnetic film. This is because the stray field energy will increase with increasing film thickness whereas the DMI energy is usually independent of the film thickness. Therefore the DMI strength can be quantified by calculating the stray field energy for a wall structure with a given film thickness [135, 136].

Another experimental method is to initialize a big magnetic bubble domain by using a pulsed perpendicular magnetic field. These magnetic bubbles, due to the presence of interfacial DMI, preserve a fixed spin chirality (namely, big skyrmions) [160, 168]. Subsequently, by applying in-plane magnetic fields that tend to align spins inside the domain wall towards the external in-plane magnetic field ($H_x$) direction, one can observe asymmetric bubble expansion, as shown in Figs. 6(C)-(D). First, from the occurrence of a minimum domain wall velocity in the positive and negative in-plane field ($H_x$) regime (*i.e.*, external in-plane field compensates the DMI effective field and chiral Néel wall transform to Bloch wall), one can identify the chirality of domain wall or bubbles as being left-handed or right-handed. Second, from the value of in-plane fields with a minimum domain wall velocity, one can determine the effective DMI field ($H_{DMI}$), which results in the determination of DMI strength based on the following [159, 160, 167, 169]:

$$D_{DMI} = \mu_0 H_{DMI} M_s \gamma_{dw} \qquad (9)$$

where $\mu_0$ is the vacuum permeability and $\gamma_{dw} = \pi\sqrt{\mathcal{A}/K_{eff}}$ is the domain wall width. $K_{eff}$ is the effective perpendicular magnetic anisotropy, $\mathcal{A}$ is the spin-wave exchange stiffness constant. This method is limited to the system with a strong PMA [167].

In systems with a relatively weak PMA, where typically labyrinthine domains are present, one cannot initialize big chiral bubble domains and therefore cannot utilize the asymmetric bubble expansion method to determine the strength/sign of interfacial DMI. In these systems, due to the competition between the interfacial DMI, magneto-static, Zeeman energy terms, one can use the evolution of domain width and hence the domain wall surface energy as a function of perpendicular magnetic field ($H_z$) to determine the strength of DMI [42]. These two complementary methods however, strongly depend on the precise determination of the material specific parameters such as effective perpendicular magnetic anisotropy $K_{eff}$, spin-wave stiffness $\mathcal{A}$. In particular, a precise determination of $\mathcal{A}$ in ultra-thin magnetic film is crucial, which is challenging. The value of $\mathcal{A}$ in these 2-dimensional films can vary significantly from its counterpart in thicker film (or bulk materials) [170]. This can produce a substantial uncertainty in estimating the strength of interfacial DMI [171].

Beyond stabilizing chiral domain walls and Néel-type magnetic skyrmions, it is theoretically suggested that interfacial DMI can also tilt the magnetization vector of the side edge of a film or device away from the perpendicular orientation [172, 173], which results in a symmetric magnetic domain nucleation [160]. Using this intrinsic feature for patterned asymmetric triangular microstructures [161], Han *et al.*, have demonstrated a significant shift of the magnetic hysteresis loop measured in a perpendicular magnetic field ($M$ *vs.* $H_z$) configuration with a fixed in-plane bias field ($H_x$), as shown in Fig. 7. This observation can be attributed to the preferential nucleation of magnetic domain wall due to the edge magnetization tilting, as a result of the DMI boundary condition, as shown in Fig. 7. From the shifted magnetic hysteresis loops, both the strength and sign of interfacial DMI can be determined. The DMI value obtained from this method is comparable with those values acquired using BLS technique. In contrast, such a hysteresis loop shift is absent in symmetric square microstructures, which validates the above method. This approach, which is based on a half-droplet model and a systematic study of the shape variation, can be easily employed by magnetometry measurements using vibration sample magnetometer (VSM), MOKE magnetometer, superconducting quantum interference device (SQUID) magnetometer or other magnetization measurement techniques. This method is thus of great importance for rapid material screening and for providing optimized DMI parameter that enables scaling the size of magnetic skyrmions ultimately down to a few nanometers. The above method, however, also strongly depends on the precise determination of spin wave stiffness $\mathcal{A}$ in ultra-thin FM films, which thus introduces a source of uncertainty in determining the strength of interfacial DMI.

Finally, it has been shown that this limitation, requiring knowledge of $\mathcal{A}$, can be overcome by studying instead the dependence of the DW velocity when apply only a field along the (out-of-plane) easy axis [174]. For systems without DMI (typically symmetric trilayer such as Pt/Co/Pt [174]), the velocity is expected to first increase linearly, but then show a decrease due to the breakdown of stationary motion after surpassing the so-called Walker field $H_W$. Note that the Walker field $H_W$ is also influenced by damping $\alpha$ and DW width $\gamma_{dw}$. In systems with DMI, $H_W$ is increased and instead a velocity plateau is first reached at $H_W$ [131], as shown in Fig. 8. At the Walker field the DW velocity is given by:

$$v_W = \frac{\pi}{2}\gamma\frac{D_{DMI}}{M_S} \qquad (10)$$

where $\gamma$ is the gyromagnetic ratio. Determination of this velocity thus allows a straightforward calculation of $D_{DMI}$ once $M_S$ is independently measured.

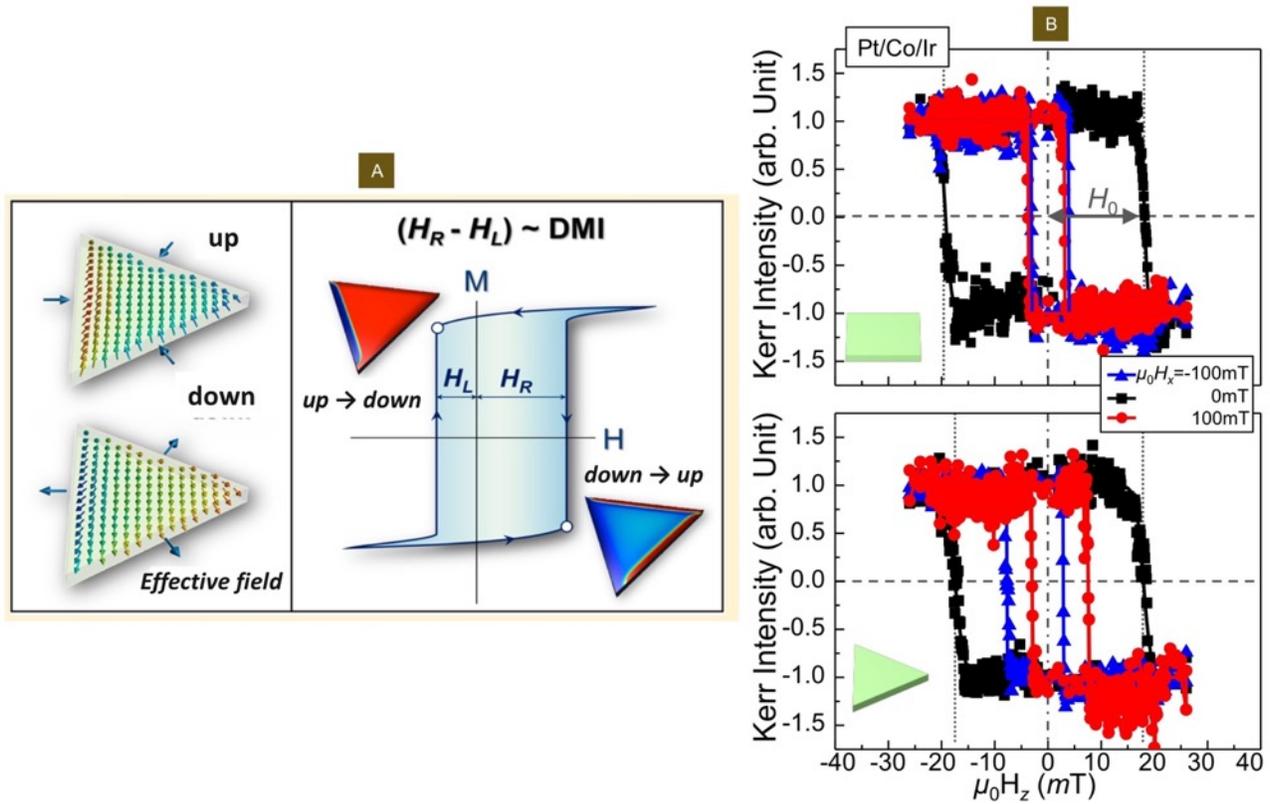

Fig. 7 Determination of the interfacial DMI based on the hysteresis ($M$ $vs.$ $H_z$) measurements in the asymmetric microstructures. Due to the presence of DMI, the edge magnetization is tilted away from the perfect perpendicular direction which creates a preferential nucleation of domains and results in a shifted hysteresis loops, shown in (A). This has been suggested via micromagnetic simulations and experimentally confirmed by using Kerr magnetometry measurements, shown in (B). It is noted that the hysteresis loop shit is absent in the symmetric square-shaped microstructure which thus validates this method. Reproduced with permission [161]. Copyright 2016, American Chemical Society.

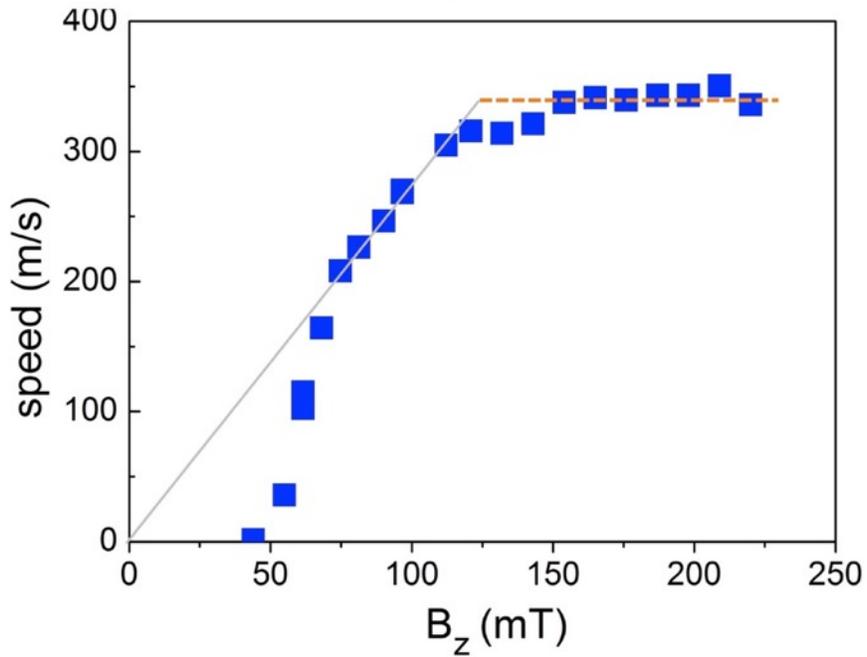

Fig. 8 The experimentally determined domain wall velocity as a function of out-of-plane field $B_z$ in an asymmetric Pt/Co/AlO$_x$ trilayer with an interfacial DMI. At the lowest fields, the velocity increases exponentially while within the creep regime. At higher fields, a linear increase of DW velocity is observed as highlighted by the gray line which indicates ta steady DW velocity. This is followed by a saturation of the DW velocity at the Walker field $H_W$, marked as the orange dashed line. Reproduced with permission [174]. Copyright 2016, EPLA.

**2.4 Unique advantages of magnetic skyrmions in heterostructures**

Typical inversion asymmetric magnetic multilayers are based on heavy metal/ultra-thin ferromagnet/insulator (HM/FM/I) heterostructures [12], as discussed previously. In the presence of strong SOC arising from the HM, the broken interfacial inversion symmetry produces an appreciable interfacial non-collinear DMI contribution, as schematically shown in Fig. 9A. Here, the DMI vector $\boldsymbol{D}_{DMI}$ lies in the film plane acting as an "effective" in-plane field that energetically favors non-collinear chiral Néel-type domain walls/skyrmions with a fixed chirality over the Bloch-type domain walls, shown in Fig. 9B [31, 39, 131, 175, 176]. A slice across the center of the Néel-type skyrmions is further illustrated in Fig. 9B (lower part), from which one clearly sees that the spin rotates cyclodially as a Néel-type domain wall, rather than rotating along a helical spiral as in a Bloch-type skyrmion. The spin chirality is defined by placing a mirror onto the center spin of the chain. Only by using a mirror operation, the spin states on the left can overlap the ones on the right. This cannot be achieved via a translational operation, shown in Fig. 9B (lower part). Although the real-space spin configurations appear differently for Bloch or Néel-type skyrmions, their topological properties are nevertheless equivalent and exhibit the same unit topological charge of $Q = \pm 1$. Namely, the Bloch skyrmion observed in non-centrosymmetric bulk materials can be generated by a 90° rotation from the Néel-type skyrmion. Thus these two types of skyrmions share the same spin topology since they can be continuously deformed into each other. Since the spin configuration on the unit sphere resembles a hedgehog, Néel-type skyrmions are also named "hedgehog" skyrmions, as shown on the right of Fig. 9B.

The strong SOC of HM can facilitate the formation of Néel-type skyrmions in the interfacial asymmetric systems. On the other hand, the associated strong SOC produces a spin dependent scattering of the incident conduction electrons that results in a spin imbalance in the perpendicular direction, as shown in Fig. 9C. Namely, the spin-up and spin-down electrons were scattered preferentially [11, 12, 69, 177, 178]. This is also known as the electronic spin Hall effect (SHE) [12, 13, 177], as schematically illustrated in Fig. 9C. At the HM/FM bilayer, the accumulation of spins at the interface subsequently transfer their spin angular momenta to the adjacent ultra-thin FM layer. Consequently, it gives rise to current-induced spin-orbit torques (SOTs), including field like torque and anti-damping like torque [69, 151, 179-183]. These spin torques, in turn, result in an efficient electrical manipulation of skyrmions [184, 185], shown in Fig. 9D. This material system is thus more technologically appealing for enabling functional skyrmionic logic or memory devices at room temperature [175]. However, it has been challenging to employ electric currents and its induced SOTs in these asymmetric multilayers to create or manipulate magnetic skyrmions at room temperature [186-188]. Until recently, this has been a bottleneck for realizing electrically reconfigurable skyrmionic logic or memory devices.

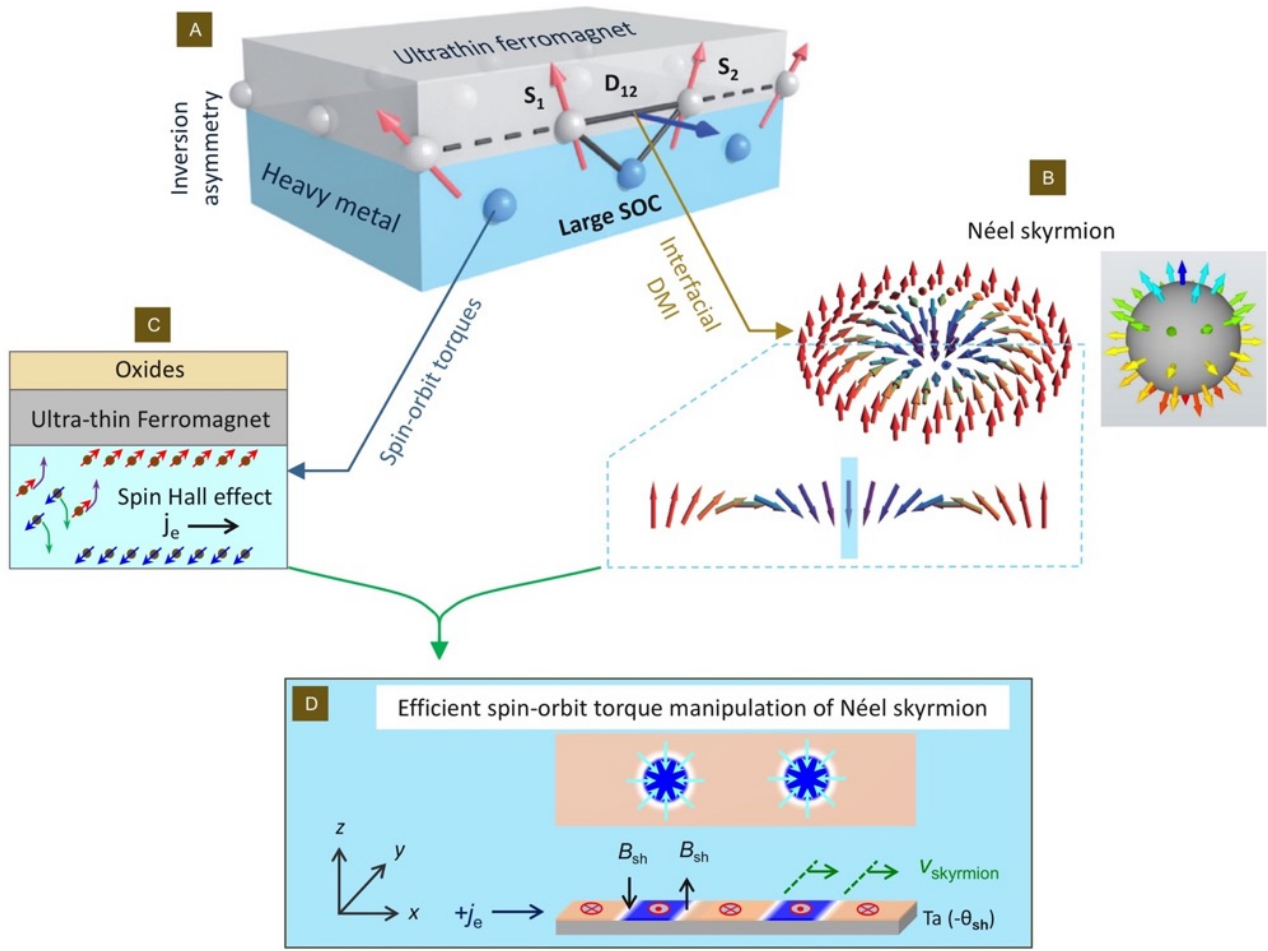

Fig. 9. Uniqueness of skyrmions stabilized in magnetic heterostructures with a broken inversion symmetry. (A) illustrates the inversion asymmetric bilayer composing of a HM/FM bilayer where the strong SOC of the HM layer, together with the ultra-thin FM, result in a non-collinear spin chirality determined by the interfacial DMI vector. Reproduced with permission [32]. Copyright 2013, Nature Publishing Group. As a result of this interfacial DMI contribution, Néel-type skyrmions can be stabilized, shown in (B). Its topological representation is further projected onto a unit sphere as a hedgehog shape. A slice across the center of the Néel-type skyrmion, is shown to reveal the presence of spin chirality by placing a mirror in the center of the spin chain. Only by performing a mirror operation can spins on the left and on the right, can be overlapped which thus underlies the name of chiral magnetic skyrmion. Reproduced with permission [189]. Copyright 2015, Nature Publishing Group. (C) illustrates the accompanied spin Hall effect. The strong SOC of HM layer creates a spin dependent scattering that splits the spin-up and spin-down electrons preferentially, which leads to a spin accumulation at the interface, enabling electrical control of the adjacent magnetization dynamics. An oxidized layer can be deposited on top of HM/FM bilayer for protecting stack from further oxidization that degrades the desired properties, and also for breaking inversion symmetry. The marriage of spin-orbit torques and interfacial DMI stabilized skyrmion consequently results in an efficient electrical manipulation of Néel-type magnetic skyrmions. The electron current ($j_e$) induced effective spin-orbit fields ($B_{sh}$), beared by spins on the right and on the left, act constructively, producing a motion of skyrmion with velocity v$_{skyrmion}$ along the same direction, as illustrated in (D). The motion of the skyrmion follows the electron current direction in the case of a negative spin Hall angle ($-\theta_{sh}$), together with left-handed chirality of Néel-type skyrmion. Reproduced with permission [41]. Copyright 2015, AAAS.

## 3. Current developments in thin-film skyrmions

Since the observation of chiral domain wall in inversion asymmetric multilayers, there has been considerable progress in the study of thin-film based Néel-type skyrmions. In the following, some of the recent advancements will be discussed, including writing and deleting a single skyrmion using spin-polarized tunneling currents at low temperature, observation of room-temperature skyrmion by transforming from chiral band domains via electrical currents, room-temperature sub-100 nm skyrmions in multilayers, prototypical skyrmion race-track memory, Hall effect of topological charge – skyrmion Hall effect, high frequency gyrotropic dynamics of magnetic skyrmions, artificial skyrmions stabilized by interlayer coupling in thin films, and novel spin-resolved imaging techniques.

### 3.1 Writing and deleting a single skyrmion

Chiral spin structures stabilized by interfacial DMI were first spatially resolved using spin-polarized scanning tunneling microscopy (SP-STM). Using a conventional STM with a magnetic tip results in spin dependent tunneling currents. This spin dependent tunneling current strongly depends on the relative magnetization orientation between the tip and sample, which thus enables spatial mapping of nanoscale spin structures down to atomic resolution [190]. In a Fe/Pd bilayer grown onto a single crystalline Ir (111) substrate, Romming *et al.* observed a magnetic field induced transition from chiral domain walls into Néel-type skyrmions [53, 190, 191], as shown in Fig. 10. The size of these skyrmions is approximately 5 nm, which is much smaller than typical Bloch-type magnetic skyrmion in bulk chiral MnSi and FeGe magnets (≈ 50 nm) [38, 58], thus suggesting the great potential for application in future ultra-dense skyrmion memory technology. The small size is due to the following facts: in the 2-dimensional limit, the strong interfacial DMI of the Ir/Fe (2 monolayer) interface of approximately $D_{DMI}$ ≈ 2 mJ/m$^2$ competes directly with the exchange interaction. Concurrently, the size of magnetic skyrmion evolves as a function of perpendicular magnetic field. Thus the size can be further reduced to 3 nm at high magnetic fields of approximately 2 T. Note that these skyrmions were observed at T = 4.2 K. It is currently not clear that magnetic skyrmions of size 3-5 nm in this material system can also exist at room temperature. A recent study of temperature-induced increase of spin spiral periods indirectly suggested the room-temperature instability of non-collinear spin structures of ultrathin Fe grown on Ir (111) substrate [192] and concluded that a layer dependent competing interactions should be considered.

More interestingly, by passing a vertical current from the STM tip, these nanoscale skyrmions can be addressed locally through spin transfer torques [53]. Namely, nanometer-sized skyrmions can be generated and deleted both locally and reversibly on demand by a spin-polarized tunneling current (of amplitude 1 nA) from the STM tip. While the exact microscopic picture of how a local tunneling spin current creates and annihilates skyrmion is not yet clear, it provides a local control mechanism for future skyrmion based memory and logic devices. For example, similar to SP-STM manipulation of skyrmions, one could build similar magnetic tunneling junction (MTJ) stacks to locally address nanoscale skyrmions. Likewise, chiral magnetic structures can also be written and deleted with an electric field applied between the STM tip and sample surface, as was demonstrated for Fe on Ir(111) [193].

The nanometer size of the skyrmions in these material systems should significantly increase the topological contribution to the topological Hall effect, which are inversely proportional to the size of the skyrmions. Thus, the topological Hall effect should in principle be amplified and can be used in future for electric readout of magnetic skyrmions. Note that while SP-STM provides an unprecedented spin-resolved spatial resolution and inspires a few novel skyrmion devices concepts, it is, however, extremely challenging to extend this technique to ex situ patterned devices where current driven dynamics of skyrmions can be investigated. In parallel, writing and deleting a single skyrmion can also be realized by using either ultrafast laser pulses [194-196] or ultrafast magnetic field pulses [197]. This manipulation scheme could potentially bridge the interesting connection between skyrmion dynamics and ultrafast magnetism.

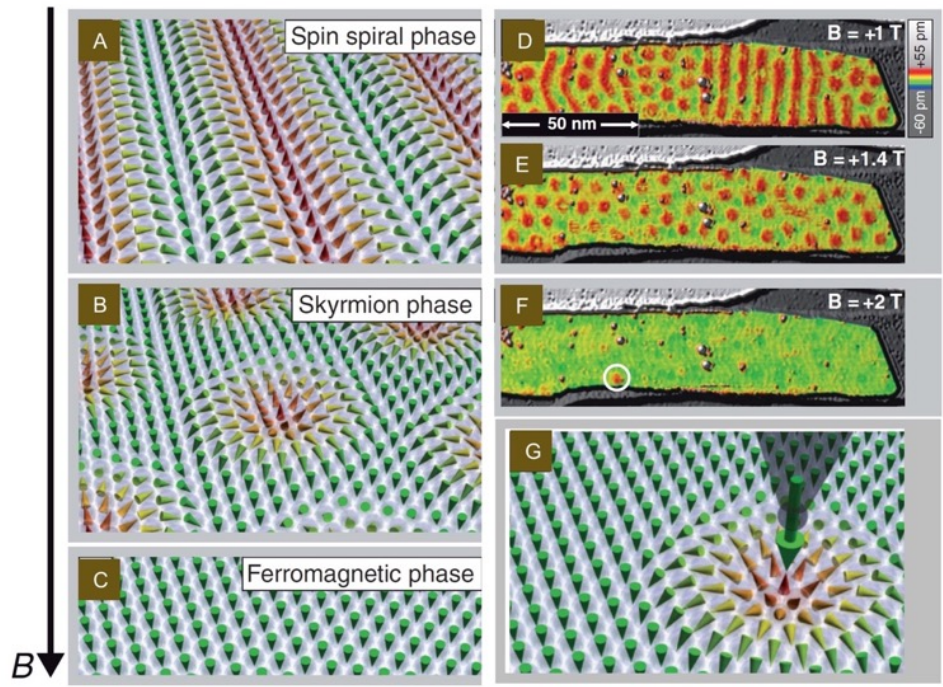

Fig. 10. Detection and addressing of nanometer scale skyrmions in a trilayer of Ir/Fe/Pd by using a spin-polarized STM. With increasing perpendicular magnetic fields, the transition from chiral domain, to Néel-type skyrmion, and then fully saturated stated are observed by micromagnetic simulations, shown in (A)-(C), and are experimentally confirmed by SP-STM imaging, shown in (D)-(F). This enables the size of skyrmions as a function of field to be studied. (G) illustrates the writing and deleting scheme using a spin-polarized tunneling current from the STM tip. Reproduced with permission [53]. Copyright 2009, AAAS.

## 3.2 Blowing magnetic skyrmion bubbles

The previous section discussed how robust nm-scale Néel-type skyrmions can be created and stabilized at low temperature (4.2 K) and high magnetic field (1 T). However, for practical spintronics application, it is desirable to stabilize them at room temperature and under small, ideally zero magnetic field. While above it was already mentioned that spin-orbit torques can be used to move skyrmions, here it will be shown how SOT can in fact be utilized to create skyrmions in a process that resembles the formation of soap bubble when air is blown through a suspended thin film of soapy water.

In inversion asymmetric multilayers composed of at least a heavy metal and a ferromagnetic exhibiting PMA, an electrical current flowing in the heavy metal produces a SOT that acts on domain walls within the ferromagnet, as illustrated in Fig. 11(A). The resultant effective spin Hall field $\vec{B}_{sh}$ arising from the field-like torque can be written as follows [151, 152]:

$$\vec{B}_{sh} = B_{sh}^0\big(\hat{m}\times(\hat{z}\times\hat{j}_e)\big) \tag{11}$$

where $\hat{m}$ is the spin vector, $\hat{z}$ is the unit vector normal to the film plane, $\hat{j}_e$ is the electron current direction. The prefactor $B_{sh}^0 = (\hbar/2|e|)\cdot(\theta_{sh}J_c/t_f M_s)$, where the reduced Planck constant $\hbar = h/2\pi$, $e$, is the electronic charge, $t_f$ is the thickness of the FM layer, and $M_s$ is the saturation volume magnetization. The spin Hall angle is defined as $\theta_{sh} = J_s/J_c$, which is the ratio between spin current density ($J_s$) and charge current density ($J_c$) flowing in the heavy metal. This effective spin Hall field has been experimentally shown to result in efficient motion of chiral domain walls in narrow wires crated from these multilayered structures [151, 180]. In a wide wire, that is a wire wide enough to allow for the formation of band domains, the magnitude of $\vec{B}_{sh}$ will depend on the orientation of the domain wall relative to $\hat{j}_e$. The two extreme cases can be illustrated by considering a straight stripe domain with a fixed left-handed chirality exposed to a homogeneous current flowing along the long axis, as illustrated in Fig. 11 (A), the symmetry of *Eq*. (11) leads to a vanishing SOT torque acting on the side walls. In other words, if the spin vector $\hat{m}$ inside the wall is parallel to $\hat{z}\times\hat{j}_e$ as is the case in region ② and region ③ of Fig. 11(B), $\vec{B}_{sh} = 0$ and the domain wall does not move. However, the end of the band domain (marked as region ①) is moving with a maximum velocity $v_{dw}$ due to the maximized $\vec{B}_{sh}$ which would give rise to an elongation of the chiral band domain if the opposite end is pinned. These effects have been experimentally illustrated in a relatively wide wire of Ta(5nm)/Co$_{20}$Fe$_{60}$B$_{20}$(CoFeB)(1.1 nm)/TaO$_x$(3 nm), shown in Fig. 11(C) and 11(D), where randomly oriented band domains line up along the current direction after applying a current.

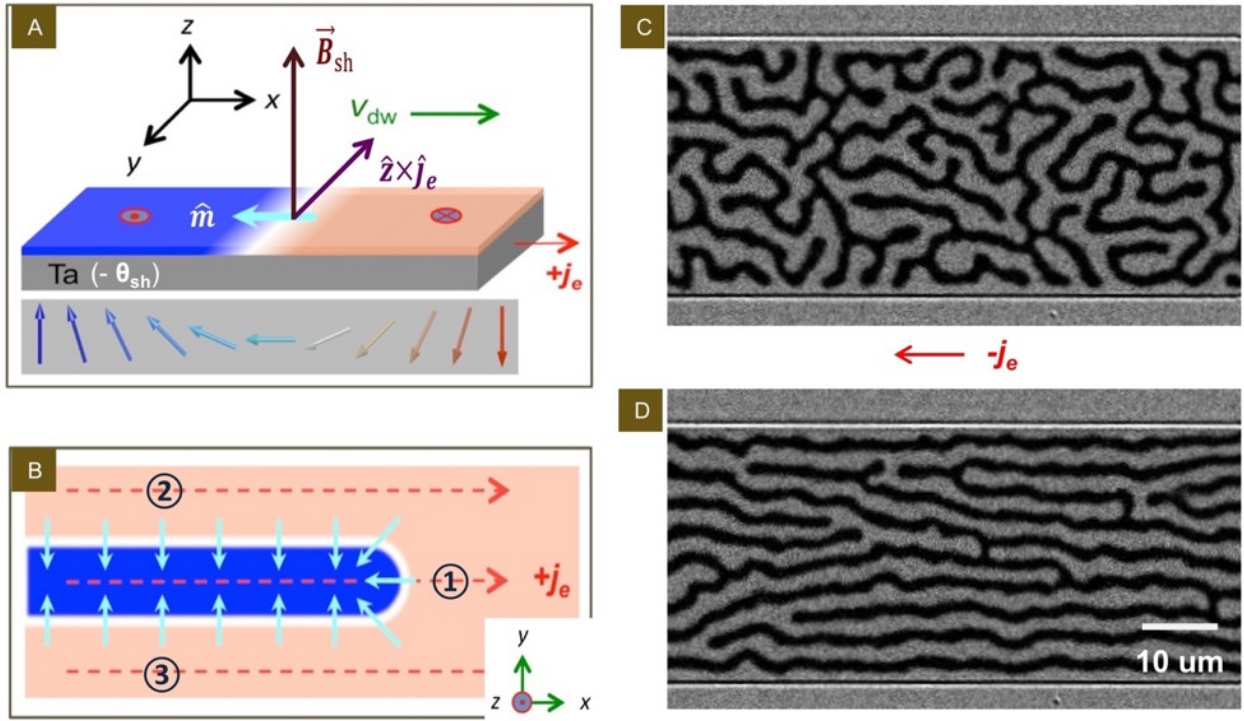

Fig. 11. Motion of chiral domain walls driven by SOTs. (A) is the illustration of how SOT acts on a left-handed (marked by the spin vector $\hat{m}$) chiral domain wall in a HM/FM bilayer. Here a HM with a negative spin Hall angle (-$\theta_{sh}$) is assumed for the under layer. The direction of chiral domain wall motion $v_{dw}$ follows the electron current direction $\hat{j}_e$ that results in the elongation of chiral stripe domain, regime ① shown in (B). Reproduced with permission [41]. Copyright 2015, AAAS. While the side wall with spin vector perpendicular to the current does not move due to a zero SOT effective field, shown in region ② and region ③. This theoretical picture was tested experimentally in a Ta/CoFeB/TaO$_x$ trilayer, shown in (C)-(D). These wires have a width of 40 μm and a length of 300 μm. A polar MOKE microscope in differential mode was used for imaging experiments at room temperature. The lighter area corresponds to a negative perpendicular magnetization orientation and darker area corresponds to a positive orientation. Shown in (C) is the typical labyrinthine domain configuration in the system with a weak perpendicular magnetic anisotropy. By applying a negative electron current $-\hat{j}_e$, it is observed that the disordered maze domain configuration was gradually reconfigured, forming a parallel stripe domain state. This is consistent with the predicted picture that the side of domain wall does not experience a SOT.

Introducing a geometrical constriction into this wide wire results in an interesting and complex scenario, as shown in Fig. 12A. It creates a spatially convergent/divergent current distribution along the *y* axis - $j_y$ around the narrow neck, as shown in Fig. 12B. Consequently, inhomogeneous effective forces - $\boldsymbol{F}_{sh}^y$ on the DWs arising from the spin Hall field are created along the *y* axis, which locally expand the chiral band domain, as shown in Fig. 12C. Following the continuous expansion of the band domain, due to the outward "pressure" from the effective spin Hall field, the surface tension inside the DW increases, which results in breaking the band domains into circular bubble domains, Fig. 12D. Note that this method differs from the recent theoretical study showing the conversion of a pair of domain walls into a skyrmion [198].

This mechanism for creating skyrmions was successfully demonstrated in Ta(5nm)/Co$_{20}$Fe$_{60}$B$_{20}$(CoFeB)(1.1 nm)/TaO$_x$(3 nm) trilayers at room temperature [41]. Figure 12 E and F show MOKE image of the device before and after a current pulse. On the right side of the device many skyrmions can be seen after applying an electric current. This process resembles how soap bubbles develop out of soap films upon blowing air through a straw. On the other hand, there exists a clear distinction between the formation of soapy bubble by blowing air through a straw and the skyrmion bubble formation through a geometrical constriction. In the former, the size of soapy bubbles is directly controlled by the inner diameter of straw and the amount of air trapped inside. In the later, the size of skyrmion bubbles is determined by the internal competition between the magnetostatic, exchange and DMI, which is independent on the width of geometrical constriction.

Due to the interfacial DMI in the trilayer, spin structures of the newly formed circular domains maintain a well-defined (left-handed) chirality and therefore have a unit topological charge characterized by the skyrmion number $Q$ = 1 [199]. These $Q$ = 1 topologically stable skyrmions, once being created are mobile at a low current density and move very efficiently following the electron current direction. Note that the dynamical creation of magnetic domain walls by electrical currents has not yet been observed. The high mobility of the skyrmions is due to the opposite direction of effective SOTs at their opposite sides. For the low current density of $j_e$ < 1 ×10$^5$ A/cm$^2$, a stochastic motion of skyrmion is observed, indicating the influence from random pinning.

Reversal of the current polarity results in the formation of skyrmions at the left side of device as opposed to the positive current. This directional dependence suggests that spatially divergent currents and SOTs are most likely responsible for transforming the chiral band domains into magnetic skyrmion bubbles. Interestingly, these skyrmions do not merge and in fact repel each other, indicating their topological protection as well as dipolar repulsion. Utilizing micromagnetic simulations that include SOTs, Lin [170] and Heinonen *et al.* [171] have demonstrated this conversion from chiral band domains into magnetic skyrmions due to diverging and inhomogeneous currents. Similar micromagnetic simulations have also been reported by Liu *et al*. [200]. Furthermore, Liu *et al.* have performed a topological density analysis and revealed that the creation of a topological skyrmion is accompanied with the nucleation and annihilation of an anti-skyrmion. This is independently confirmed by studying the dynamical creation of skyrmion-antiskyrmion pair by in-plane electrical currents [201]. In particular, Stier et al., has demonstrated that antiskyrmion is not a stable solution for a given Zeeman energy and DMI, and annihilated due to Gilbert damping [201]. These studies can provide further insights into future experimental investigations.

Experimentally, the dynamical generation of magnetic skyrmion by electrical currents has also been demonstrated in other material systems where the skyrmion size approaches 100 nm [202, 203]. Consequently, inhomogeneous currents provide a practical pathway for dynamical generation of magnetic skyrmions that can be extended into other relevant material systems, even potentially to chiral bulk magnets, as suggested by numerical simulation studies [204].

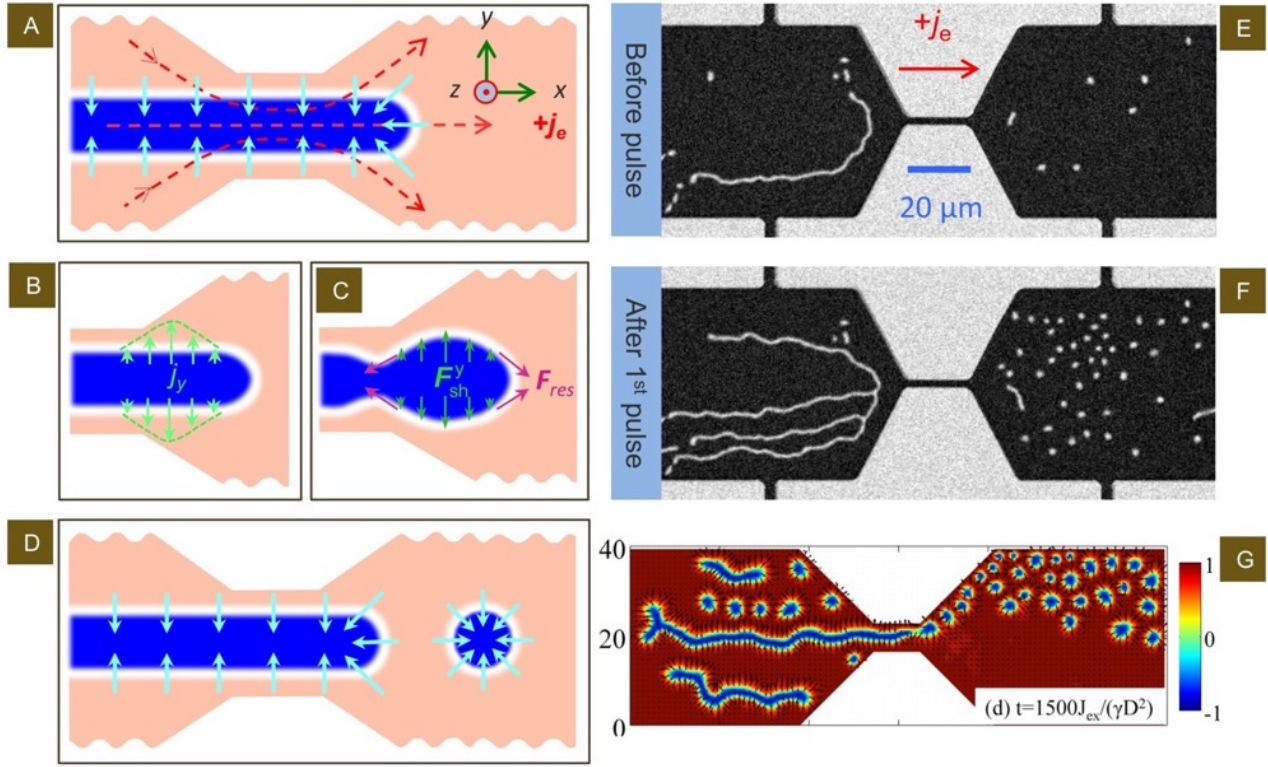

Fig. 12. Transformation from chiral band domain into magnetic skyrmions enabled by the SOT-induced magnetization instability. (A) illustrates the superposition of a left-handed chiral band domain and an inhomogeneous current distribution that is produced by a geometrical constriction. The current distribution around the narrow neck along the y-axis is further illustrated in (B). The inhomogeneous spin orbit effective field along the y-axis locally expands the chiral stripe domain, shown in (C). Ultimately a magnetic skyrmion is formed, following the increase of surface tension inside the domain wall, shown in (D). This hypothesis was experimentally tested in a Ta/CoFeB/TaO$_x$ trilayer with a geometrical constriction of 3-μm width and of 20-μm length in the center of which polar MOKE microscopy images are shown in (E)-(F). Before applying a current and with a perpendicular magnetic field of $B_\perp$ = +0.5 mT, sparse magnetic band and bubble domains were observed at both sides, as shown in (E). By applying a pulse current of amplitude $j_e$ = +5×10$^5$ A/cm$^2$ (normalized by the width of device – 60 μm) and of duration 1 s, it is observed that the band domains started to migrate, and subsequently form extended band domains on the left side, which converged at the left side of constriction. Interestingly at the right side of the device the band domains are transformed into skyrmion bubbles immediately after passing through the constriction, as shown in (F). Reproduced with permission [41]. Copyright 2015, AAAS. Shown in (G) are micromagnetic simulation results showing the generation of magnetic skyrmion by transforming the chiral band domain. This simulation study confirms the important role of current induced magnetization instability in the process of skyrmion formation, and also validates the schematics shown in (A)-(D). Reproduced with permission [204]. Copyright 2016, American Physical Society.

### 3.3 Moving skyrmions in wires

As demonstrated through micromagnetic simulations, shown in Fig. 13A, magnetic skyrmions can overcome the pinning potentials created by the structural defects [32]. Thus they can be driven efficiently along a nano-track by a current induced SOT in HM/FM/I asymmetric trilayers. This indicates the potential for implementing nanoscale magnetic skyrmions as information carriers in nonvolatile spintronic technology such as skyrmion race-track memory. Similar to the well-known domain wall racetrack memory, the presence or absence of magnetic skyrmions along the racetrack can be used to encode binary bits, 1 and 0, respectively. The memory operation is then enabled by applying a current to move the skyrmions along the racetrack (also known as bit line).

In the following, the experimentally determined motion of a single magnetic skyrmions in a single wire, similar to the proposed racetrack device concepts is discussed. As previously discussed and shown in Fig. 12, a geometrical constriction facilitates the formation of magnetic skyrmions, but it also impedes a systematic study of SOTs induced motion of skyrmions. This is due to the very efficient production of skyrmions through the constriction under the high current densities, which makes it difficult to identify the behavior of individual skyrmions. At the same time, to achieve faster motion of skyrmion and hence fast operation of skyrmions, the maximum current density that can be applied is limited by the width of the constriction.

This can be solved by using a "four-terminal" device that separates the skyrmion generation line and motion line. Fig. 13B shows the electrical control of a single skyrmion along such a skyrmion motion line. From the images shown in Fig. 13B, it can be seen that the skyrmion is displaced by about 10 μm for each 100 μs current pulse of $5 \times 10^5$ A/cm$^2$. This corresponds to a velocity of 0.1 m/s. Furthermore, the displacement is approximately identical for subsequent current pulses, neglecting the current shunting effect from the Hall leads. This indicates that for these higher current densities random pinning is inconsequential.

Very recently, magnetic skyrmions of smaller diameter (≈ 200 nm) were also stabilized and driven along a nanoscale racetrack with a velocity up to 120 m/s, at a current density of approximately $5 \times 10^7$ A/cm$^2$ in a multilayer made of Pt/Co/Ta trilayer [42, 151, 205, 206]. The advantage of choosing this material system will be elaborated in the next section. It should be emphasized here that while significant progress has been made, including establishing individual magnetic skyrmions onto a prototypical racetrack, and deterministic creation of magnetic skyrmions at defects by using SOTs [207], an electrical readout by using (topological) Hall effect or other means has not yet been reported. This is one major obstacle towards functional skyrmionic devices.

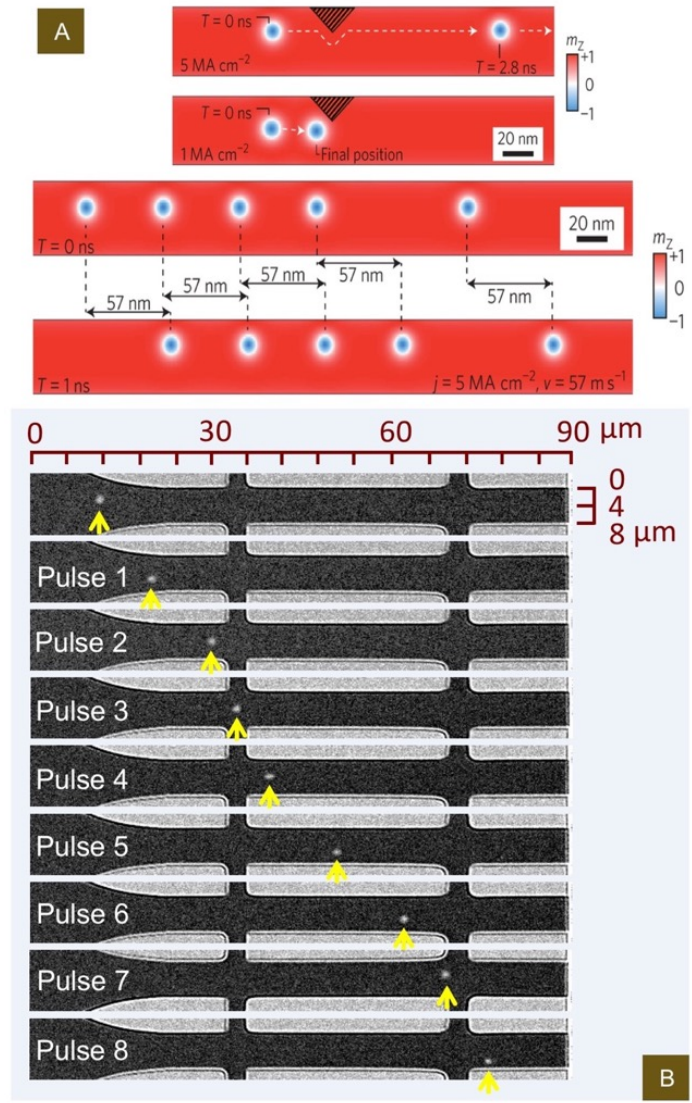

Fig. 13. Performance of magnetic skyrmions in a racetrack. (A) shows micromagnetic simulation results with stabile 10-nm Néel-type skyrmions. In a nanoscale racetrack, the coherent motion of a row of magnetic skyrmions with a velocity of 57 m/s was deterministically produced using a train of current pulses with a current density of $5 \times 10^6$ A/cm$^2$. Reproduced with permission [32]. Copyright 2013, Nature Publishing Group. Experimental demonstration using Polar MOKE microscopy images of the motion of a single magnetic skyrmion along a wire in a Ta/ CoFeB/TaO$_x$ trilayer, shown in (B). It can be seen that the magnetic skyrmion, with a diameter ≈ 1 μm, is displaced by about 10 μm for each 100-μs long current pulse of amplitude $5 \times 10^5$ A/cm$^2$, which yields a velocity of 0.1 m/s. Reproduced with permission [208]. Copyright 2016, AIP Publishing.

## 3.4 Magnetic skyrmions in asymmetric multilayers

In order to make magnetic skyrmions truly attractive for the next generation energy-efficient and ultra-high density data storage, and to reveal the room-temperature emergent topological physics, efficient manipulation and size miniaturization of magnetic skyrmions are extremely important. In this section, practical strategies for materials optimization of asymmetric multilayers will be discussed first. Through a comprehensive review of the role of each layer and their interfaces in the aforementioned magnetic heterostructures, it can be concluded that interfacing FM between two different HM layers is the most advantageous route for providing complementary spin-orbit torques and enhanced interfacial DMI strengths, as schematically shown in Fig. 14.

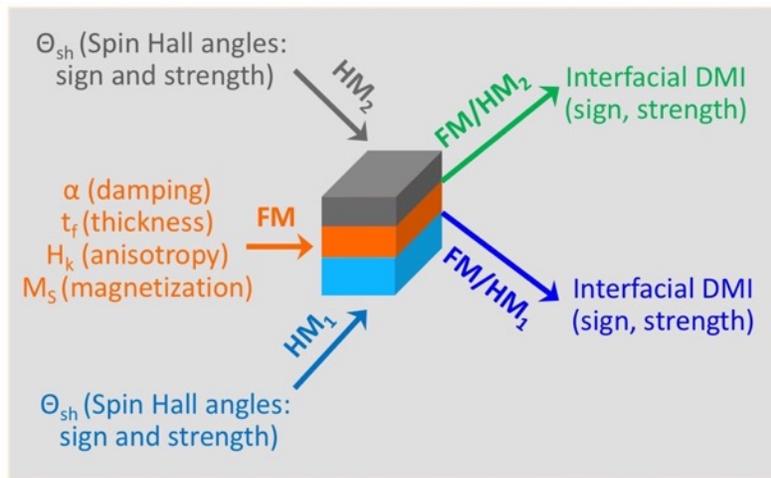

Fig. 14. Strategy for material optimization enabled by understanding the controlling factors of each layer and their interfaces in heavy metal1/ultra-thin ferromagnet/heavy metal2 ($HM_1$/FM/$HM_2$) multilayers.

In these systems, the non-magnetic (typically $5d$) transition metal HM with a large spin Hall effect injects a perpendicular spin current ($\bm{j}_s$) into the adjacent FM layer, enabling an efficient electrical manipulation of the magnetization dynamics. Hence, the sign and amplitude of the spin Hall angle ($\theta_{sh}$) of the HM layer can be adjusted with the appropriate choice of heavy metal. The HM/FM interface directly dictates the strength/sign of the interfacial DMI through orbital hybridization. Guided by first principle calculations, both the sign and strength of interfacial DMI component can also be optimized via proper designs of the stack structure. Material specific parameters of the FM layer such as damping parameter ($\alpha$), thickness ($t_f$), magnetic anisotropy ($K$), and saturation magnetization ($M_s$) that strongly influence magnetization dynamics can also be tailored. In HM/FM/I trilayer, the insulating layer mainly behaves as a capping layer that prevents stack from further oxidization, and may also contribute (partially) to the presence of PMA [14] and potentially modulate the interfacial DMI [209]. A comprehensive understanding of the underlying physics behind these various parameters in the stack structure thus provides guiding principles for optimizing skyrmion materials as will be discussed below.

To optimize the energy efficient skyrmion dynamics, one can choose different heavy metals with a larger spin Hall angle ($\theta_{sh}$) such as tungsten (W) with $\theta_{sh} \approx$ -35% [210] instead of Pt or Ta, or other exotic materials such as topological insulators with $\theta_{sh}$ in the range of 300 - 40000% (noted that both the determination and the physical origin of this extremely large spin Hall angle are still under debate [211]). However, making an ultra-thin ferromagnet with an interfacial dominant PMA on top of these materials is experimentally challenging due to different growth dynamics.

Choosing an ultra-thin FM layer with a low intrinsic magnetic damping parameter is a promising direction for efficient skyrmion operation. In this regard, CoFeB [69] or Heusler alloys [212, 213] are perhaps better choices than Co films. Indeed, very recently, the stabilization of magnetic skyrmions in Ta/Co$_2$FeAl/MgO hetero-structures with approximately 100 nm diameter is reported [214]. However, an efficient electric control of skyrmions in this material system is yet to be shown. Nevertheless, one should consider that additional interfacial spin scattering could occur at the HM/FM interface regardless of the material choices, which will significantly enhance the effective magnetic damping parameters and consequently reduce the operation efficiency [215]. For example, a damping parameter of $\alpha$ = 0.5 has been suggested for a magnetic multilayer [Pt$_{3.2nm}$/CoFeB$_{0.7nm}$/MgO$_{1.4nm}$]$_{14}$ [96], where the enhancement is most likely due to spin pumping and the strong spin sink effects of Pt [216].

Another practical approach for realizing efficient manipulation of Néel-type skyrmions is to replace the capping insulator layer with a different HM layer (denoted as HM$_2$) that exhibits an opposite sign of spin Hall angle as compared with the seeding HM layer (denoted as HM$_1$), as illustrated in Fig. 14. Such HM$_1$/FM/HM$_2$ trilayer systems may still maintain a PMA and an interfacial inversion symmetric breaking, and thus can potentially host a skyrmion phase. Since top and bottom HM layers have opposite signs of the spin Hall angle, this trilayer system accommodates complimentary spin-orbit torques at the opposite interfaces of the FM layer, as is schematically illustrated in Fig. 15A. The resultant additive SOTs acts constructively on the magnetization dynamics and thus results in an efficient electrical manipulation and translation motion of magnetic skyrmions [151, 205, 206]. Experimentally, by utilizing different signs and strengths of the spin Hall angle from Pt ($\theta_{sh} \approx$ +10%) and Ta ($\theta_{sh} \approx$ -20%), Woo *et al*. have demonstrated the complementary SOTs induced perpendicular magnetization switching in a Pt/Co/Ta trilayer with an effective spin Hall angle of $\theta_{sh} \approx$ 34% [205].

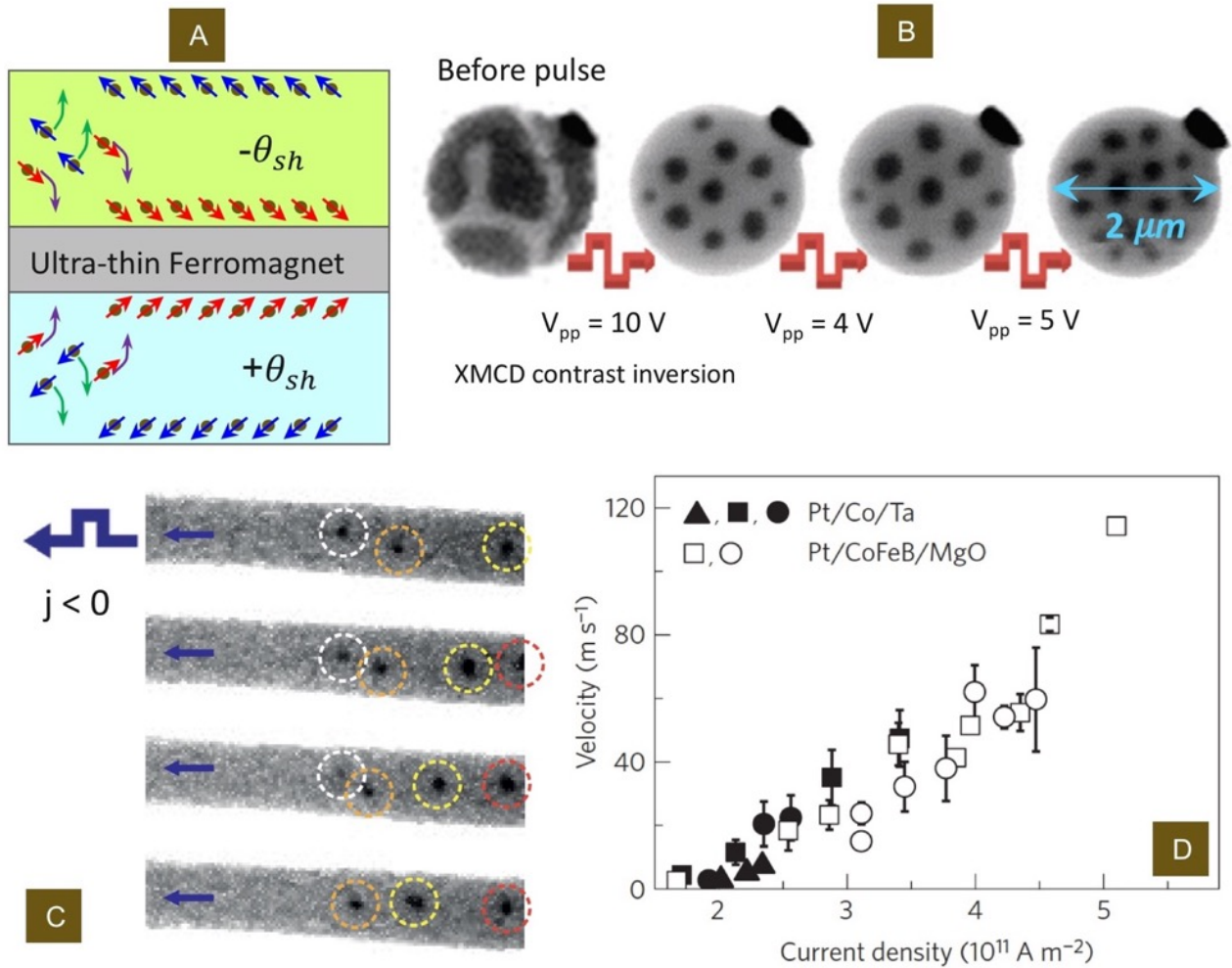

Fig. 15. Stabilization of 200 nm skyrmions in interfacially asymmetric $(Pt_{3nm}/Co_{0.9nm}/Ta_{4nm})_{15}$ trilayer based multilayers. (A) illustrates the origin of complimentary spin orbital torques by taking advantage of opposite spin Hall angles. (B) are the corresponding XMCD images that capture the transformation from chiral band domains into magnetic skyrmions enabled by pulsed magnetic fields. The size of the skyrmions is estimated to be ≈ 200 nm. (C) is the electrical current induced motion of skyrmions along a nanoscale racetrack. (D) is the velocity vs. current density dependence, with a skyrmion velocity up to 120 m/s at a current density of $j_e ≈ 5 \times 10^7$ A/cm$^2$. Reproduced with permission [42]. Copyright 2016, Nature Publishing Group.

Furthermore, in a [Pt$_{3nm}$/Co$_{0.9nm}$/Ta$_{4nm}$]$_{15}$ multilayer with a large DMI of value $D_{DMI} = 1.3 \pm 0.2 \ mJ/m^2$, Woo *et al.* have successfully observed magnetic skyrmions of diameter $\approx 200$ nm at room temperature. Since the skyrmion state in this multilayer is not the ground state, Woo *et al.*, applied pulsed magnetic fields from an "Ω" shaped loop antenna that transformed the virgin band magnetic domains into magnetic skyrmions [42]. Due to geometrical confinement, the resultant skyrmions in the nanodisk form a hexagonal lattice, as shown in Fig. 15B. However, the size of the magnetic skyrmions differs from each other, which might be related to spatial variations of the DMI and magnetic anisotropy, as well as pinning of the domain wall boundary. In contrast to the expected weak pinning effect due to the topological protection, a strong pinning effect is observed which directly indicates that Co thin-film is not a good material choice for realizing efficient current driven dynamics of magnetic skyrmion. Nevertheless, Woo *et al.* have observed a skyrmion velocity up to 120 m/s with the current density at the level of 5 × 10$^7$ A/cm$^2$.

The stronger DMI generally leads to smaller periodic of chiral spin textures (skyrmion size) (see section 2.1). It is important to note that flipping the interface, *e.g.,* reversing the growth order of the stacking, results in the sign change of the interfacial DMI. This fact is associated with the orientation of the DMI vector at interfaces [31]. Therefore, building binary multilayer is usually expected to suppress the total DMI. On the other hand, asymmetric trilayer provide additional degrees of freedom to tailor the interface DMI without DMI cancellation, *e.g.*, boosting the DMI by various combinations of the DMI at interfaces adjacent to the FM layer, or by repeating the stacking number. The influence of the asymmetric trilayer stacking on the effective DMI was first tested in [Fe/Ni/Cu]$_4$ multilayers, where the enhancement of the DMI was found due to the increased number of interfaces with DMI, resulting in the stabilization of chiral Néel-type spin textures in thicker films [136]. Boosting the total DMI by combining two FM/HM interfaces with different DMI strength was also proposed, and experimentally demonstrated in [Co/Ir/Ni]$_n$ multilayer grown on Cu(111) single crystal [217].

Among all the choices, HM$_1$/FM/HM$_2$-based asymmetric trilayer may potentially be the most efficient way to enhance the total DMI, due to the wide range of possible choices of the DMI parameters at FM(3$d$)/HM(5$d$) interfaces. In this case, the enhancement of the DMI, which could help to minimize the skyrmion size, can be achieved by selecting two interfaces (FM/HM$_1$ and FM/HM$_2$) with opposite signs of the DMI, because the DMI at FM/HM$_1$ interface will reverse its sign, so that both HM$_1$/FM interface (flipped) and FM/HM$_2$ interface contribute to the same sign of the DMI, as shown in Fig.16 A. In addition, choosing the opposite signs of spin Hall angles between HM$_1$ and HM$_2$ could boost the motion of magnetic skyrmion. The Pt/Co/Ta trilayer is thus not a perfect material choice since although the spin-orbit torques are additive, the two DMI contributions cancels each other partially. This is because the Pt/Co interface exhibits a stronger DMI of approximately 1.4 mJ/m$^2$, as compared with the interfacial DMI at the Ta/Co interface which is much smaller than 0.5 mJ/m$^2$ and of the same sign. It is recently suggested that both the sign and amplitude of interfacial DMI at the Ta/CoFeB interface can be complex as a result of B diffusion into Ta layer [218, 219].

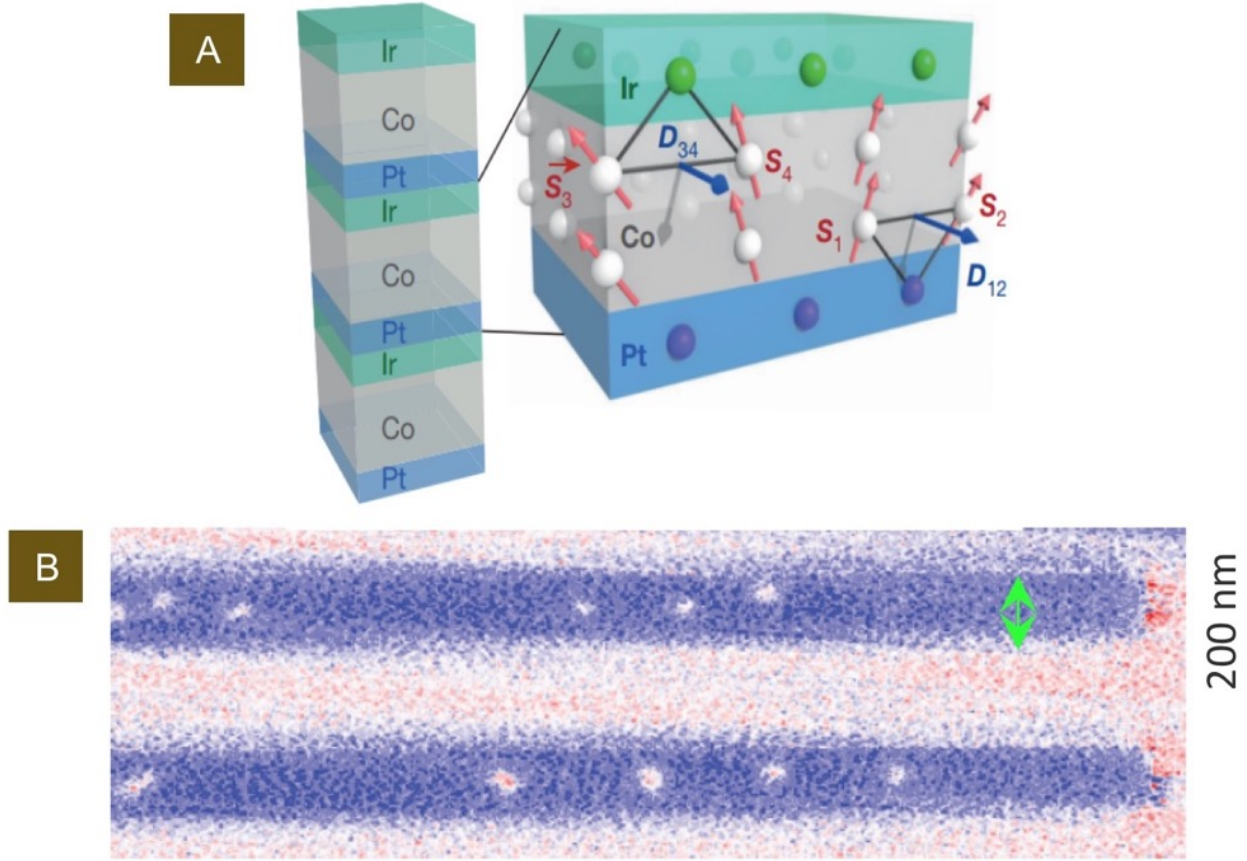

Fig. 16. Stabilization of nanoscale skyrmions in interfacially asymmetric trilayer based multilayers. (A) illustrates the working mechanism of interfacial DMI at the Pt/Co/Ir interfaces. Due to the opposite signs of the interfacial DMI vectors, an additive interfacial DMI is observed, as shown in (A). These additive interfacial DMI components can be utilized to stabilize room temperature skyrmion of size approaching 50 nm, as shown in (B).  Reproduced with permission [43]. Copyright 2016, Nature Publishing Group.

By noting this important factor, in accessing the additive DMI from the opposite interfaces of an ultrathin Co layer, in a Pt/Co/Ir trilayer based magnetic multilayer, room temperature magnetic skyrmions of approximately 50 nm size have been experimentally stabilized, as shown in Fig. 16 B. This occurs as a direct result of the opposite sign of interfacial DMI between the Co/Pt and the Co/Ir interfaces [43, 203], adding them together at the opposite interfaces of the Co layer, which thus boosts the combined strength of interfacial DMI to approaching 2 mJ/m$^2$. Furthermore, guided by first principles calculations, in a Pt/Co/Fe/Ir multilayer, even stronger DMI is theoretically proposed [143] and experimentally confirmed [220]. Although the Pt/Co/Ir trilayers offer an additive interfacial DMI that helped to minimize the size of skyrmions, it however, fails to provide efficient electrical manipulation due to the negligible spin Hall effect of the Ir layer. In this regard, an optimized material system is not yet available, which requires further investigations in this direction in the future. In this regard, a multilayer made of Pt/CoFeB/W might be more advantageous by providing simultaneously both complimentary SOTs [210, 221] and interfacial DMIs [44, 222]. Note that a proper design and precise control of interfacial DMI in magnetic heterostructures can be much more complicated, since interfacial DMI depend strongly on the details of the local microstructure as well as the growth dynamics of ultra-thin films. This probably explains the recent observation of Néel-type skyrmion in the nominally inversion symmetric [Co/Pd]$_{X=5}$ multilayer [223].

Another important, yet unanswered question, is the following: what is the dominant factor that leads to the skyrmion phase in trilayer-based multilayers? In PMA systems, the formation of periodic band and bubble domain is closely related to the effective magnetic anisotropy when it competes with the exchange and dipolar interactions, *i.e.*, smaller effective magnetic anisotropy leads to a narrower domain size [224]. One clear evidence can be found in an interfacially asymmetric Fe(wedge)/Ni(10.6 monolayer) grown onto Cu(001) single crystal in which densely packed magnetic bubbles were identified by using PEEM [225]. Note that these magnetic bubbles may well be skyrmions due to the presence of interfacial asymmetry that could potentially establish a chiral DMI in the system, as suggested by a SPLEEM experiment performed in a similar material system [136]. The monotonically decreasing size of the bubble domains (and hence effective magnetic anisotropy) as a function of increasing the Fe layer thickness in the range between 3.4 ML and 3.8 ML can thus be useful in the future for miniaturizing skyrmions in magnetic multilayer systems. In this sense, a comprehensive examination of the evolution of domain patterns in the perpendicularly magnetized system can be beneficial for future skyrmion materials optimization [224, 226].

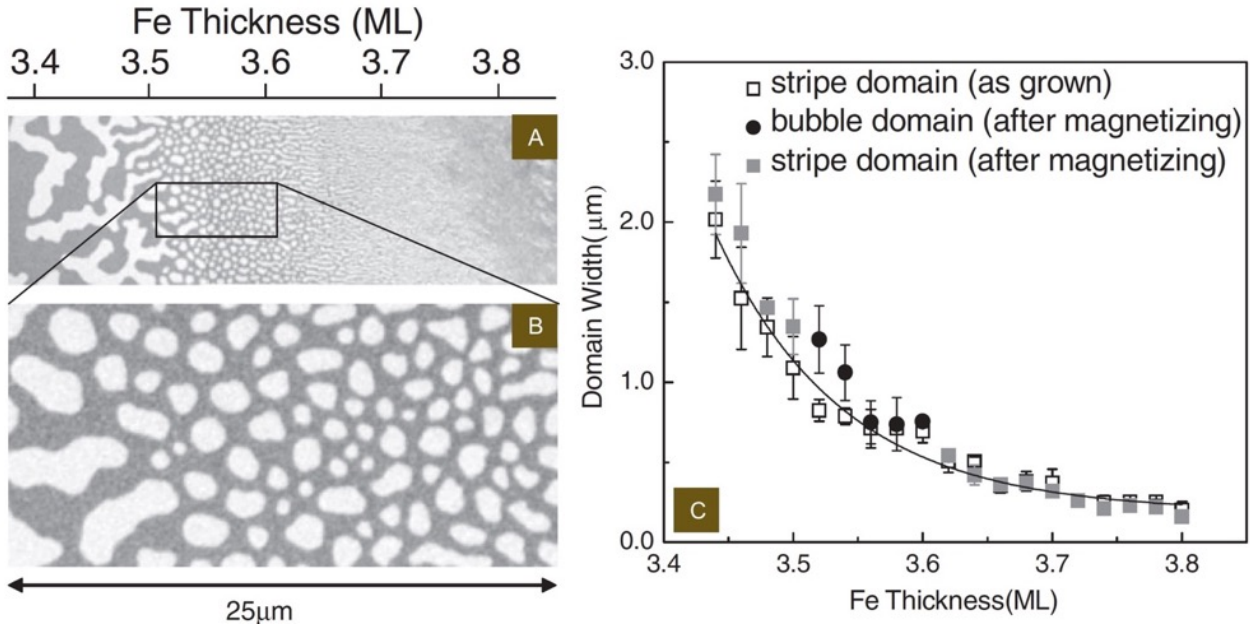

Figure 17. Evolution of magnetic domain states after applying a 100 mT in-plane field in an asymmetric Fe(wedge)/Ni(10.6 monolayer)/Cu(001) system. (A) and (B) The bubble domain phase can be seen for Fe thicknesses in the range of 3.4 ML - 3.8 ML, as revealed by a PEEM experiment. (C) shows the domain width evolution as a function of the Fe thickness, in which the size of bubble varies monotonically from 2 μm to approximately 200 nm. Reproduced with permission [225]. Copyright 2007, American Physical Society.

Another example is the multilayer $[HM_1/FM/HM_2]_{X=1}$ with a single repetition (X = 1) or smaller repetition number (X < 4), where the system typically exhibits a strong PMA and no skyrmion phase [223]. The multi-domain or skyrmion phase appears following the increase of layer repetition. A clear evidence can be seen in $[Co_{0.4nm}/Pt_{0.7nm}]_X$ multilayers, where Hellwig *et al.* have revealed an evolution of complex domain state from big band domain into nanoscale labyrinthine and bubble domains as a function of repetition number X. The occurrence of this behavior can be attributed to the competition between magnetostatic energy and domain wall energy [227, 228]. In addition, an evolution of the domain width as a function of repetition (total thickness) is also found with a minimum width approaching 100 nm at X ≈ 20, as shown in Fig. 18. The stripe and bubble domains observed in this nominally inversion symmetric system may well be chiral due to the complex growth dynamics of Pt grown on Co, and vice versa [227]. A detailed theoretical investigation of the underlying competing interactions in magnetic multilayers will hopefully answer the above critical question and provide clear guiding principles for optimizing room-temperature skyrmion materials.

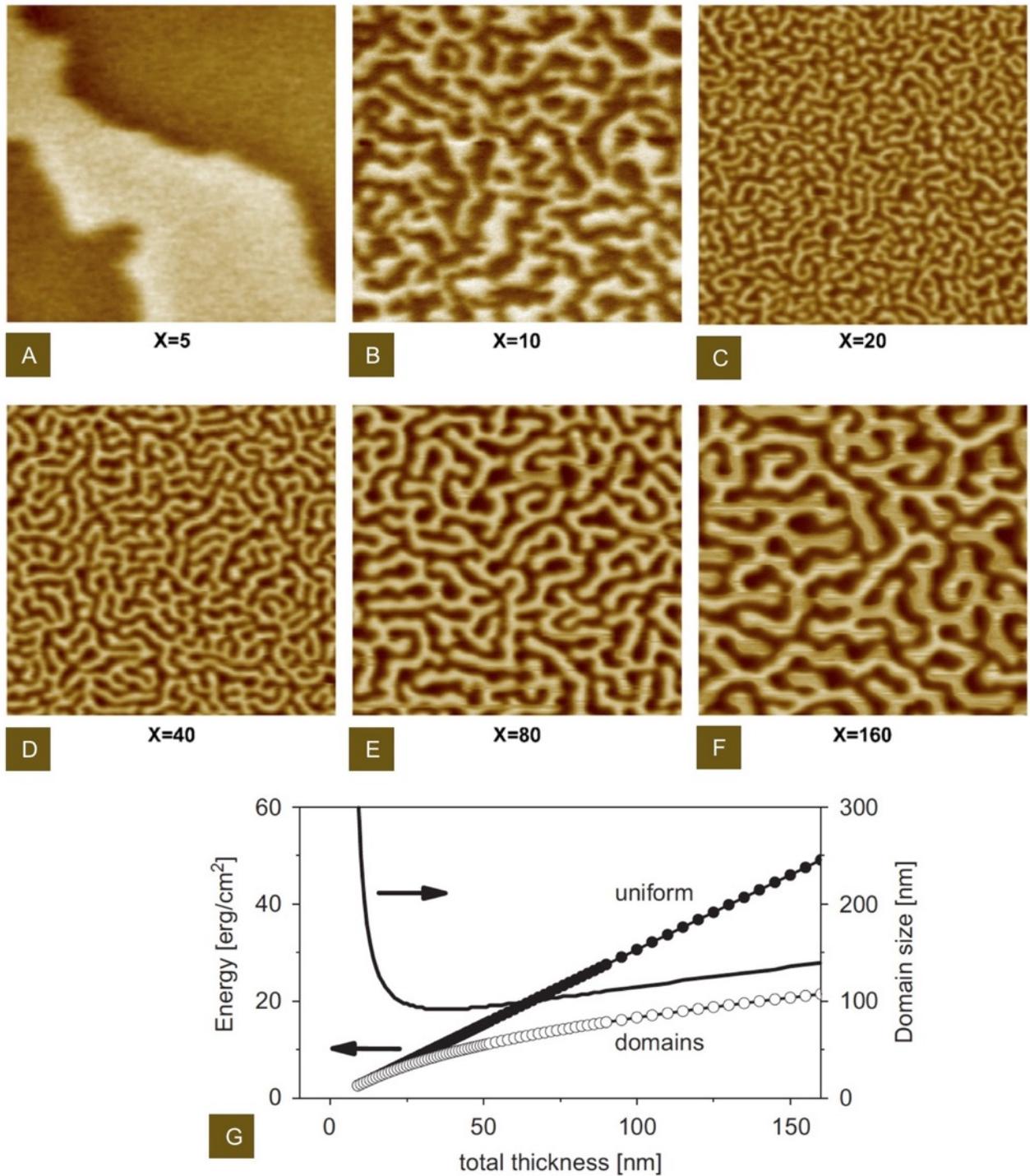

Fig. 18. Evolution of complex domain patterns in $[Co_{0.4nm}/Pt_{0.7nm}]_X$ multilayers. Shown in the upper panel are the corresponding magnetic force microscopy images acquired after *ac* demagnetization. The size of each figure is 5 μm$^2$. X denotes the number of repetition. The calculated magnetostatic energy versus domain wall energy is shown in the lower panel of (G), from which a minimum domain size is found around X ≈ 20. Reproduced with permission [228]. Copyright 2007, Elsevier.

## 3.5 Characteristics of topological trivial bubbles

As previously mentioned, the detailed topology of spin textures distinguishes whether a magnetic bubble is a topologically non-trivial skyrmion ($Q = 1$) or a topologically trivial magnetic bubble ($Q = 0$). It will be also intriguing to investigate the current driven dynamics of topological trivial bubbles [229], where the spin topology is illustrated in Fig. 19A. Due to the competition between long-range dipolar and short-range exchange interaction, a system with weak PMA typically undergoes a spin reorientation transition with in-plane magnetic fields that is manifested by a band-to-bubble domain phase transition. Such an in-plane field induced magnetic bubble state is established by sweeping the magnetic field from $B_\parallel$ = +100 mT to $B_\parallel$ = +10 mT in the same material system as before [41] – Ta(5 nm)/CoFeB(1.1 nm)/TaO$_x$(3 nm) trilayer. These in-plane magnetic field-induced magnetic bubbles are shrinking and vanishing in the presence of a positive electron current density, see Figs. 19(B)–(F), or elongate and transform into stripe domains in the presence of negative electron current density, see Figs. 19(H)–(L). This observation is in stark contrast with the current driven motion of topological nontrivial skyrmion ($Q = 1$), thus directly indicates the different spin structures surrounding these field induced bubbles, and thereby different topological skyrmion numbers.

For the magnetic bubbles induced by in-plane field the corresponding topological skyrmion number is determined to be $Q = 0$, since the spin structures within the DWs follow the external magnetic fields, as schematically illustrated in Fig. 19(A). This can also be derived by using the spherical integral of the skyrmion number, *Eq*. (3), which shows a solid angle of zero degree thus a zero skyrmion number. Due to the same direction of the spin Hall effective fields given by the reversed DW orientations, topologically trivial $Q = 0$ magnetic bubbles therefore experience opposite forces on the DWs at opposite ends. This leads to either a shrinking or elongation of the bubbles depending on the direction of currents, which is consistent with the experimental observations in a Ta(5 nm)/CoFeB(1.1 nm)/TaO$_x$(3 nm) trilayer. It also provides an alternative explanation of the in-plane electrical current induced perpendicular magnetization switching in the presence of in-plane fields, in addition to the coherent rotation of macro spin model [69, 230, 231]. Note that bubble-shaped spin textures have also been observed in the GdFeCo single layer of thickness 10 nm [232], in which the spin transfer torque induced dynamics of these magnetic bubbles are quite similar to the aforementioned topologically trivial bubbles (expansion upon applying current), as expected from the absence of interfacial DMI.

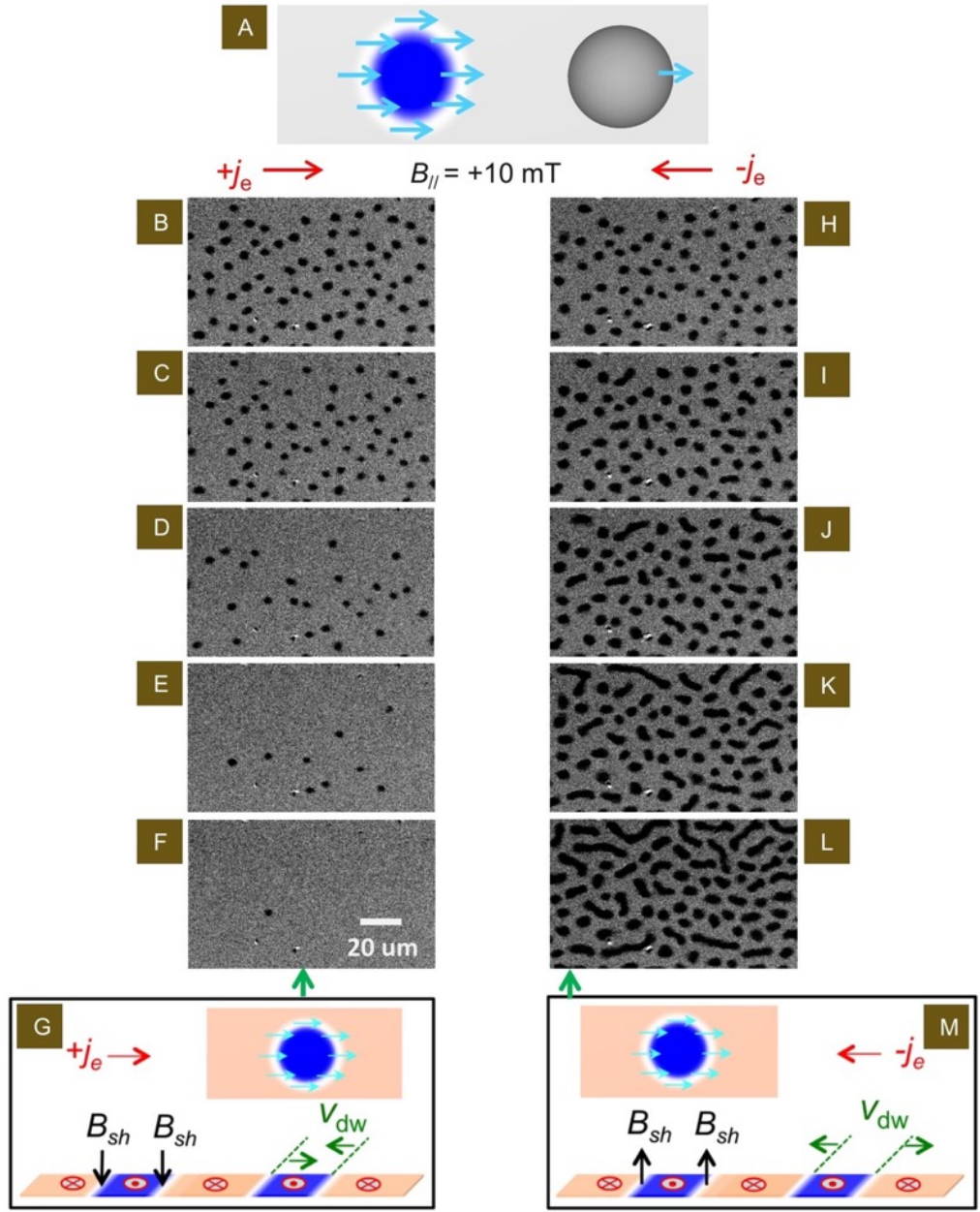

Fig. 19. Characteristics of topologically trivial bubbles. (A) Illustration of the spin profile of a topologically trivial bubble with a skyrmion number $Q = 0$ as can be created by applying an in-plane field. (B)-(L) Polar MOKE microscopy images of magnetic bubbles created by in-plane magnetic field $B_\parallel = +10$ mT in a Ta/ CoFeB/TaO$_x$ trilayer. Depending on the direction of the current, these magnetic bubbles are either subjecting to shrinking or expansion. (B) – (F) present the shrinking of bubbles following the increasing current density from $j_e = + 5 \times 10^4$ A/cm$^2$ to $+ 2.5 \times 10^5$ A/cm$^2$ in $5 \times 10^4$ A/cm$^2$ steps. (H) – (L) corresponds to the expansion of bubble for current from $j_e = - 0.5 \times 10^5$ A/cm$^2$ to $- 2.5 \times 10^5$ A/cm$^2$ in $5 \times 10^4$ A/cm$^2$ steps. These results directly indicate a varying chirality of the spin texture along the domain wall that is stabilized by the in-plane magnetic field, which therefore bears varying orientations of spin Hall effective fields, and implies $Q = 0$ bubbles. Reproduced with permission [41]. Copyright 2015, AAAS.

## 3.6 Hall effect of topological charge – skyrmion Hall effect

The well-known Hall effect in conductors describes the transverse deflection, and eventually accumulation of charged particles (electrons or holes for examples in conductors) in the presence of a perpendicular magnetic field, as a result of the Lorentz force. Besides the ordinary Hall effect there are other related phenomena including anomalous Hall effect [79], quantum Hall effect [233], spin Hall effect [12, 13, 177] and quantum anomalous Hall effect [234]. Research on this topic has largely enriched our fundamental understanding of the charge and spin transport, especially in low dimensional quantum materials. On the other hand, it is also interesting to reveal if other types of "charged" particle, especially particle with a topological charge, exhibit similar features as the conventional Hall effect, namely deflection and accumulation upon uniform driving. Magnetic skyrmions with well-defined topological charge $Q = \pm 1$ thus serve as perfect candidates to test this hypothesis. Research along this direction reveals the intriguing topological transport of magnetic skyrmions, but also confirms the spin topology for magnetic skyrmions. Furthermore, since the topological charge is an intrinsic quantity of magnetic skyrmions, it could enable a direct observation of the deflection, and accumulation of magnetic skyrmions in real space upon applying a uniform driving mechanism.

Assuming a Néel skyrmion as a rigid point-like quasiparticle (namely no distortion of its shape during motion), the translational motion that is driven by the SOT from the spin Hall effect can be studied by a modified Thiele equation [95, 235]:

$$\boldsymbol{G} \times \boldsymbol{v} - 4\pi\alpha \boldsymbol{\mathcal{D}} \cdot \boldsymbol{v} + 4\pi \boldsymbol{\mathcal{B}} \cdot \boldsymbol{j}_c = \boldsymbol{0} \qquad (12)$$

Here $\boldsymbol{G} = (0, 0, -4\pi Q)$ is the gyromagnetic coupling vector. $\boldsymbol{v} = (v_x, v_y)$ is the skyrmion drift velocity along the $x$ and $y$ axis, respectively. $\alpha$ is the magnetic damping coefficient, and $\boldsymbol{\mathcal{D}} = \begin{pmatrix} \mathcal{D}_{xx} & \mathcal{D}_{xy} \\ \mathcal{D}_{yx} & \mathcal{D}_{yy} \end{pmatrix}$ is the dissipative force tensor, in which $\mathcal{D} = \mathcal{D}_{xx} = \mathcal{D}_{yy} = \frac{1}{4\pi} \int \frac{\partial \mathbf{m}}{\partial x} \cdot \frac{\partial \mathbf{m}}{\partial y} dx dy$, and $\mathcal{D}_{xy} = \mathcal{D}_{yx} = 0$. The tensor $\boldsymbol{\mathcal{B}}$ quantifies the efficiency of the spin Hall spin torque over the 2-dimensional spin texture of the skyrmion, $\boldsymbol{j}_c = \boldsymbol{j}_s / \theta_{sh}$ is the electrical current density flowing in the heavy metal, $j_s$ is the spin current density, $\theta_{sh}$ is the spin Hall angle of the heavy metal.

The first term in *Eq*. (12) is the topological Magnus force that results in the transverse (gyrotropic) motion of magnetic skyrmions with respect to the laterally homogeneous driving current. This term thus acts equivalently to the Lorentz force for electric charge, and results in a Hall-like response of magnetic skyrmions, including skyrmion deflection and accumulation. The second term is the dissipative force that is linked to the intrinsic magnetic damping of a moving magnetic skyrmion, and the third term is the driving force from the spin-orbit torque. Based on the Thiele equation, a micromagnetic simulation study of the dynamics of magnetic skyrmion was conducted and presented in Fig. 20A [95].

By examining the definition of the topological charge: $Q = 1/4\pi \int \mathbf{m} \cdot (\partial_x \mathbf{m} \times \partial_y \mathbf{m}) \, dxdy$, it is found that $Q$ is an odd function of the magnetization vector $\mathbf{m}$, and therefore reverses its sign upon inversion of the spin textures. The sign reversal of magnetization $\mathbf{m}$ and concomitant topological charge can be done by reversing the polarity of the perpendicular magnetic fields. This sign reversal thus leads to the opposite direction of the topological Magnus force due to the fact that $\mathbf{G} \times \mathbf{v}$ is linked to the sign of topological charge since $\mathbf{G}$ = (0, 0, -4πQ). This is thus equivalent to the sign reversal of the Lorentz force of charge carriers in conductors upon reversal of magnetic field. As a result, the topological Magnus force produces the following real-space skyrmion transport phenomena: (1) the development of a transverse motion with respect to a uniform lateral driving current, and (2) a macroscopic accumulation of skyrmions at the edges of devices orthogonal to the current direction. These two interesting aspects will be discussed, and have been experimentally confirmed. This behavior consequently resembles the electronic Hall effect of electrons or holes in conductors in the presence of a perpendicular magnetic field, Figs. 20 B – C; therefore this behavior of topological charged particle motion is called the *skyrmion Hall effect*, Figs. 20 D – E. However, note that in the conventional electric charge Hall effect, reversing perpendicular magnetic fields from $(0, 0, B_z)$ to $(0, 0, -B_z)$, does not change the sign of the associated charge carriers, but instead, reverses the sign of the Lorentz force from $\mathbf{v} \times (0, 0, B_z)$ to $\mathbf{v} \times (0, 0, -B_z)$. This is not the case in the skyrmion Hall effect where the sign of topological charge does reverse which serves as the difference between the conventional charge Hall effect and the skyrmion Hall effect from the topological charge.

Upon applying a uniform electrical current along the $x$ direction $\mathbf{j}_c = (j_x, 0)$, the resultant skyrmion velocity along the $x$ and $y$ axes can be computed respectively by solving Thiele's equation shown in *Eq.* (12) [235-237]:

$$v_x = \frac{-\alpha \mathcal{D}}{Q^2 + \alpha^2 \mathcal{D}^2} B_0 j_x, \quad v_y = \frac{Q}{Q^2 + \alpha^2 \mathcal{D}^2} B_0 j_x \qquad (13)$$

Here $B_0$ is a constant that can be estimated based on the detailed spin configuration. This leads to the expression for the ratio of in-plane velocity components $(v_y/v_x)$ to be written as follows:

$$\frac{v_y}{v_x} = -\frac{Q}{\alpha \mathcal{D}} \qquad (14)$$

where $\alpha$ is the damping parameter. For example, for the previously mentioned Ta/CoFeB system ≈ 0.02. From *Eq.* (13), for a fixed direction of driving current $j_c = (j_x, 0)$, it can be seen more clearly that the ratio of $v_y/v_x$ changes its sign upon inverting the sign of topological charge $Q$, which is consistent with the previous discussion of the sign change of the topological Magnus force. The skyrmion Hall angle, representing the deflection angle of the skyrmion motion relative to the current direction, is defined as:

$$\Phi_{sk} = tan^{-1} \left( v_y / v_x \right) \qquad (15)$$

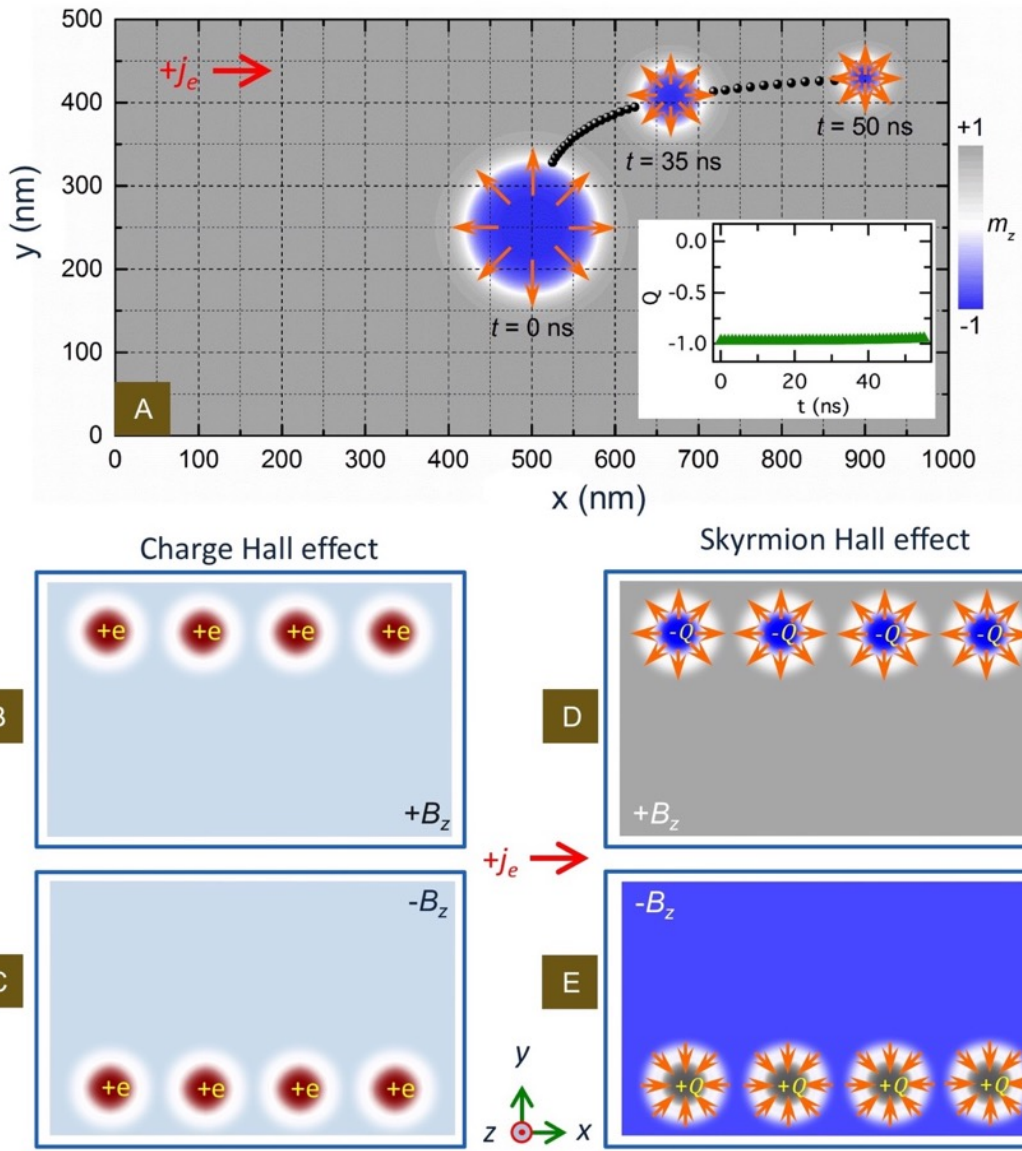

Figure 20. Illustration of Hall effects for electronic charge and topological charge. (A) Micromagnetic simulation result of a Néel skyrmion driven by a positive electron current $+j_e$ along a wire, from which one can see the presence of transverse velocity ($v_y$) with respect to the current along the $+x$ direction. Note that during the motion along the edge of the device, the size of the skyrmion shrinks due to the dipolar repulsive force from the edge. Blue color corresponds to a negative perpendicular (out of plane) magnetization ($-m_z$), Gray color corresponds to a positive magnetization ($+m_z$). A constant topological charge $Q = -1$ during the motion of skyrmion is found and shown in the inset. (B) – (E) Comparison between the electronic Hall effect and skyrmion Hall effect. (B) – (C) illustrate the Hall effect for holes with a unit electronic charge of $+e$, that accumulate at the opposite edges of a device upon reversal of the magnetic field directions. (D) – (E) illustrate the skyrmion Hall effect, for which the reversal of the magnetic field direction reverses the sign of the topological charge from $Q = -1$ to $Q = +1$ leading to the accumulation of skyrmions of opposite topological charges at opposite edges of the device. Reproduced with permission [95]. Copyright 2017, Nature Publishing Group.

The value of the dissipative force tensor $\mathcal{D}$ depends on the evolution of the magnetic moments within the skyrmion, and as such depends on the size of the skyrmion. Assuming a simple linear change of the angle of the spins inside the Néel domain wall along the radial direction ($\rho$) as the spins transition from "up" in the core to "down" outside the skyrmions, or vice versa, the dissipative force tensor element $\mathcal{D}$ can be reduced to:

$$\mathcal{D} = d\pi^2 / 8\gamma_{dw} \tag{16}$$

Thus, it is clear to see that $\mathcal{D}$ is proportional to the skyrmion diameter $d$ and depends inversely on the domain wall width $\gamma_{dw} = \pi\sqrt{\mathcal{A}/K}$.

Together with *Eq.* (16), one can easily see that for small Néel skyrmions without a homogenously magnetized core ($d = \gamma_{dw}$) $\mathcal{D} \approx 1$ and thus $v_y/v_x \gg 1$, results in mostly transverse motion and thus a skyrmion Hall angle close to 90 degrees, while for larger skyrmions with a uniformly magnetized core ($d \gg \gamma_{dw}$) the perpendicular motion is less pronounced with $v_y/v_x \approx 1$ (*i.e.* a smaller skyrmion Hall angle). In either case, this should lead to the accumulation of skyrmions at the device edge, as indeed was demonstrated by micromagnetic simulations shown in the Fig. 20A. Similarly, by changing the sign of topological charge, the skyrmion Hall angle also changes its sign. As previously discussed, according to the definition of topological charge $Q$ which integrates from zero to infinity in spherical coordinates, the spin topology of the skyrmion spin texture is independent of the size. This is shown in the inset to Fig. 20A, where the size of the skyrmion changes from 160 nm to 50 nm, but the calculated topological charge remains constant at $Q = -1$. Note that during the motion along the edge, the size of skyrmion shrinks due to the dipolar repulsive force from the edge.

Experimentally the skyrmion Hall effect should have been observed in the same system previously discussed in section 3.2, for which the dynamic generation of Néel skyrmions with electrical currents was demonstrated. In those initial experiments, a transverse motion was not conclusively observed [41, 42, 95], in contrast to the theoretical modeling [204]. However, subsequent studies show a current dependence of the skyrmion Hall effect and angle [95]. Individual polar-MOKE images taken after applying a relatively low current density $j_e = +1.3 \times 10^6$ A/cm$^2$ induced are shown in Figs. 21 A – F for $Q = -1$ skyrmions and in Figs. 21 G –L for $Q = +1$ skyrmions. The red arrow refers to a positive electron motion direction ($+j_e$) from left to right. By comparing the trajectories shown in Figs. 21 F for the $Q = -1$ skyrmion and Fig. 21 L for the $Q = +1$ skyrmion, a stochastic motion of magnetic skyrmions is clearly seen, without a net transverse component. By increasing the current density to $j_e = +2.8 \times 10^6$ A/cm$^2$, it is observed that the direction of the skyrmion motion develops a well-defined transverse component, which is exemplified by a straight and diagonal trajectory. Figs. 21 M – R correspond to a $Q = -1$ skyrmion, and Figs. 21 S – X to a $Q = +1$ skyrmion. The opposite sign of slopes in Figs. 21 R and X are consistent with the opposite sign of the topological Magnus force, that gives rise to opposite directions for the transverse motion, as suggested by *Eq*. 12. By reversing the electron current direction, the direction of motion is also reversed.

When the applied current density is larger than $|j_e| > 8 \times 10^6$ A/cm$^2$, a saturation of the skyrmion Hall angle is observed and shown in Fig. 22 (A). Shown in regime I($+j_e, -Q$) is the data for the positive electron current density ($+j_e$) and positive perpendicular magnetic fields ($-Q$). For $H_\perp$ = +0.54 mT, a saturation of the skyrmion Hall angle $\Phi_{sk} \approx 32 \pm 2°$ is observed (skyrmion of diameter $d = 800 \pm 300$ nm), and $\Phi_{sk} \approx 28 \pm 2°$ is observed for $H_\perp$ = +0.48 mT (with a larger diameter $d = 1100 \pm 300$ nm). Namely, by varying the strength of magnetic field and hence the size of skyrmion bubbles, a size dependent skyrmion Hall angle is demonstrated. This trend agrees with the dependence of $\mathcal{D}$ on $d$ given by *Eq*. (16). A negative electron current direction ($-j_e$) reverses the direction of skyrmion motion that leads to a negative saturation skyrmion Hall angle, as shown in regime III($-j_e, -Q$). In the presence of negative perpendicular magnetic field (-$H_\perp$) with positive topological charge (+$Q$), such a trend is reversed, as summarized in the regime II($-j_e, +Q$), and regime IV($+j_e, +Q$), respectively. This picture is consistent with the opposite sign of topological Magnus forces. Furthermore, these experimental observations are thus consistent with the absence of skyrmion Hall effect in the earlier study using a driving current density $j_e < 10^5$ A/cm$^2$ [41].

The observed current density dependence is however, inconsistent with the simple theoretical prediction given in *Eq*. (14), which suggests a constant value of $v_y/v_x$ that is given by $1/\alpha\mathcal{D}$ and is independent of the driving current. One possible reason for this apparent discrepancy is the presence of pinning that affects the skyrmion motion. Such pinning may originate from random defects/disorders in the sputtered films, as suggested by the experimentally observed threshold depinning of skyrmions and stochastic motion at low driving currents. In the presence of random defects, a recent study of the dependence of skyrmion Hall angle on the driving force reproduced the experimentally observed behavior and suggested the saturation of skyrmion Hall angle should occur once a strong driving is reached [238-240].

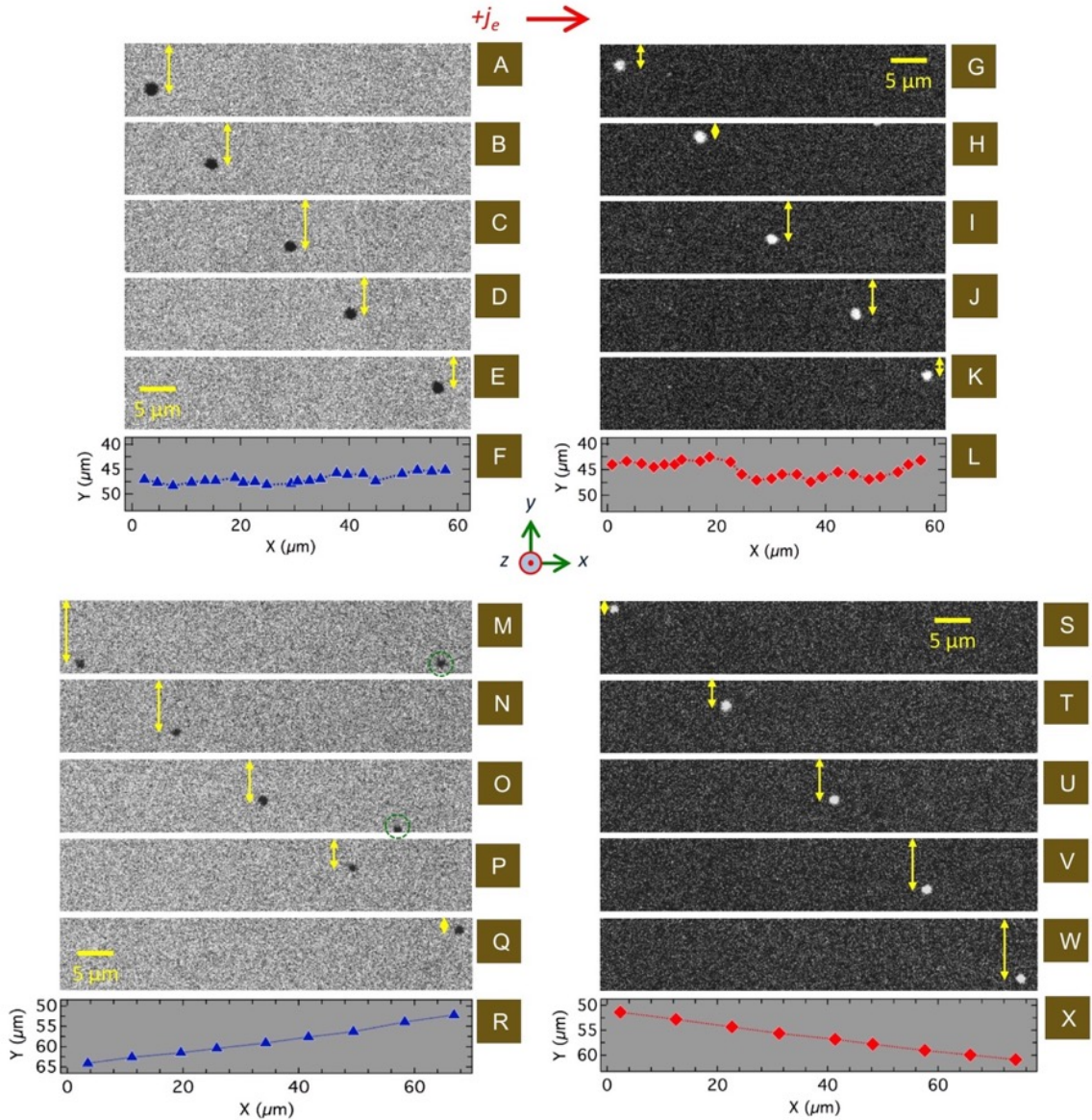

Figure 21. Polar MOKE microscopy images of pulsed current driven skyrmion motion in a wire patterned from a Ta/ CoFeB/TaO$_x$ trilayer. All experiments were performed using using 50 μs current pulses. (A) - (E) Snapshots of ($Q = -1$) skyrmion motion captured after applying successive current pulses of amplitude $j_e = +1.3 \times 10^6$ A/cm$^2$ in an external perpendicular magnetic field $H_\perp$ = +0.48 mT. The skyrmion trajectory summarized in (F), shows no net transverse motion along the $y$ direction. (G) - (K) Snapshots of ($Q = +1$) skyrmion motion at $j_e = +1.3 \times 10^6$ A/cm$^2$ with $H_\perp$ = -5.2 Oe. The stochastic trajectory is shown in (I), again showing no net transverse motion. (M) - (Q) Snapshots of ($Q = -1$) skyrmion motion at $j_e = +2.8 \times 10^6$ A/cm$^2$ with $H_\perp$ = +0.54 mT. Its nearly straight and diagonal trajectory is shown in (R), indicating the presence of transverse motion along $+y$ direction. The size of the skyrmion is slightly smaller as compared to (A) - (E) due to the larger perpendicular magnetic field. Two other skyrmions that moved into the frame, marked as green circle, were not studied. (S) - (W) Snapshots of ($Q = +1$) skyrmion motion at $j_e = +2.8 \times 10^6$ A/cm$^2$ with $H_\perp$ = -0.52 mT. Again, there is a nearly straight and diagonal trajectory shown in (X). However, the slope is opposite, indicating the presence of transverse motion in opposite direction (along $-y$ direction). Reproduced with permission [95]. Copyright 2016, Nature Publishing Group.

In order to draw a close link between the conventional charge Hall effect and the Hall effect from topological charge, the accumulation of skyrmions at the edge of device has been demonstrated experimentally as well, as shown in Figs. 22 B – C which is also consistent with the theoretical calculation given in Fig. 12 G. By applying a pulse train of amplitude $j_e = -6 \times 10^6 \text{ A/cm}^2$, the accumulation of ($Q = -1$) skyrmions is observed at the lower edge of the device, see Fig. 22 (B). Again, reversal of topological charge results in the accumulation of skyrmions ($Q = +1$) at the upper edge, Fig. 22 (C). This observation resembles the charge accumulation of Hall effect in conductors. In the future, one could electrically probe the skyrmion accumulation due to the skyrmion Hall effect, from which the reciprocity between the topological Hall effect and the skyrmion Hall effect can be established. For the topological Hall effect, it should be beneficial to have smaller skyrmions with consequently larger emerging magnetic field [57].

Using a time-resolved X-ray transmission electron microscopy, Litzius *et al.*, have also observed the skyrmion Hall effect in a [Pt/CoFeB/MgO]$_{15}$ multilayer with skyrmions that are approximately 200 nm in diameter. While both experiments reveal qualitatively the same behavior – deviation of skyrmion motion from the homogeneous driving current direction, Litzius *et al.* attributed the reduced values of the observed skyrmion Hall angles compared to theoretical exectations to the complex internal mode excitations of skyrmions in combination with a field-like torque [96]. These authors performed extensive micromagnetic simulations in support of a deformation of skyrmion profile induced by the contribution from a field-like torque. Note that these micromagnetic simulations do not provide a quantitative description of the current dependence of the skyrmion Hall angle and suggest that the deformations would be more important at low driving forces. This indicates the physical origin that underlies the skyrmion Hall effect can be potentially very complex. However, one should notice that the contribution of field-like toque, as compared to the antidamping-like torque, is typically small. The influence of this internal excitation mode to the dynamics of magnetic skyrmion, apparently, should be explored in the future by using further spatial/temporal resolved imaging technique. To some extent, this could clarify the complex (multiple) origins that may be controlling the skyrmion Hall effect. Note that a recent time-dependent X-ray pump-probe imaging study performed by Woo *et al.* suggested that a magnetic skyrmion in the large medium does not experience inertial effect and thus only very small amount of deformation should happen in a [Pt/CoFeB/MgO]$_{20}$ multilayer [241].

Nevertheless, by changing the sign of the topological charge, the sign of the electric current, both experiments have revealed a strong similarity between the conventional Hall effect of the electronic charge and the Hall effect due to the topological charge [42, 95]. In addition, these experiments in different material systems also demonstrate the common topological nature of magnetic skyrmions, regardless of the varying size of magnetic skyrmions from μm to 100nm scale. Furthermore, these results suggest the important role of defects for understanding the detailed dynamics of magnetic skyrmions. These observations also indicate that the topological charges of magnetic skyrmions, in combination with the current induced spin Hall spin torque, can be potentially integrated to achieve novel functionalities, such as topological sorting.

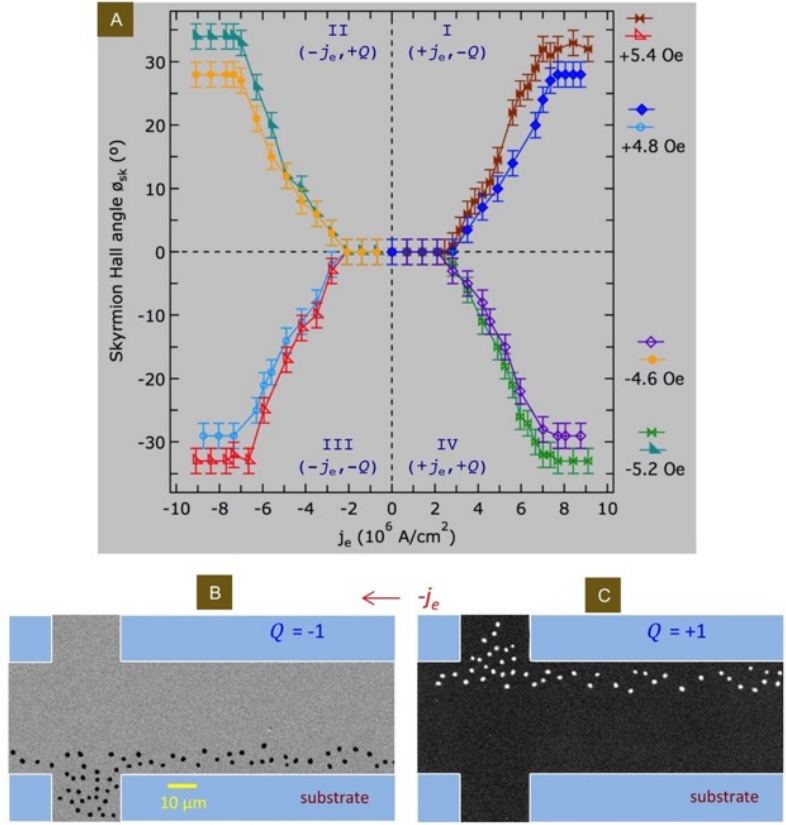

Figure 22. Skyrmion Hall angle and accumulation. **(A)** Phase diagram of the skyrmion Hall angle as a function of the current density and sign of the topological charge, obtained by tracking the motion of several tens of skyrmions in wires of dimension 80 (width) ×100 (length) μm$^2$ patterned from a Ta/CoFeB/TaO$_x$ trilayer. In the low current density regime, the skyrmion Hall angle $\Phi_{sk}$ exhibits a linear dependence. Further increase of current density $|j_e| > 8 \times 10^6$ A/cm$^2$ results in the saturation of the skyrmion Hall angle. By alternating the sign of driving electron current density ($\pm j_e$) and the sign of topological charge ($\pm Q$), a phase diagram for the four different regimes was determined. Namely, for a negative topological charge (under positive magnetic fields), regime I($+j_e, -Q$) with positive $\Phi_{sk}$ and regime III($-j_e, -Q$) with negative $\Phi_{sk}$ were identified by changing the polarity of electron current. For skyrmions with a positive topological charge (under negative magnetic fields) a positive $\Phi_{sk}$ in regime II($-j_e, +Q$), and negative $\Phi_{sk}$ in regime IV($+j_e, +Q$) were detected. The decrease of the skyrmion Hall angle from $|\Phi_{sk}| \approx 32 \pm 2°$ to $|\Phi_{sk}| \approx 28 \pm 2°$ is demonstrated by increasing the skyrmion diameter from $d = 800 \pm 300$ nm (+0.54 mT/-0.52 mT) to $d = 1100 \pm 300$ nm (+0.48 mT/-0.46 mT). (B) Polar MOKE microscopy images demonstrating skyrmion ($Q = -1$) accumulation at the edge of the device (of width 60 μm and length 500 μm). This is achieved by repetitively applying 50 current pulses of duration 50 μs at a frequency of 1 Hz with a current density $j_e = -6 \times 10^6$ A/cm$^2$ and an applied field of +0.54 mT. (C) Reversing the magnetic field from positive (+0.54 mT) to negative (-0.52 mT) leads to the accumulation of skyrmions with positive topological charge $Q = +1$ at the opposite edge. Reproduced with permission [95]. Copyright 2016, Nature Publishing Group.

## 3.7 High frequency dynamics of magnetic skyrmion

The previous section discussed how the topological charge influences the translational motion of magnetic skyrmions. This can be largely explained within a rigid particle model containing gyrotropic forces dependent on the sign of the topological charge. Beyond this, it is also interesting to investigate the rich high frequency characteristics of magnetic skyrmions, including inertial effects [114, 242], effective mass, gyrotropic motion [243], breathing and chiral edge modes [244], and the (mutual) interaction with high frequency spin waves [157, 245].

Inertial effects are visible only at frequencies comparable to or faster than those intrinsic to the system, which are in the Gigahertz regime for magnetic skyrmions. In contrast to the magnetic vortices that exhibit gyrotropic motion on circular orbits, magnetic skyrmions are expected to result in more complex hypocycloidal trajectories [243, 246, 247]. This dynamical mass enhanced inertia can be revealed by using a spatial resolved pump-probe x-ray transmission microscope upon brief magnetic field gradient excitation [114]. By utilizing this technique, the gyrotropic trajectory of the center of mass can be captured with a high accuracy ($\approx$ 3 nm), as shown in Fig. 23 A. Through studying the inertia effect of a single Néel-type skyrmion in a nanodisk made of Pt/CoB bilayer, the effective mass of magnetic skyrmion spin structure can be determined to be $m = 2.0(4) \times 10^{-7}$ kg/m$^2$ which is two orders of magnitude larger than that of a straight domain wall. This surprisingly large mass of magnetic skyrmions tailors their associated dynamical responses and can be attributed to the intrinsic breathing mode resulting from the strong dipolar energy in the confined nanostructures. On the other hand, Woo *et al.*, recently conducted a time-dependent X-ray pump-probe imaging study in which they suggested that a magnetic skyrmion in a large medium does not experience inertial effect and the observation of enhanced mass reported in Ref. [114] could extrinsically arise from the interaction between the skyrmion and geometrical boundary [241].

Another interesting question is what are the high frequency excitations of magnetic skyrmions and how do they interact with other elementary magnetic excitations, such as spin waves. Indeed for the skyrmion lattice phase of bulk skyrmion materials these excitations have been observed with broadband inductive measurements [248, 249]. Additionally, these dynamic modes have also been observed with direct optical, time-resolved spectroscopy [250]. Independent of whether the materials are metallic, semiconducting, or insulating, the excitations show common features in these skyrmion materials, which therefore demonstrate their unique topological nature. Compared to the helical, conical or field polarized states, the skyrmion lattice phase has distinct excitations at significantly lower frequencies. There are three basic excitations: breathing, clockwise and anticlockwise radial modes [251, 252]. These three modes have very different field dispersions. The breathing mode decreases in frequency with increasing field, while the counter-clockwise radial mode increases, and the clockwise radial mode is mostly independent of the magnetic field.

Individual skyrmions in chiral bulk magnets also have breathing, clockwise and anticlockwise radial modes [253], besides the lowest energy translational mode. But they also contain additional symmetry breaking excitations [254, 255]. In particular, there can be several higher order modes at energies below the breathing modes. Of these quadrupolar and sextupolar distortions are the energetically most favorable ones. The magnetic field dependence of these internal modes can be seen in Fig. 23 B. At low fields the size of skyrmion is bigger which could accommodate more internal modes. With reducing fields some of these modes become gapless. This indicates that individual skyrmions become unstable at sufficiently low magnetic fields.

Skyrmion-like magnetization dynamics may even occur in materials and systems that do not necessarily have stable static magnetic skyrmions. Interestingly, even in nanocontacts without dipolar interaction or DMI, micromagnetic simulations have suggested that spin transfer torques can stabilize dynamical skyrmion modes [256]. Moreover, if DMI exists in such a system, then the in-plane rotations of the spins in these dynamical skyrmions will result in strong breathing mode excitations, which in turn might be efficient and tunable sources for microwave generation. Indeed, electrical measurements in Co/Ni multilayers have suggested the presence of such a dynamical skyrmion state [257]. Furthermore, under specific excitation conditions it may also be possible to have continuous transitions from non-topological solitons to topological skyrmion states [258, 259], which may result in the incoherent emission of spin waves or the formation of transient skyrmion spin textures.

Lastly, an interesting problem is how other magnetic excitations, such as magnons may interact with skyrmions or other chiral spin structures. Theoretical investigations show that magnons scatter of individual skyrmions, which can be understood as becoming enhanced due to the formation of a magnon-skyrmion bound state [245, 255]. Due to the topological non-trivial spin texture of skyrmions, similar to the case of electrons, magnons will experience an emergent magnetic field that results in the sideway deflection of these magnons, *i.e.*, the topological magnon Hall effect [245, 260]. Similarly, the conservation of spin momentum will result in the skyrmion being displaced in both transverse and longitudinal directions with respect to the incoming magnon momentum. Namely, a translational skyrmion motion and skyrmion Hall effect can be created by a magnonic spin transfer torque. Both the asymmetric scattering of the magnons and the resultant direction of skyrmion displacement is shown for a numerical simulation in Fig. 23 C. The magnon-skyrmion interaction is expected to be maximized when the wavelength of the magnons matches the geometric size of the skyrmions. Thus, these theoretical predictions open up interesting perspectives for using magnons for skyrmion manipulation [261]. Skyrmion motion, enabled by thermally excited magnon flows even in magnetic insulators, is one of the examples [71]. Indeed, rotational motion of skyrmion lattices in both metallic and insulating bulk skyrmion materials have been explained by inadvertent temperature gradient [72]. In addition, similar to the optical case, a concept of a spin-wave fiber in which spin wave are guided by the chiral domain wall has been theoretically suggested [262]. And more recently, control of spin-wave refraction using periodically patterned skyrmion arrays has been reported which suggested that magnetic skyrmions can also be used as building blocks of magnonic devices [263].

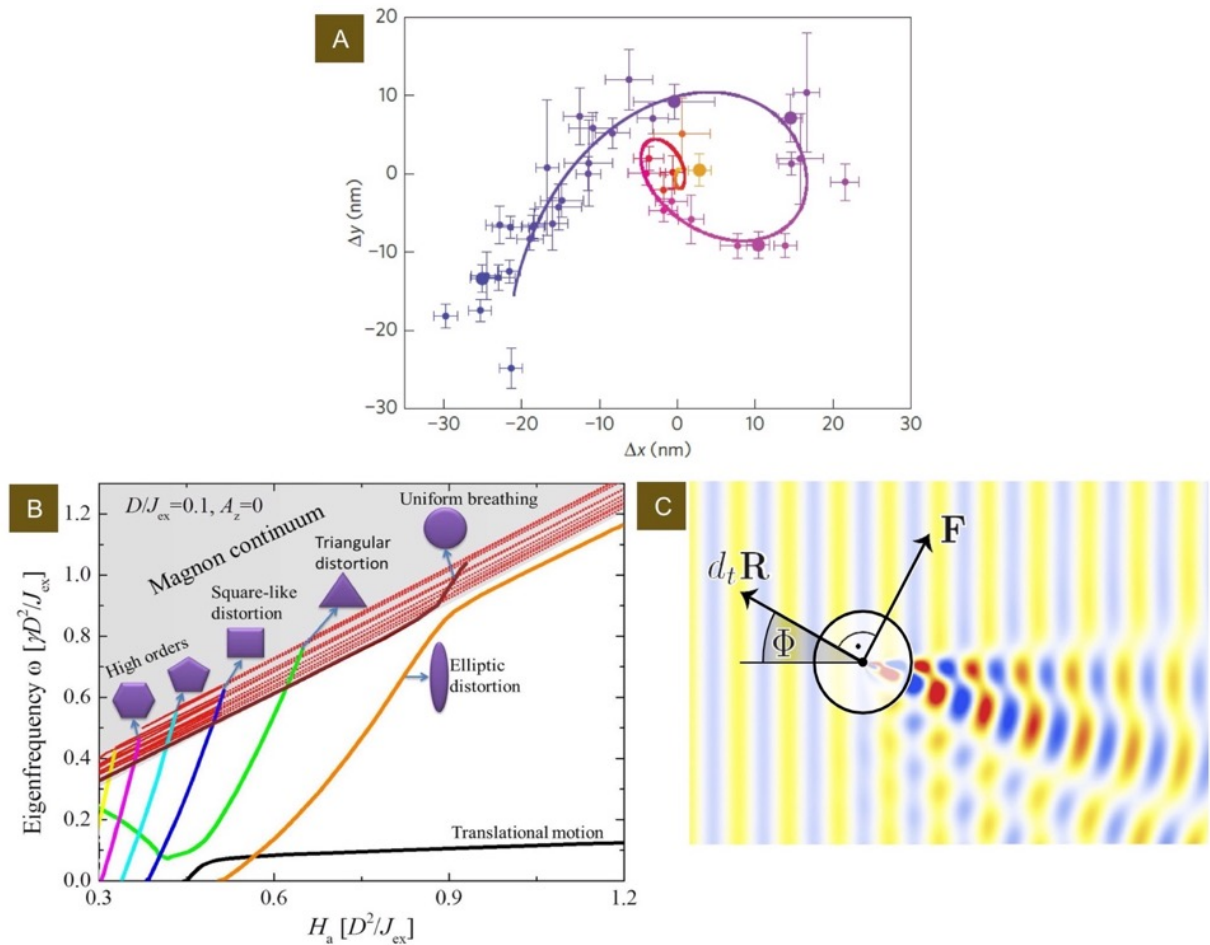

Fig. 23. (A) Gyrotropic trajectory of a single magnetic skyrmion. Reproduced with permission [114]. Copyright 2016, Nature Publishing Group. (B) Magnetic field dependence of different dynamic eigenmodes of individual magnetic skyrmions. The mode corresponding to the translation motion is gapped because a discrete spin model is used in the calculations. Reproduced with permission [254]. Copyright 2014, American Physical Society. (C) Magnon scattering by skyrmions. In this simulation magnons are emitted on the left-hand side and impinge on a skyrmion. Clearly, the magnons are asymmetrically scattered with a preferential direction towards the bottom of the image. Conversely, the scattering results in a force $F$ acting on the skyrmion, which results in its displacement in the direction $dR/dt$. Reproduced with permission [255]. Copyright 2014, American Physical Society.

## 3.8 Artificial skyrmions stabilized by interlayer coupling in thin films

In this section, artificial skyrmions created in patterned nanostructures will be discussed. In contrast to the DMI stabilized Néel-type skyrmions in asymmetric multilayers and Bloch-type skyrmions in bulk chiral magnets, artificial skyrmions can also be created at room temperature by fabricating nanodots with a magnetic vortex state on perpendicularly magnetized films [115-118]. The formation of Bloch-type skyrmions in these hybridized multilayers is a result of imprinted vortex spin texture from nanodots to the perpendicular underlayer, through an exchange coupling at the interface, see Figs. 24 A and 24 B. Magnetic vortices have been well studied in nanostructures with a suitable aspect ratio (dot diameter versus thickness) [264, 265]. Note that the in-plane curling spin configuration within a vortex could be clockwise or counter-clockwise (defined as circularity), and the core of a vortex could point either up or down, out of the plane (defined as polarity) [264]. With these two possible orientations of circularity and polarity, four possible ground states are expected to be energetically degenerate in a circular shaped vortex. In the following, experimental strategies for achieving homo-chiral spin configuration of artificial skyrmions will be discussed, *i.e.*, same circularity and polarity over arrays of asymmetric-shaped nanoelements [266, 267], (see Figs. 24A and 24B).

Experimental efforts in creating artificial skyrmions were made in several systems, including Co dots on Ni film with a PMA grown on Cu(001) [116], Co dots on [Co/Pt]$_n$ multilayers [117], and Co dots on ion-irradiated [Co/Pd]$_n$ multilayers[118, 268], where the size of Bloch-type skyrmions vary from several hundred nanometers to a few microns. The key procedure to set homo-chirality of the skyrmion lattices involves demagnetizing samples along the in-plane direction to define circularity and magnetizing samples along the out-of-plane direction to define the polarity. For example, in a system made of Co dots on [Co/Pt]$_n$ multilayers [117], the Co dots were saturated into a single domain state with an in-plane magnetic field ($\approx$ 20 mT) along the edge-cut, and then evolved to vortex states with the same circularity upon the removal of field. Consequently, the vortex core and surrounding underlayer area were set to be parallel by applying/removing a perpendicular magnetic field (+1 T) smaller than the out-of-plane saturation field (1.3 T). Finally, the surrounding underlayer area was aligned antiparallel to the core direction with a smaller negative field (-150 mT). It is worth noting that ion-irradiation has proven effective in facilitating the formation of the skyrmion state [118, 269]. In another system of Co dots on [Co/Pd]$_n$ multilayers [118], the Co/Pd underlayer was first ion-irradiated to suppress the PMA in-plane in regions directly underneath the Co dots. The Co/Pd underlayer was then saturated along the positive out-of-plane direction using a 1.5 T field. Subsequently the Co dots were saturated by an in-plane field (100 mT) along the edge-cut, and the nucleation of magnetic vortices with controlled circularity and polarity (core direction) was guided by removing the in-plane field and applying an additional negative out-of-plane field (-100 mT, for setting polarity).

To directly image the in-plane spin structures of the vortices in the nanodots, various magnetic imaging techniques such as MOKE microscopy, Magnetic Force Microscopy [117, 118], PEEM [116] as well as Scanning Electron Microscopy with Polarization Analysis (SEMPA) [118] have been used. Figure 24B shows representative SEMPA image of Co dots array, where the in-plane magnetization curling direction is highlighted by the color wheel. The image shows that all Co dots in the array form typical vortex spin structures with a uniform counterclockwise rotation sense, indicating the successful control of the circularity. These are also consistent with magnetometry signatures obtained from first-order reversal curve measurements which illustrate two distinct annihilation fields [118, 270]. On the other hand, direct imaging of the vortex core direction in the buried underlayer is technically challenging because of the small size and buried surface. Note that the polarity-setting procedure introduced above also allows the control of the core direction with respect to the underlayer magnetization, *e.g.*, parallel, antiparallel, or mixed configuration depending on how the sample is magnetized. These possible configurations lead to three magnetic states, *i.e.*, skyrmion lattice (SL), vortex lattice (VL), and mixed lattice (ML) (see the sketches in Fig. 24C), which are expected to have different perpendicular remanent magnetization as a result of the vortex core direction. Experimental magnetization curves by sweeping the field reveals clear differences (see Fig. 24 D), where skyrmion/vortex lattice configuration has the smallest/largest magnetization, and mixed lattice configuration with randomly aligned skyrmion or vortex configuration, showing a curve in between skyrmion and vortex states.

To probe the all-important imprinting of the vortex spin texture into the perpendicularly magnetized underlayer to form the skyrmions, polarized neutron reflectometry (PNR) has been utilized, which is sensitive to structural and magnetic depth profiles in thin films [271, 272]. For the skyrmion system with Co dots on top of ion-irradiated Co/Pd, the structural depth profile reproduces the designed sample structure (Fig. 24 C); interestingly, the magnetic depth profile which only comes from in-plane magnetization, not only captures the in-plane moments from the Co dots, but also reveals an extension of the in-plane magnetization into the irradiated Co/Pd region whose thickness corresponds to the designed ion-irradiation depth. These results directly demonstrate the existence of artificial skyrmions at the interface between the irradiated Co/Pd film and the Co dots, caused by the imprinting of the vortex spin texture from the Co dots, which was also confirmed by anisotropic magnetoresistance measurements and micromagnetic simulation [118].

The possibility to control the polarity of the artificial skyrmions thus offers an opportunity to explore topological properties of skyrmions. For example, the skyrmion number $Q$ can be determined by the core polarity direction relative to the surrounding underlayer magnetization direction. In case of the skyrmion configuration discussed above with the core polarity opposite to the underlayer magnetization, skyrmion number is $Q = 1$, whereas that in a vortex configuration with the core parallel to the underlayer has $Q = 0$ at the interface region [116, 118]. A physical manifestation of this topological quantization is that a skyrmion state with $Q = 1$, is topologically protected from transforming continuously into another spin configuration, such as a single domain. This topological effect was observed during the annihilation of skyrmion or vortex state into a single domain state. The idea was examined in a system of Co dots on perpendicularly magnetized Ni film, where skyrmion and vortex configuration can be controlled by switching the magnetization of the surrounding Ni underlayer while the core of Ni vortex is pinned [116]. It was found that the in-plane annihilation field of the vortex state is significantly smaller than that of the skyrmion state (see Figs. 24 E and F), indicating a topological effect of the magnetic skyrmions in the core annihilation process.

These artificial skyrmions in thin-film nanostructures are an interesting alternative platform that exhibit the Bloch-type configuration, and are stable at room temperature, even in the absence of an external magnetic field. The regular lattices are also ideal to explore topological characters of such skyrmion systems. Indeed, there have been very active research efforts in this direction, from demonstration of the skyrmion state [115-118], to their dynamic properties [273-277], topological characteristics [116, 269, 276, 278], and novel design of skyrmion based spin-ice system [279]. Since the vortex state in the dots is determined by the aspect ratio of the dot, further scaling down of the artificial skyrmion size is possible. In most of the early studies of artificial skyrmions, the magnetic nanodots on top, while playing an important role in defining the skyrmion state, also hinders the movement of skyrmions once they are formed. Active efforts are underway, *e.g.*, to achieve planar skyrmions without the protruding dots [118, 280], towards eventually mobilizing artificial skyrmions. Very recently, a concept of hybrid skyrmion structure has been proposed by patterning magnetic nanodisks onto a B20 skyrmion material, in which the enhanced stability of skyrmion state and suppression of skyrmion Hall effect have been revealed through a micromagnetic simulation study [281].

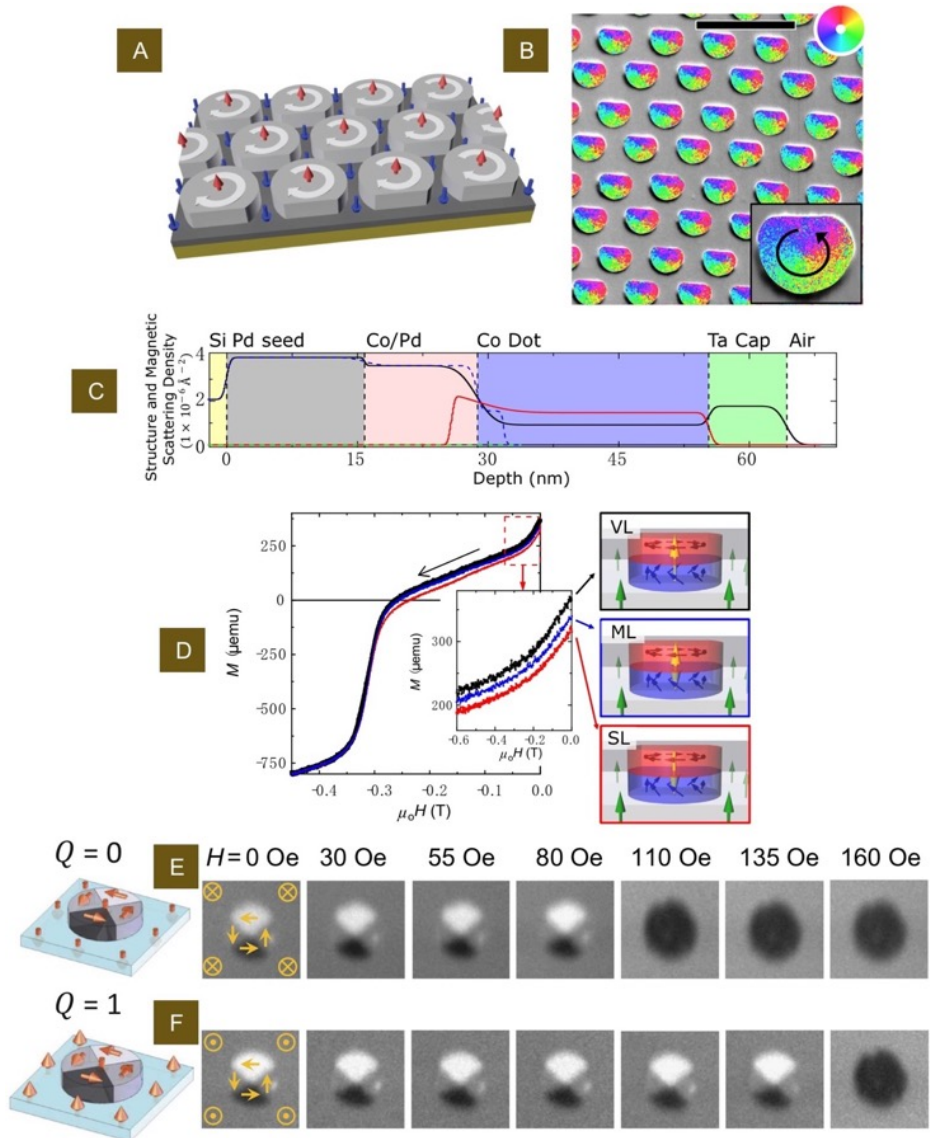

Fig. 24. Artificial skyrmions. Panel A shows a schematic of an artificial skyrmion lattice. Ordered arrays of magnetic nanodots are prepared on top of a perpendicularly magnetized film. The arrows represent the magnetization direction of local moments. Reproduced with permission [115, 117]. Copyright 2016, American Physical Society. Panel B shows SEMPA image of a remanent-state Bloch skyrmion lattice after saturating the dot. Color highlights the in-plane magnetization direction, see color wheel at top right. The scale bar is 2 μm. Inset shows the magnetization curling direction. Reproduced with permission [118]. Copyright 2015, Nature Publishing Group. Panel C shows depth profiles obtained by PNR on the same artificial skyrmion lattices as shown in panel B. The magnetic depth profile (red) extends from the Co dots into the top portion of the irradiated Co/Pd film. Panel D shows magnetization curves, with the field sweeping from zero to negative saturation, for the hybrid structure shown in panel B prepared into the SL (red), VL (black) and ML (blue) states at remanence. Schematic illustrations (right) show the spin structure of the SL, VL and ML states. Reproduced with permission [270]. Copyright 2014, Nature Publishing Group. Panel E shows PEEM images of the vortex state (upper row) and skyrmion state (lower row) at remanence after applying an external in-plane field. The skyrmion state ($Q$ =1) has a stronger annihilation field. Reproduced with permission [116]. Copyright 2014, Nature Publishing Group.

## 3.9 Other novel spin sensitive imaging techniques

In this section, several promising spin sensitive imaging techniques will be reviewed, including Lorentz TEM, SPLEEM and PEEM. Note that SP-STM has also been extensively utilized to quantify the skyrmion spin textures, as has been discussed in the section 3.1. MOKE microscopy can quantify the motion of larger magnetic skyrmions (with sizes above 300 nm), but it cannot be used to study smaller magnetic skyrmion. Representative examples of MOKE studies have been presented in section 3.2. Very recently, a single electron spin in diamond with a nitrogen vacancy center was used as a probe to assess the vectorial spin profile of skyrmion, as well as the associated spin topology [138]. While magnetic force microscopy (MFM) has been extensively utilized to study the current or magnetic field driven dynamics of noncollinear spin textures, including domain walls and magnetic skyrmions, it cannot reveal the spin chirality of chiral domain walls and skyrmions.

### 3.9.1 Lorentz transmission electron microscopy

As mentioned at the beginning of this article, spatially resolved Lorentz TEM has been extensively utilized as a very powerful tool to directly characterize the spin topology of Bloch-type skyrmions in many bulk materials [38, 55, 56, 58, 61, 282]. Subsequently it was used for investigating Néel-type skyrmions that were stabilized by interfacial DMI in magnetic heterostructures. In the Lorentz TEM mode, the magnetization ($m$) of a sample gives rise to a phase shift - $\Delta\Phi(x,y)$ of the electron wave that can be imaged by defocusing the objective lens of the microscope [283]. Observing a non-zero phase shift - $\Delta\Phi(x,y)$ allows to resolve the in-plane $m(x,y)$ components of the spin structure. The associated magnetic image can be analyzed based on a transport of intensity equation (TIE), which calculates the distribution of magnetization components along the $x,y$ directions. However, in the out-of-focus mode, Lorentz TEM cannot provide magnetic contrast for Néel-type spin textures, including chiral Néel domain walls [139] and chiral Néel skyrmions [223]. In these cases the electrons are deflected along the domain wall structures resulting in a zero phase shift $\Delta\Phi(x,y) = 0$. By tilting the sample by a finite angle (*e.g.*, 30°), the projection of the out-of-plane magnetization into the plane perpendicular to the electron propagation enables a magnetic contrast to be revolved, as shown below in Fig. 25. However, the chirality of Néel-type spin textures cannot be directly identified by this tilting method. Thus, while Lorentz TEM can be used to differentiate the type of skyrmions of being Bloch-type or Néel-type, it cannot be used to identify the chirality of Néel-type skyrmion [223]. Innovative room-temperature spin imaging techniques with the capability of quantifying 3-dimension arbitral spin textures and their (field/current-driven) dynamics are thus in high demand in the future.

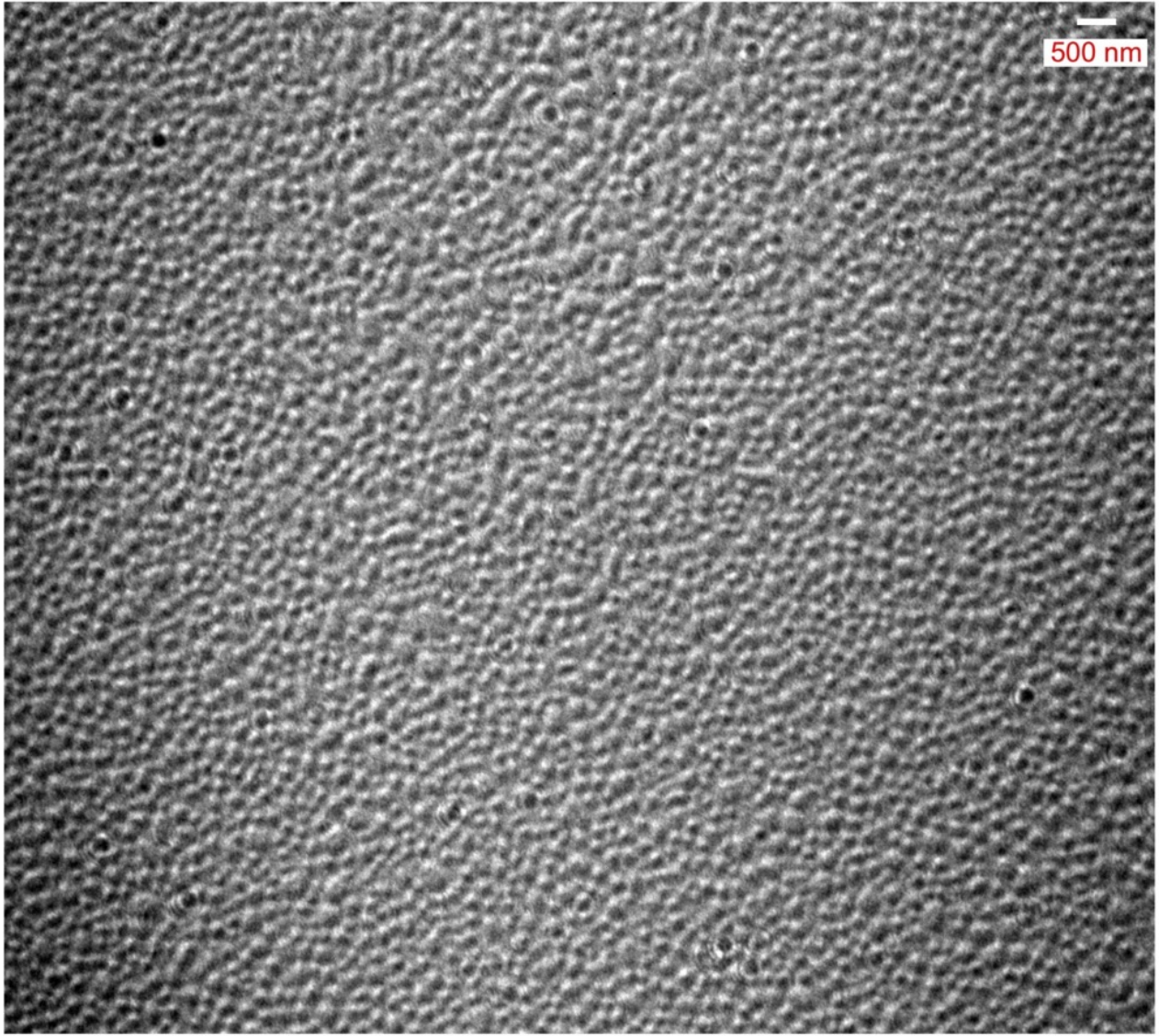

Fig. 25. Lorentz TEM imaging data acquired in a [Pt/Co/Ir]$_8$ multilayer. Note that these densely packed Néel-type skyrmions were stabilized at room temperature in the absence of magnetic fields. A multilayer of Ta(2nm)/[Pt(1.5nm)/Co(1nm)/Ir(1nm)]$_8$/Pt(1.5nm) was grown by using dc magnetron sputtering under a 3 mTorr Ar pressure onto $2 \times 2 \text{ mm}^2$ TEM grids with 50 nm thick silicon nitride (SiN) membrane widow. The zero-field Lorentz TEM experiments were performed in a JEOL 2100F instrument with a spherical aberration corrector. The corresponding out-of-focus Fresnel imaging mode was used with a defocus value of 6.4 mm. The sample was tilted 30 degree along the diagonal direction to gain magnetic contrast (manuscript in preparation by Wanjun Jiang, et al.).

### 3.9.2 Spin-polarized low energy electron microscopy

SPLEEM [284] and X-ray Magnetic Circular Dichroism–Photo Emission Electron Microscope (XMCD-PEEM) are high resolution magnetic imaging techniques that are based on spatial imaging of electrons deflected or emitted from the sample [135, 136, 147]. In particular, SPLEEM is capable of quantifying the arbitrary orientation of spin states. Room-temperature stabilization of magnetic skyrmions revealed by SPLEEM is shown in Fig. 26A. This can be done by tuning the spin polarization direction of the incident electron beam, which is only sensitive to the component of the magnetization parallel or anti-parallel to the electron spin-polarization direction. Therefore, by using electrons with spin polarization direction along three orthogonal directions, one can independently image spin components along x, y, and z directions, and map out three-dimensional spin structures on the sample surface. This enables the unambiguous determination of the chirality, if there is any. Furthermore, it can provide high spatial resolution, possibly down to a few nm. With the development of aberration correction techniques, the lateral resolution could be improved, possibly to about 2 nm. Due to the low energy of the incidence electron beam, which is typically of the order of a few electron volts, SPLEEM is very surface sensitive. In some cases, the magnetic contrast of buried magnetic films is also accessible, even after been exposed to air. Such versatility allows possible studies of ex situ prepared crystalline samples. On the other hand, as a surface sensitive imaging technique, it is still challenging to use SPLEEM to characterize ex-situ patterned nanoscale devices and to reveal the dynamics of chiral spin textures driven by electrical currents and magnetic fields.

### 3.9.3 Photoemission electron microscopy

Another interesting technique is synchrotron based PEEM. This technique can resolve element specific spin configuration with spatial resolution of the order of 20 nm. The measured magnetic contrast is only sensitive to the projection of magnetization along the X-ray beam direction. By tuning the incident X-ray beam direction with respect to the sample surface, the contributions of the out-of-plane and in-plane magnetization components are weighed differently, as shown in Fig. 26B. Focusing on the in-plane spin structure, this technique can thus be used to identify the chiral feature of a ferromagnetic Néel-type skyrmion based on XMCD. XMCD–PEEM can also image antiferromagnetic spin structures based on X-ray Magnetic Linear Dichroism (XMLD), *e.g.*, in-plane spin structures of antiferromagnetic vortex in patterned structures can be mapped out, which enables opportunities to experimentally explore antiferromagnetic skyrmions [285]. This technique can be extended to patterned nanoscale devices (racetrack device for example), in which the current or field driven skyrmion dynamics can be studied [44]. For field driven experiments, note that both SPLEEM and PEEM are electron based microscopes, therefore an additional deflector in the electron lens will be required to balance the Lorentz force that strongly affects the imaging condition in the presence of magnetic field.

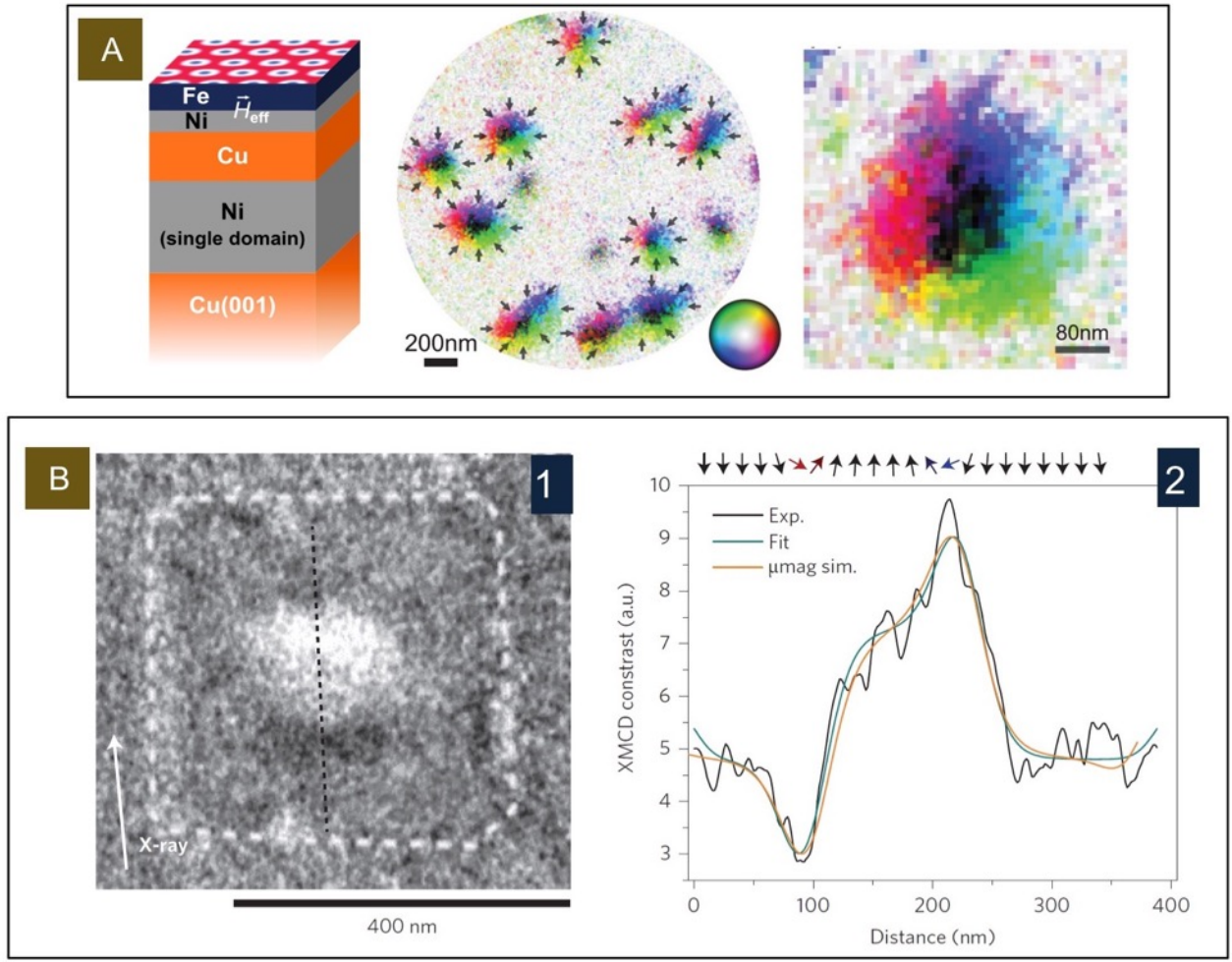

Fig. 26. Novel spin resolved imaging techniques. (A) SPLEEM imaging results performed in a Ni/Fe bilayer grown onto Cu(001)/Ni/Cu substrate, the virtual magnetic field of the Ni layer from the substrate stabilizes zero field magnetic skyrmions around 180 nm in size. Reproduced with permission [147]. Copyright 2015, AIP Publishing. The spin profile of these skyrmions is continuous rotating rather than having a big uniformly magnetized inner core. (B) XMCD-PEEM imaging results of a Pt/Co/MgO nanostructure hosting a single skyrmion. A slice across the skyrmion spin texture is shown on the right. By comparing with the simulation results, one can identify the (in-plane) spin chirality of the skyrmion to be left-handed. Reproduced with permission [44]. Copyright 2016, Nature Publishing Group.

## 4. Perspectives

Since the observation of chiral spin structures in interfacial asymmetric heterostructures, tremendous progress has been made studying thin-film based magnetic skyrmions [286]. For example, by properly designing the stack structure to maximize the strength of the interfacial DMI, 0.5 nm magnetic skyrmions can be stabilized in an Fe/Ir(111) bilayer, as shown in Fig. 27A. Interestingly, due to the onset of four-spin antiferromagnetic exchange interactions, these magnetic skyrmions are forming a square lattice, rather than the typically observed triangular lattice. However, these measurements were performed at cryogenic temperature and under high magnetic field. On the other hand, the ultimate limit down to atomic size points to the great potential of employing magnetic skyrmion for ultra-dense data storage that goes beyond the scaling limit of the existing CMOS technology. Concomitantly, skyrmions at the few nanometer scale could significantly magnify the strength of the emergent magnetic field (since $\langle b_z \rangle$ is proportional to $1/d^2$, $d$ denotes the diameter of skyrmion), which consequently enables emergent topological physics to be studied at room temperature. Enhanced topological transport, in particular the topological Hall effect, is also beneficial for efficient electrical readout of magnetic skyrmions, which is a pivotal step towards the future of skyrmion based electronics.

Searching for novel chiral spin structures in Dirac materials emerges as another trend, which provides a unique perspective for studying the interplay between real-space spin topology and momentum-space band structure topology. Studies along this direction may bring intriguing novel concepts to the condensed matter physics community. It is well known that topological insulators exhibit a spin-polarized surface state due to strong spin-orbit interaction. The surface magnetism is mediated by an indirect Ruderman-Kittel-Kasuya-Yosida (RKKY) interaction which contains a DMI component that can be utilized to stabilize chiral spin structures, as theoretically suggested [287, 288]. Extending the broken interfacial inversion symmetry into heterostructures made of thin-film topological insulator multilayers could naturally introduce an interfacial DMI component that in principle stabilizes chiral Néel-type spin textures. Experimentally, the observed topological Hall effect in $Cr_x(Bi_{1-y}Sb_y)_{2-x}Te_3/(Bi_{1-y}Sb_y)_{2-x}Te_3$ hetero-structures serves as an indirect manifestation of chiral Néel-type skyrmions [289]. By comparing Fig. 27(B)-2 (pure 7 nm thick magnetic topological insulator without signature of topological Hall contribution) and Fig. 27(B)-4 (2-nm magnetic topological insulator interfaced with 5-nm nonmagnetic topological insulator with a substantial topological Hall component), Yasuda *et al.*, concluded that the spin-polarized surface state mediates the interfacial DMI that stabilizes chiral skyrmion spin textures. The phase diagram is further summarized in Fig. 27(B)-5. Note that due to the related low temperature, weak magnetic moments of the magnetically doped topological insulator thin film, a direct spin resolved imaging of these chiral Néel-type spin structures has so far been unavailable. In fact, since these spin textures are of Néel-type, a direct spin imaging using Lorentz TEM could be even more challenging.

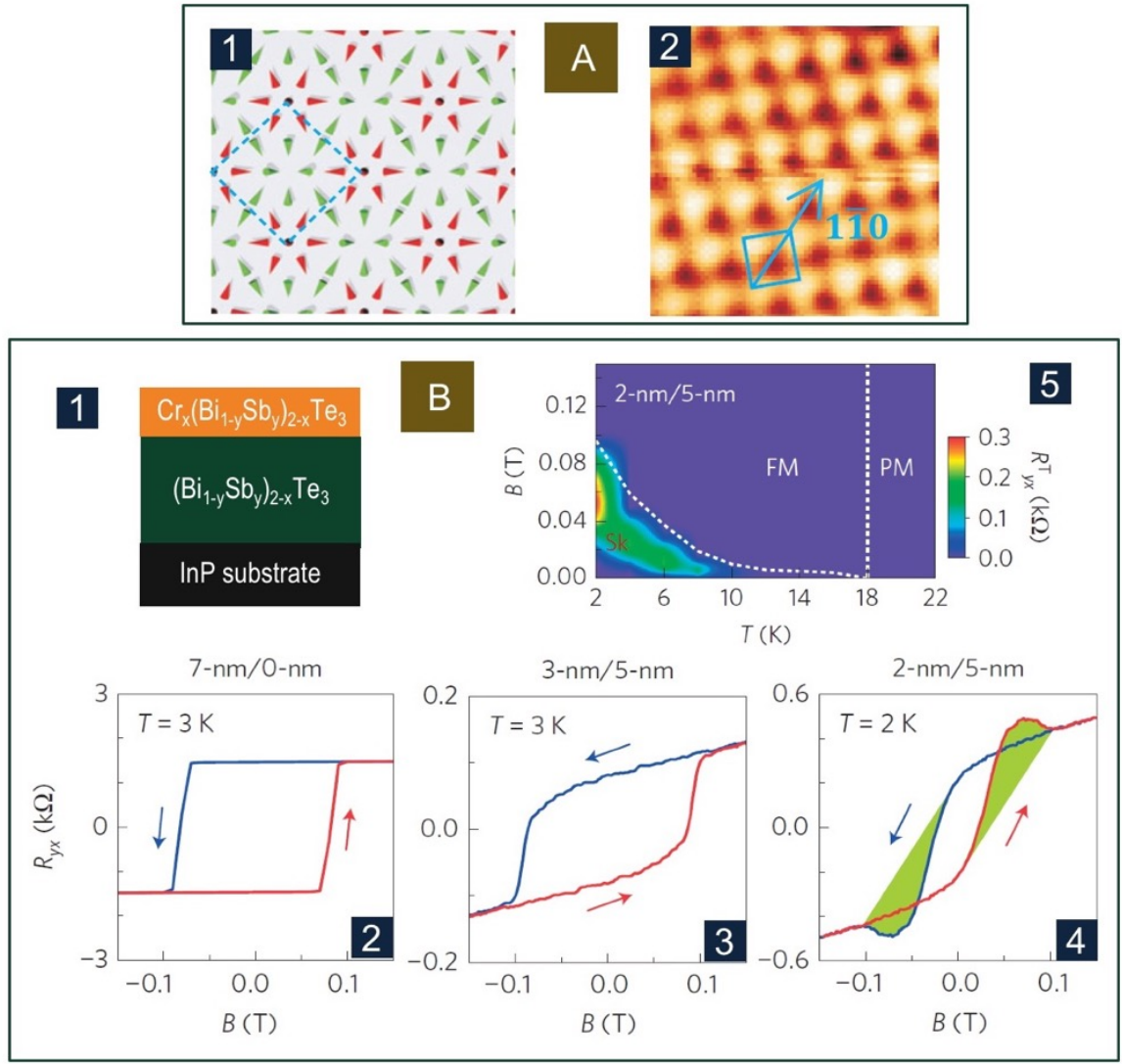

Fig. 27. Stabilization of atomic scale magnetic skyrmions in a Fe/Ir(111) bilayer. A-1 illustrates the onset of a four-spin antiferromagnetic exchange interaction that facilitates the formation of a square skyrmion lattice, in sharp contrast to the hexagonal skyrmion lattice in the B20 compounds. Reproduced with permission [124]. Copyright 2011, Nature Publishing Group. (B) Shows topological transport in an inversion asymmetric heterostructure consisting of a layer of nonmagnetic topological insulator and a layer of magnetically doped topological insulator. Shown in 2 is the anomalous Hall measurement in a 7 nm thick $Cr_x(Bi_{1-y}Sb_y)_{2-x}Te_3$ layer. Shown in 3 is a bilayer of thickness 3nm/5nm and of composition $Cr_x(Bi_{1-y}Sb_y)_{2-x}Te_3/(Bi_{1-y}Sb_y)_{2-x}Te_3$ respectively. Shown in 4 is a bilayer of thickness 2nm/5nm and of the same composition. The measured Hall signal shows the onset of a topological Hall contribution that signifies the chiral spin textures. The associated phase diagram is summarized and presented in the part 5. Reproduced with permission [289]. Copyright 2016, Nature Publishing Group.

Interfacial DMI stabilized Néel-type skyrmions have also been proposed in oxide heterostructures made of $5d$ transition metal oxides such as $SrRuO_3$-$SrIrO_3$ bilayer [290]. In this material system, $SrIrO_3$ act as an effective heavy metal layer that provides a strong spin-orbit interaction and hence an origin of the interfacial DMI. In this experiment, it is found that the magnitude of the topological Hall effect decreases with the increasing thickness of ferromagnetic $SrRuO_3$ layer (hence the increase of bulk contribution). This picture is consistent with interfacial-DMI stabilized chiral magnetism. However, a spin space imaging of these chiral spin texture in $SrRuO_3$-$SrIrO_3$ bilayer is still missing. In addition, it has also been theoretically suggested that Néel-type skyrmions can also be stabilized in $LaAlO_3$/$SrTiO_3$ two-dimensional systems with a Rashba spin-orbit coupling when the effective anisotropy is of the easy-plane type [291]. Experimental signatures, such as topological Hall effect in this later material system, are not yet available.

In addition, skyrmions in antiferromagnetic (AFM) materials currently attract intense attention [237, 292]. Skyrmions in AFM materials are not influenced by the skyrmion Hall effect due to the coupled motion of magnetic skyrmions with opposite topological charge from two sublattices, which is thus insensitive to the stray field and free from edge scattering, as shown in Fig. 28A. Their experimental realization in real materials might be challenging, particularly for systems containing room-temperature AFM skyrmions. On the other hand, synthetic AFM skyrmions appears as an alternative theoretical proposal [293]. The experimental realization of synthetic AFM skyrmions can be done by utilizing a antiferromagnetically coupled magnetic hetero-structure such as Co/Ru/Co trilayers [247]. Equivalently, the coupled motion of skyrmions in separated magnetic layers results in a net zero topological Magnus force and thus absence of skyrmion Hall effect [236], as shown in Fig. 28B. However, due to the presence of AFM coupling that results in a zero-net magnetic moment, a direct spin-space imaging of AFM skyrmion by MOKE, Lorentz TEM, SPLEEM could be challenging. One possible solution could be to use an element specific technique such as XMLD-PEEM. While both topological Hall and skyrmion Hall effects are precisely cancelled in an ideal material system with periodic AFM skyrmions (which is however, not experimentally demonstrated yet), it is theoretically proposed that a topological spin Hall effect could occur as a result of the alignment of spins with skyrmion spin textures [294]. By using the spin-to-charge interconversion phenomena (spin Hall effect and inverse spin Hall effect [12, 13]), this spin-polarized version of topological Hall effect can be potentially detected and further used for electric probing of AFM skyrmions [294, 295].

Note that, very recently, experimentally magnetic skyrmions in ferrimagnet GdFeCo have been studied, where magnetic moments from two sublattices were not completely compensated. However, the reversal of moments (hence chirality) across the compensation point, and the absence of a skyrmion Hall effect are not established yet in this material [296]. Note that a sign reversal of transverse motion of magnetic bubbles across the compensation point in a (Tb/Co)$_7$ multilayer is revealed by MOKE imaging [297]. The dynamics of magnetic skyrmions in ferrimagnet around the compensation point have also been theoretically demonstrated [298].

For efficient electrical manipulation of magnetic skyrmions, voltage or electric field control of magnetism is a more promising pathway [6, 193, 209, 299-303]. This aspect – electrical control of topological spin textures, also beautifully illustrates the synergies between two different aspects of modern magnetism. For example, in a typical inversion asymmetric HM/FM/I trilayer, it has been theoretically suggested and experimentally demonstrated that the application of electric field across the stack could control both the speed and the direction of motion of magnetic skyrmions. Furthermore, by tailoring the device configuration via electric voltages, novel spintronic devices such as transistor and multiplexer have been proposed theoretically. Very recently, it has also been theoretically suggested that application of electric field could linearly enhance the interfacial DMI strength in a Pt/Co/MgO trilayer [209]. Note that, the microscopic picture of this interesting phenomenon is not completely clear yet, but maybe related to the enhanced orbital hybridization at the *3d - 5d* interface. Nevertheless, electric field could serve as one more degree of freedom for effectively engineering interfacial DMI physics. It is worth noting that, a substantial interfacial DMI has been recently observed in an AFM/FM bilayer made of IrMn/CoFeB [304]. Utilizing the exchange bias generated by an AFM layer, it is then possible to investigate the formation of magnetic skyrmions in the absence of applied field at room temperature.

Stabilization of chiral domain walls/skyrmions can also be extended to magnetic multilayers that do not contain a heavy metal layer. A recent example is the stabilization of chiral domain wall in graphene/Co bilayer [305]. This is in stark contrast to the Fert-Levy mechanism [37] where the DMI strength generally scales with the spin-orbit coupling strength of the adjacent heavy metal layer. Graphene with a finite spin-orbit coupling is conceptually not expected to introduce significant DMI that affects spin chirality. In a recent experiment by H. Yang et al., the presence of a graphene layer on a Co film deposited on Ru(0001) single crystal flips the sign of DMI in Co/Ru(0001) system, suggesting that graphene-induced interfacial DMI can have a similar magnitude as one at interfaces with heavy metal elements. The proposed dominating mechanism at graphene/Co interface is Rashba-type DMI [306, 307], and the DMI strength can be estimated as $d = 4\mathcal{A}\alpha_R m_e/\hbar^2$, where $\mathcal{A}$ is the exchange stiffness, $\alpha_R$ is the Rashba coefficient, $m_e$ is effective electron mass. This picture is further supported by the agreement between the theoretical value from the first-principle calculation and experimentally estimated DMI values [305]. This result implies that a combination of skyrmions and distinctive electronic properties of graphene or other 2 dimensional materials might bring interesting opportunities in the near future.

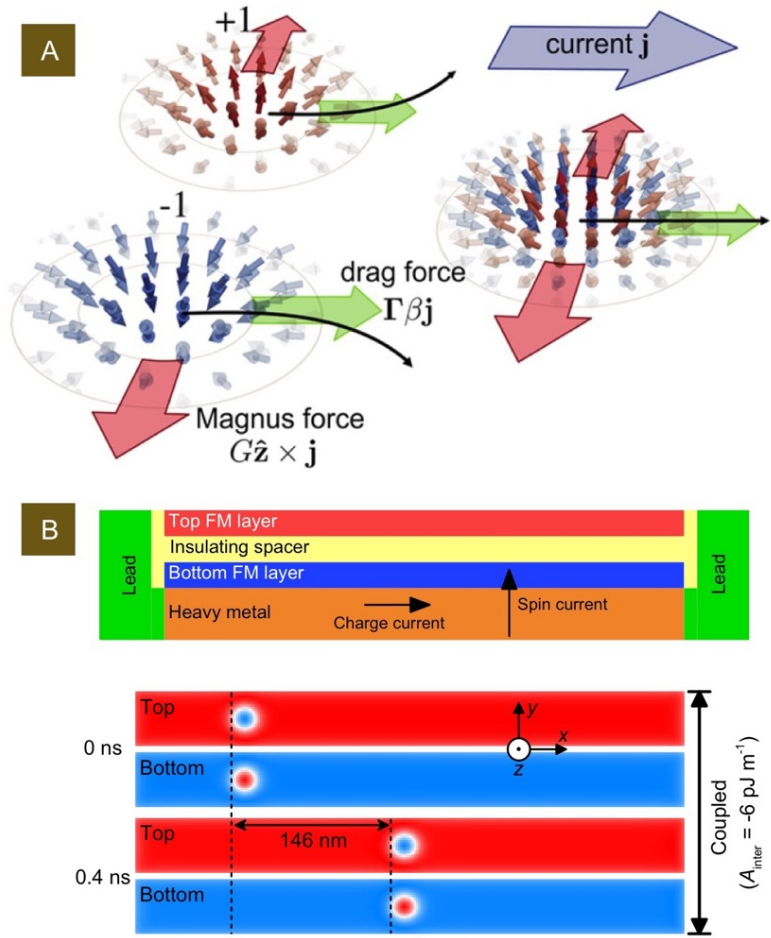

Fig. 28. Schematic illustration of the dynamics of an AFM skyrmion driven by currents. The AFM skyrmion can be envisioned as a pair of coupled FM skyrmions with opposite topological charges from two sublattices. The resultant topological Magnus force of each topological charge thus acts in opposite directions that cancel each other completely, which give rise to a linear motion of AFM skyrmion, shown in (A). Reproduced with permission [292]. Copyright 2016, American Physical Society. The dynamics of synthetic AFM skyrmions exhibit similar behavior to that shown in (B). Reproduced with permission [236]. Copyright 2016, Nature Publishing Group.

From the applied point of view, many skyrmion based innovative device concepts have been proposed such as skyrmion racetrack memory [308, 309], skyrmion-based cache memory [310], skyrmion nano-oscillators [257, 311, 312], skyrmion transistors [313], skyrmion synaptic devices [314, 315], and skyrmion logics [313, 316]. However, several important issues should be fully addressed before skyrmions can be put to use in any real applications [32, 53, 70, 235, 256, 301, 317, 318]. For reliable memory and logic performance, an on-demand skyrmion operation by electrical currents should be achieved, including generation, manipulation and annihilation [185, 186, 188]. Both experimental and theoretical efforts have been made to demonstrate a prototypical skyrmion racetrack memory. Stabilizing and driving magnetic skyrmions along the micro- or nanoscale racetrack have been realized. But a successful electrical read-out scheme has not yet been established. While magnetic skyrmions are theoretically insensitive to pinning centers due to their topological protection [238, 239, 319, 320], current densities comparable to those used to drive chiral domain walls will be required in order to achieve higher velocity above 200 m/s [26, 42]. Thus, in this respect, skyrmions possess only a limited advantage. On the other hand, one could perhaps explore the unique topological properties of magnetic skyrmion, such as topological tunneling magneto-resistive devices, topological noncollinear magneto-resistive devices [321-323]. In addition, the stability issue of a single skyrmion as information bit should also be addressed [324].

The last couple of years have witnessed a rapid development of new concepts based on magnetic skyrmions (please see review articles and references therein [40, 286, 308, 318, 325]). While their original discovery occurred in bulk materials with intrinsic chiral magnetic interactions, the realization that interfacial interactions can stabilize similar magnetic structures in magnetic multilayers opened up a broad range of new opportunities. By now these include stable magnetic skyrmions at room-temperature and above, novel pathways to skyrmion generation and manipulation, and advanced magnetic imaging capabilities that enable their investigations to length-scales well within the fundamental limits. But many open questions remain as follows:

(A) Are the topological properties beneficial or detrimental to skyrmion applications? *I.e.*, can we quantify the additional stability due to non-trivial topological charge?
(B) How energy-efficient are skyrmions in real materials and device settings? How rigid are the skyrmion spin textures under external stimuli? What is the maximum velocity of skyrmion driven by electrical current before it approaches Walker breakdown?
(C) Can the gyrotropic motion due to the topological charge be useful or is it a nuisance that would best be eliminated? Related to this, there are many other remaining fundamental science challenges. In particular, a more detailed understanding of the fundamental excitations and how they influence skyrmion dynamics is highly desirable. On the other hand, there are many applied challenges as well.

(D) Can we understand the interfacial chiral interactions to a degree that enables us to engineer them in magnetic multilayers for the desired functionality? Do random fluctuations of the microstructure matter? Can they help to reduce detrimental effects from thermal fluctuations? What are the key parameters for material optimization in magnetic multilayers? How stable skyrmion spin textures against external disturbing?

(E) Similarly, the other open frontier in skyrmion exploration is their behavior far from equilibrium. What are the limitations to their speed? To what extent do dynamical instabilities matter? Towards this end excitations with ultrafast pulses may reveal additional complexities of the competing interactions required to form magnetic skyrmions.

Going forward, the possibility to introduce chiral interactions into thin films and multilayers that already possess more complicated spin structures than ordinary ferromagnets, such as ferri- and antiferromagnets will offer distinct opportunities to test the role that topology plays for the dynamic and static stability of magnetic skyrmions. Thus, there is still a wide range of materials combinations that await to be explored. Following the rapid development of skyrmion based research, many interesting ideas and results have emerged, but not all of them could be mentioned in this review article.

To conclude, from perspectives of both the optimized materials and functional devices, this field is still young. Consorted efforts from the nanomagnetics and spintronics community are thus highly desirable in the future. With the above ideas in mind we are looking forward to the exploration of novel physics and potential applications that undoubtedly will follow the recent exploding interest in non-trivial spin textures!

**Acknowledgements**

Wanjun Jiang was supported by National Key R&D Program of China under contract number 2017YFA0206200, 2016YFA0302300, the 1000-Youth talent program of China, the State Key Laboratory of Low-Dimensional Quantum Physics, the Beijing Advanced Innovation Center for Future Chip (ICFC). Gong Chen was supported by the UC Office of the President Multicampus Research Programs and Initiatives (MRP-17-454963). Kai Liu was supported by the US NSF (DMR-1610060). Jiadong Zang was supported by the U.S. Department of Energy (DOE), Office of Science, Basic Energy Sciences (BES) under Award No. DE-SC0016424. Work carried out at the Argonne National Laboratory was supported by the U.S. Department of Energy, Office of Science, Basic Energy Science, Materials Science and Engineering Division. We wish to thank collaboration with: Andrew L. Balk, Julie A. Borchers, Tingyong Chen, Xuemei Cheng, Chia-Ling Chien, Mairbek Chshiev, Alexandre A. C. Cotta, Haifeng Du, Rafeal Dunin-Borkowski, Albert Fert, Peter

Fischer, Frank Fradin, Victor Galitski, Dustin Gilbert, Jung Hoon Han, Olle Heinonen, Song Jin, Benjamin Matthias Jungfleisch, Sang Pyo Kang, Brian J. Kirby, Hee Young Kwon, Tianping Ma, Waldemar A. A. Macedo, Brian Maranville, Arantzazu Mascaraque, Maxim Mostovoy, Naoto Nagaosa, Sergey A. Nikolaev, Alpha N'Diaye, John E. Pearson, Amanda Petford-Long, Charudatta Phatak, Daniel T. Pierce, Ziqiang Qiu, Adrián Quesada, Henrik Ronnow, Andreas Schmid, Edmar A. Soares, Mingliang Tian, Yaroslav Tserkovnyak, Yoshinori Tokura, Pramey Upadhyaya, John Unguris, Kang L. Wang, Qiang Wang, Xiao Wang, Changyeon Won, Yizheng Wu, Hongxin Yang, Guoqiang Yu, Xiuzhen Yu, Sheng Zhang, Wei Zhang, Xichao Zhang, Yuheng Zhang and Yan Zhou. Authors also wish to acknowledge fruitful discussion with Stefan Blügel, Collin Broholm, Vincent Cros, Haifeng Ding, Shi-Zeng Lin, Chang Liu, J. P. Liu, Christopher Marrows, Ivar Martin, Charles Reichhardt, Jing Shi, Cheng Song, André Thiaville, Oleg Tchernyshyov, Yayu Wang, Zhenchao Wen, Seonghoon Woo, Jiang Xiao, Jinxing Zhang, Guoping Zhao and Heng-An Zhou.